%% file: main.tex
\documentclass[nofootinbib,longbibliography,superscriptaddress,eqsecnum,tightenlines,11pt,notitlepage]{revtex4-1}

\pdfoutput=1

\usepackage{hyperref}
\usepackage{graphicx}
\usepackage{amsmath,amssymb,amsfonts,amsthm,stmaryrd,mathtools,bm}
\usepackage{multirow,bigdelim}
\usepackage{dsfont}
\usepackage{color}
\usepackage[dvipsnames]{xcolor}
\usepackage{soul}

\usepackage{tikz}
\usepackage{tikz-3dplot}
\usepackage{tikzit}
\usepackage{braket}
\input{Style.tikzstyles}
\usetikzlibrary{calc}
\usetikzlibrary{shapes.geometric,patterns,positioning} 
\usepackage{blkarray}
\usepackage{tensor}
\usepackage[normalem]{ulem}

\makeatletter
\newcounter{subsubsubsection}[subsubsection]
\renewcommand\thesubsubsubsection{\thesubsubsection .\@alph\c@subsubsubsection}
\newcommand\subsubsubsection{\@startsection{subsubsubsection}{4}{\z@}%
                                     {-3.25ex\@plus -1ex \@minus -.2ex}%
                                     {1.5ex \@plus .2ex}%
                                     {\centering\normalfont\small\textit}}
\newcommand*\l@subsubsubsection{\@dottedtocline{3}{10.0em}{4.1em}}
\newcommand*{\subsubsubsectionmark}[1]{}
\makeatother
\newcommand\scalemath[2]{\scalebox{#1}{\mbox{\ensuremath{\displaystyle #2}}}}

\newtheorem{Theorem}{Theorem}[section]

\def\be{\begin{equation}}
\def\ee{\end{equation}}
\def\ba{\begin{eqnarray}}
\def\ea{\end{eqnarray}}

\def\sl2c{\text{SL}(2,\mathbb{C})}

\newcommand{\dif}{\text{d}}

\newcommand\nn{\nonumber}
\newcommand{\q}{\quad}

\newcommand{\blue}{\color{blue}}
\newcommand{\red}{\color{red}}

\renewcommand{\d}{{\mathrm d}}

\newcommand{\tr}{{\text{tr}}}

\usetikzlibrary{patterns}
\usetikzlibrary{matrix,decorations.pathreplacing}
\pgfkeys{tikz/mymatrixenv/.style={decoration={brace},every left delimiter/.style={xshift=8pt},every right delimiter/.style={xshift=-8pt}}}
\pgfkeys{tikz/mymatrix/.style={matrix of math nodes,nodes in empty cells,left delimiter={(},right delimiter={)},inner sep=1pt,outer sep=4pt,column sep=8pt,row sep=8pt,nodes={minimum width=20pt,minimum height=10pt,anchor=center,inner sep=0,outer sep=0}}}
\pgfkeys{tikz/mymatrixbrace/.style={decorate, thick, decoration={raise=3pt}}}

\newcommand*\mymatrixbraceleft[4][m]{
    \draw[mymatrixbrace] (#1.east|-#1-#3-1.north east) -- node[right=5pt] {#4} (#1.east|-#1-#3-1.south east);
}

\newcommand*\mymatrixbracebottom[4][m]{
    \draw[mymatrixbrace] (#1.south-|#1-1-#2.north east) -- node[below=5pt] {#4} (#1.south-|#1-1-#3.north west);
}

\begin{document}

\title{ Spin-foams as semi-classical vertices:\\ gluing constraints and a hybrid algorithm }

\author{Seth K. Asante}
\email{seth.asante@uni-jena.de}
\affiliation{Theoretisch-Physikalisches Institut, Friedrich-Schiller-Universit\"at Jena, Max-Wien-Platz 1, 07743 Jena, Germany}
\author{Jos\'{e} D. Sim\~{a}o}
\email{j.d.simao@uni-jena.de}
\affiliation{Theoretisch-Physikalisches Institut, Friedrich-Schiller-Universit\"at Jena, Max-Wien-Platz 1, 07743 Jena, Germany}

\author{Sebastian Steinhaus}
\email{sebastian.steinhaus@uni-jena.de}
\affiliation{Theoretisch-Physikalisches Institut, Friedrich-Schiller-Universit\"at Jena, Max-Wien-Platz 1, 07743 Jena, Germany}

\begin{abstract}


Numerical methods in spin-foam models have significantly advanced in the last few years, yet challenges remain in efficiently extracting results for amplitudes with many quantum degrees of freedom. In this paper we sketch a proposal for a ``hybrid algorithm'' that would use both the full quantum amplitude and its asymptotic approximation in the relevant regimes. As a first step towards the algorithm, we derive a new representation of the partition function where each spin-foam vertex possesses its own coherent data, such that it can be individually asymptotically approximated. We do this through the implementation of gluing constraints between vertices, which we study numerically. We further derive an asymptotic expression for the constraints for arbitrary boundary data, including data for which there are no critical points. From this new representation we conjecture an intermediate quasi-geometric spin-foam regime describing a superposition of semi-classical vertices glued in a non-matching way via the gluing constraints.

\end{abstract}

\maketitle


\section{Introduction}

Among the multitude of existing approaches to the problem that is constructing a quantum theory of gravity, spin-foam models \cite{Perez:2012wv,Baez2000} stand out as a non-perturbative and background independent path integral formulation closely related to canonical loop quantum gravity \cite{carlobook,thomasbook}. Over recent years the field has seen a resurgence in interest particularly directed towards the numerical aspects of its models \cite{Dona:2017dvf,Dona:2019dkf,Gozzini:2021kbt,Asante:2020qpa,Han:2020npv,Han:2021kll,Huang:2022plb,Bahr:2016hwc,Allen:2022unb}, having as main motivation the need to elevate the abstract theory to a computationally viable framework capable of making consistent and physically meaningful predictions. This objective is substantially more challenging than one might expect at first sight, and the difficulty is rooted in the intrinsic computational complexity of the spin-foam state-sum
. The present work intends to contribute to the effort of making spin-foams computationally more viable in different regimes of the models. 

As previously mentioned, spin-foam models constitute a class of theories for quantum gravity based on a path-integral approach. While the different theories differ in a number of key technical aspects \cite{Perez:2012wv}, they all share a number of common features: spin-foams are formulated over a 2-dimensional cell-complex (thought of as a dual object to discrete space-time), to which one assigns group-theoretic data, usually related to unitary and irreducible representations of the symmetry group of the underlying classical gravity theory. The cell complex is commonly taken to be the Poincaré dual of a triangulation of the classical spacetime manifold, but generalizations to other types of cell-complexes exist \cite{Kaminski:2009fm}. One can then assign a state-sum to this combinatorial and group-theoretic construction, and that state-sum defines the spin-foam amplitude characterizing the system. 

Currently, one of the most studied and best understood 4d spin-foam model is perhaps the Engle-Pereira-Rovelli-Livine/Freidel-Krasnov (EPRL/FK) simplicial model \cite{Engle:2007qf,Engle:2007uq,Engle:2007wy,Freidel:2007py}, both in Euclidean and Lorentzian signatures. Its particularity resides in the type of group-theoretic data assigned to the cell-complex, which forms a subset of unitary irreducible representations of either Spin(4) or SL$(2,\mathbb{C})$, respectively, obtained via the so-called ``simplicity constraints''.
Heuristically, in the Lorentzian case these constraints are imposed by choosing a time-like normal for each tetrahedron, rendering it space-like, which singles out a subgroup isomorphic to SU$(2)$. This construction has the additional advantage of recovering, at the boundary of the spin-foam, similar states to the ones characterizing the kinematical Hilbert space of loop quantum gravity. 

The fundamental amplitudes of the EPRL/FK model, and in particular the vertex amplitude assigned to a dual 4-simplex, is well understood in different regimes. In the case of small representation labels, which we frequently refer to as the deep quantum regime, this amplitude is generically defined as a contraction of invariant tensors, i.e. intertwiners, 
which can only be efficiently computed numerically. There has recently been a concentrated effort to systematically improve and optimize such numerical algorithms \cite{Dona:2017dvf}, with a main focus on the Lorentzian SL$(2,\mathbb{C})$ EPRL model \cite{Dona:2019dkf,Gozzini:2021kbt,Dona:2022dxs}. 
These developments culminated in the creation of the efficient \verb|sl2cfoam-next| package \cite{Gozzini:2021kbt}, which e.g. unlocked GPUs as computational resources, allowed for a comparison of the full vertex amplitude with its asymptotic formula \cite{Dona:2017dvf,Dona:2019dkf,Gozzini:2021kbt} and facilitated numerical studies of 2-complexes with multiple simplices, further clarifying the so-called ``flatness problem'' \cite{Dona:2020tvv}. 

In addition to these numerical studies, our understanding of the spin-foam vertex amplitude is complemented analytically through an asymptotic approximation for sufficiently large spins. It was shown for a multitude of different spin-foam models and space-time signatures that the vertex amplitude possesses critical points, expressed in terms of coherent boundary data, which dominate in the limit of large representations \cite{Barrett:1998gs,Baez:2002rx,Conrady:2008mk,Barrett2009a,Barrett2010,Bahr:2015gxa,Kaminski2018a,Liu2019,Simao:2021qno,Dona2020}. Among other cases, these critical points correspond to geometric 4-simplices, for which the amplitude oscillates with a  Regge-type action - a discrete action of general relativity \cite{Regge:1961px}. This robust result is a strong indication that (discrete) gravity might be recovered from spin-foams in an appropriate limit, and it crucially provides a much more tractable analytical expression for the vertex amplitude. These results were used in different contexts, e.g. 2-complexes with multiple simplices, where all variables of the spin-foam - including bulk representations - were treated using asymptotic expansions \cite{Han:2017xwo,Han:2018fmu}. This regime is considered to be semi-classical as it is dominated by the solutions to the critical point equations of all variables, up to small deviations away from these points\footnote{Such deviations have been called complex critical points in some of the literature \cite{Han:2021kll}.}. 

At this stage it is not clear how wide the gap is between the quantum regime and the semi-classical one beyond a single vertex amplitude, i.e. when it is that the semi-classical regime becomes a valid approximation for large 2-complexes. A few years ago, effective spin-foam models \cite{Asante:2020qpa,Asante:2020iwm,Asante:2021zzh,Asante:2021phx,Dittrich:2022yoo} were proposed in the hope that they could serve as a tool that would help bridge that gap. The key idea is to consider each vertex as semi-classical, taking the shape of a 4-simplex. Since spin-foams are defined with area degrees of freedom and not edge lengths, neighbouring 4-simplices might not glue properly along tetrahedra, which is understood as a consequence of metric discontinuities or torsion \cite{Dittrich:2022yoo, Asante:2018wqy}. Shape matching is enforced weakly by so-called gluing constraints, assumed to be Gaussians peaked on equal shape of tetrahedra formulated in terms of dihedral angles, ultimately deriving from to the weak implementation of simplicity constraints. The authors of such models studied triangulations of several simplices and identified regimes in which they recover the classical solution of (length) Regge calculus \cite{Asante:2020iwm}.

{\centering \noindent\rule{6cm}{0.05pt} \\~\\}

This paper concerns itself with probing the intermediate regime between the full spin-foam quantum amplitude and the semi-classical one, with a special focus on $\text{SU}(2)$ BF theory and on the Lorentzian EPRL/FK model. We term this interpolating regime \textit{quasi-geometric} (in contrast with the quantum pre-geometric and the semi-classical geometric regimes), since it is dominated not only by critical points, but also by neighbourhoods of those points - where the usual geometric interpretation does not strictly apply. We shall provide throughout the paper numerical evidence supporting the existence and importance of this domain. 

Having argued for the relevance of the quasi-geometric regime, the problem arises of how to correctly take it into account when performing spin-foam calculations. To this end we sketch a proposal for a \textit{hybrid algorithm} which, when evaluating spin-foam amplitudes, would use the full quantum amplitude at sufficiently small spins, and transition to an asymptotic approximation at larger spins whenever possible. In order to make use of the usual single-vertex asymptotic amplitude in the context of a general 2-complex, we moreover restructure the spin-foam partition function via the inclusion of \textit{gluing constraints}, analogous objects to their aforementioned namesakes from effective spin-foams; such constraints benefit the analysis by disentangling the boundary data of each individual vertex. 

The main focus of this article, as a first step towards the eventual construction of the hybrid algorithm, is the study of the gluing constraints we introduce.  Our analysis is both numerical and analytical: we provide a numerical characterization of the gluing constraints for both the Lorentzian EPRL/FK and $\text{SU}(2)$ BF models, and, in order to understand the structure of the constraints, we extend the usual methods of asymptotic analysis of spin-foams to provide also an asymptotic expression of the gluing constraints away from the critical points - thus obtaining a function of general boundary data. We conjecture that such an analysis can perhaps be extended to the vertex amplitude itself, and we leave the study of this possibility for the future. 


The text is organized as follows: in section \ref{hybrid_algo} we begin with a brief introduction of spin-foam models and discuss the coherent state representation of their partition function. From its structure we argue that the amplitude is not immediately suitable for efficient numerical methods and we propose to slightly rearrange it, explaining the notion of gluing constraints and sketching the idea of the hybrid algorithm. We define the gluing constraints for SU$(2)$ BF theory and for the Euclidean and Lorentzian EPRL models in section \ref{gluing_constraints}. In section \ref{asymptotics_gluing} we compute their asymptotic expansion on and away from critical points, which we compare to the full numerical results in section \ref{numerical_analysis}. We close with a general discussion in section \ref{discussion}. Technical details and conventions, like the parametrization of Euclidean tetrahedra in terms of areas and angles, are relegated to the appendix \ref{AppendixS}.

\section{spin-foams in a nutshell: coherent state representation and a hybdrid algorithm} \label{hybrid_algo}


In the following we give a broad and rather conceptual overview of spin-foam models \cite{Perez:2012wv} in a model-agnostic manner, as we will consider different theories throughout this paper. spin-foam models define a transition amplitude for gravity using representation labels of a gauge group $\mathcal G$, which color a 2-dimensional cell-complex $\Delta^*$ dual to a triangulation $\Delta$ of the underlying space-time manifold $\mathcal M$. See table \ref{tab:dual} for a dictionary between the  2-complex and its dual triangulation. In four dimensions one assigns to every face dual to a triangle a label $\chi_t$ of the unitary irreducible representation $D^{\chi_t}$ of $\mathcal G$.
\begin{table}[h!]
\centering
\begin{tabular}{  |c|c| } 
\hline 
\,\,\,2-complex \, $\Delta^*$ \,\,\, & \,\,\, triangulation \, $\Delta$ \,\,\, \\ \hline \hline
vertex $v$ & simplex $\sigma$ \\  
edge $e$ & tetrahedron $\tau$ \\ 
face $f$ & triangle $t$ \\ 
 \hline
\end{tabular}
\caption{Cells of a 2-complex $\Delta^*$ dual to a triangulation $\Delta$ } 
\label{tab:dual}
\end{table}
One also considers holonomies from one vertex to another via assignments of group elements $g_\tau \in \mathcal G$ to edges dual to tetrahedra. Each of these edges is shared by four faces, corresponding to the four triangles of a tetrahedron in $\Delta$. Therefore each edge in $\Delta^*$ is associated with a tensor product of four representation spaces on which the single holonomy $g_\tau$ acts. Since we are considering a path integral, we integrate over all possible holonomies per edge, such that to each edge the following object is assigned:
\begin{equation}
    \mathcal{P}^{\{\chi_t\}} = \int_\mathcal{G} \dif \mu(g_\tau) \bigotimes_{t=1}^4 D^{\chi_t}(g_\tau) \, .
\end{equation}
Here $\dif \mu(g_\tau)$ denotes the Haar measure of $\mathcal{G}$. The map $\mathcal{P}$ constitutes a projector onto the gauge-invariant subspace of the tensor product of representation spaces. We represent this object pictorially by four wires and a box over the wires: the wires stand for group representations, while the box denotes the group integration, as in the diagram $\scalebox{0.55}{ \input{gluenocoh.tikz}}$. The explicit form of these projectors differs depending on the model in question.

At the vertices of the spin-foam, which are dual to 4-simplices, five edges meet. When represented pictorially as wires, we connect up the wires at the vertex according to the same combinatorial structure of tetrahedra at the boundary of a 4-simplex; wires corresponding to the same triangle get connected. Here connecting means identification and summation over indices of representation spaces. The group integrations associated to edges of $\Delta^*$ can be explicitly performed, giving rise to a sum over orthonormal invariant tensors and their duals, the intertwiners $\iota_\tau$, which are then split and associated to the vertices. Hence, for a given choice of representations and intertwiner labels, we assign a particular contraction of intertwiners to a vertex, resulting in a vertex amplitude. 
In the intertwiner basis, frequently called the spin-network basis, the partition function is then given by

\begin{equation} \label{eq:spin_foam_part_SNW}
    {\cal Z} = \sum_{ \{\chi_t\}, \{\iota_\tau\} } \prod_t d_{\chi_t} \prod_\sigma \mathcal{A}_\sigma \, , 
\end{equation}
where $\mathcal{A}_\sigma$ denotes the vertex amplitude dual to the simplex $\sigma$ and $d_{\chi_t}$ denotes the dimension of the representation $\chi_t$ (or some appropriate notion of dimension in case the representation space is infinite-dimensional).


\subsection{Coherent state representation}

Instead of performing the integration over group elements explicitly and using the thus obtained intertwiners to split the partition function into local vertex amplitudes, one may also parametrize the amplitude in a different manner by inserting resolutions of identity in the middle of wires connecting different vertices. A specially useful basis for these identities is the coherent state one, which, for the class of spin-foam models considered here, is constructed via $\text{SU}(2)$ states. In the vector space $\mathcal{H}^j$ associated to the unitary and irreducible representation of $\text{SU}(2)$, one has the relation
\be \label{Idcoh}
\mathds 1_{j} = d_{j}\int_{\rm SU(2)/U(1)}  \d h \,  | j,  h \rangle \langle j,  h | \equiv \scalebox{1}{\input{resolution.tikz}}^j\,,
\ee
where $d_{j} = 2j+1 $ is the dimension of $\mathcal{H}^j$ and $| j, h \rangle$ are coherent states, i.e. states of the form
\be
\ket{j,h}=D^j(h) \ket{\text{ref}_j} \equiv \scalebox{1}{\input{cstate.tikz}}^{j,h}\,, \quad h \in \text{SU}(2)/\text{U}(1)\,.
\ee
It is common to take the reference state $\ket{\text{ref}_j}$ to be either a maximal weight state $\ket{j ,j}$ or a lowest one $\ket{j, -j}$, though the identity holds generally. 
Although such coherent states form a basis only of the representation spaces of $\text{SU}(2)$, they turn out to induce bases for all spaces relevant for the spin-foam models studied in this article, where the gauge group is either $\text{SL}(2,\mathbb{C})$ or $\text{SU}(2)$. The construction for the special linear group is reviewed in section \ref{gluesl}.

Choosing to expand the spin-foam amplitudes in terms of $\rm SU(2)$ coherent states, the partition function generally reads
\be \label{TAmp}
{\cal Z} = \sum_{ \{\chi_t\} } \int \prod_{t} \prod_{t,\tau} d_{\chi_{t}} \d h_{t,\tau}  \prod_{\sigma }  A_\sigma\, ,
\ee
where there is a coherent state label per pair of triangle $t$ and tetrahedron $\tau$. The coherent vertex amplitude $A_\sigma$ is in turn given by 
\be \label{vertamp}
A_\sigma = \int_\mathcal{G} \prod_\tau \dif \mu( g_\tau) \prod_{\tau,\tau'}  \langle {J \,\triangleright} \chi_t,   h_{t,\tau} \, |D^{\chi_t}(g_{\tau}^{-1} g_{\tau'}) |\chi_t,  h_{t,\tau'} \rangle \equiv \scalebox{0.55}{ \input{vertex_left.tikz}} \,.
\ee
On the right hand side stands the standard graphical representation (ignoring orientations) of the coherent vertex amplitude. The lines with dots at the ends represent inner products of coherent states connecting different faces. Solid rectangles represent group integrations over the group representations, as mentioned previously. We remind the reader that the above expression is only schematic, as a number of technical details must be observed depending on the model at hand (e.g. one needs to regulate the $\text{SL}(2,\mathbb{C})$ model by removing one of the group integrations). 


The coherent state parametrization is useful for studying the asymptotics of vertex amplitudes \cite{Conrady:2008mk,Barrett2009a,Barrett2010,Barrett:2009as, Liu2019, Simao:2021qno, Kaminski2018a}, often also referred to as the semi-classical limit or ``large $j$ / large representation limit''. There one considers the partition function \eqref{TAmp} for a single vertex and fixed boundary and computes the stationary phase approximation of the integral over (several copies of) the group $\mathcal{G}$. To this end the inner product of coherent states is exponentiated to an action, whose critical and stationary points dominate if all representations $\chi_t$ are large. It is a robust and frequently derived result that these critical points correspond to the boundary data of geometric 4-simplices (among other solutions called vector geometries \cite{Barrett2009a,Dona:2017dvf}), where the associated vertex amplitude oscillates with the Regge action of this 4-simplex \cite{Regge:1961px}. This is an encouraging finding and an indication that continuum gravity might be recovered from spin-foams in a suitable limit.


We now comment on the practical matter of evaluating the coherent amplitude for multiple vertices. Each integration over the coherent state variables $\d h_t$ in the partition function \eqref{TAmp} is performed over a pair of vertex amplitudes, since for two vertices $\sigma,\sigma'$ sharing a common edge $\tau$ both vertex amplitudes $A_\sigma$ and $A_{\sigma'}$ carry the same label $h_{t,\tau}$. Computing these integrals is challenging for two reasons: the primary reason are the integrals themselves, which are multidimensional and highly oscillatory, in particular for large representations; for $\text{SU}(2)$ the labels $h_\tau$ are normalized vectors $\vec{h} \in S^2$ \cite{Livine:2007vk}, and hence there is an eight-dimensional integral per bulk edge. Clearly this gets quickly overwhelming. 
The second reason has to do with the non-locality of coherent data integrations, which makes it difficult to study the asymptotic regime of the full spin-foam amplitude for more than one vertex. As an example, consider an equilateral vertex amplitude, i.e. all the spins are equal. One dominant contribution is an Euclidean equilateral 4-simplex \cite{Dona:2017dvf}, whose five tetrahedra are also equilateral. Consider gluing this vertex to another vertex, whose remaining representations are not the same, potentially giving rise to a critical point in which the shared tetrahedron is not equilateral. In the coherent state representation the coherent data of glued vertex amplitudes must be equal, such that the dominant contribution from the integral over coherent data is determined by the overlap of both vertex amplitudes. This contribution might well come from a configuration which agrees with neither critical point of the vertex amplitudes. While accounting for this is not impossible, it requires one to compute the overlap of glued coherent vertex amplitudes, which in turn must be known sufficiently away from their critical points, and to repeat this procedure for all remaining neighbouring vertex amplitudes - but, as far as we are aware, a general asymptotic formula for the coherent vertex amplitude away from its critical points is currently not known. One could alternatively study the asymptotic amplitude of the full 2-complex, but this is only possible in a case-by-case basis. On the other hand, the insights from the asymptotic analysis of single vertex amplitudes, i.e. the dominance of critical points, suggest that it should be possible to optimize these integrations, e.g. identify regions of coherent data that more significantly contribute. In this article we present a proposal for such a method, which might eventually lead to an efficient numerical hybrid algorithm. We discuss its idea in the following.

\subsection{Idea of a hybrid algorithm} \label{hybrid_alg}

To disentangle the coherent data of neighbouring vertices, we adopt the concept of \textit{gluing constraints} from effective spin-foam models \cite{Asante:2020qpa,Asante:2020iwm,Asante:2021zzh,Asante:2021phx}, and derive these constraints directly from the actual models. Starting from the coherent state representation, we again insert a resolution of identity on each edge. 
This procedure doubles the coherent data (and integrations) per edge, such that each vertex amplitude carries its own set of data $n_\tau$. In this manner every pair of vertices in the bulk is connected by an object which encodes how well their respective boundary data fit together: the gluing constraints. Such a pair of vertices would then schematically take the form
\begin{equation}
    \scalebox{0.55}{ \input{vertex_left.tikz} \input{gluing.tikz} \input{vertex_right.tikz}}\,,
\end{equation}
with the middle diagram representing the constraints. In section \ref{gluing_constraints} we will clarify the diagram above by explicitly defining different variants of these constraints, specifically gauge-variant and gauge-invariant types (equivalent at the level of the partition function), for different symmetry groups and spin-foam models. 

Having each vertex amplitude equipped with its own set of coherent data, one can apply the well-studied asymptotic analysis \textit{per vertex}:
for sufficiently large spins - a condition the precise meaning of which must be carefully studied - the integral over coherent data should only receive contributions from (small regions around) the critical points of the vertex amplitudes, which can furthermore be approximated by the semi-classical formula on the critical points. In general, the critical points of vertices will not match, and this is accounted for in the gluing constraints. Hence we conjecture the existence of an additional regime in spin-foams models, where the vertices exhibit semi-classical, geometric features, but the overall geometry still fluctuates and shape matching of glued vertices is not strongly enforced. We thus expect that the landscape of spin-foam configurations can be split roughly into three domains:
\begin{itemize}
    \item \textit{Pre-geometric regime} - superposition of quantum geometric building blocks;
    
    For the smallest representations, spin-foam amplitudes describe genuine quantum geometric objects best understood in the orthonormal spin network basis. The vertex amplitudes must be numerically computed for all representation and intertwiner labels and summed over. This process gets more costly with increasing representation labels.
    
    \item \textit{Quasi-geometric regime} - superposition of semi-classical geometries;
    
    For large enough spins, the vertex amplitudes can be approximated by their asymptotic expansion and only small regions around critical points contribute. Therefore, we have a superposition of local critical point configurations, corresponding to degenerate, vector and Regge geometries, glued in a non-matching way. For an appropriate domain of representation labels, non-matching vertices corresponding to small neighbourhoods around critical points are non-negligible in the amplitude. 
    
    \item \textit{Geometric} regime - globally semi-classical geometries;
    
    Increasing representation labels further eventually results in only shape-matching configurations contributing to the path integral substantially.  The amplitude of the full 2-complex is dominated by the global critical points, i.e. those of the entire 2-complex. As before, for finite representations small regions around these global critical points might still be non-negligible. 
\end{itemize}

Partial evidence for the existence of a quasi-geometric regime already exist in the literature. The convergence of the semi-classical approximation of the vertex amplitude to the full coherent amplitude was shown in recent years \cite{Dona:2017dvf,Gozzini:2021kbt} for different models, as was numerically proven the exponential suppression of amplitudes without a critical point. In section \ref{sec:numerical_vertex_amplitude} we expand on this by considering configurations around critical points, finding they are progressively less suppressed the closer they are to criticality. While more evidence is required to determine when  quasi-geometry arises, we hope to use this regime to help bridge the gap between quantum and semi-classical physics and augment existing numerical algorithms. Our proposal is to develop a ``hybrid algorithm'', following the idea of the ``Chimera'' algorithm in loop quantum cosmology \cite{Diener:2013uka}, which uses the full quantum dynamics when necessary and switches over to semi-classical, less costly, algorithms as soon as those provide a good approximation. In this regime, the hybrid algorithm would compute the semi-classical amplitudes vertex-wise, and then pair-wise match them using the gluing constraints. 

\subsection{Numerical studies of coherent vertex amplitude} \label{sec:numerical_vertex_amplitude}

The purpose of this subsection is two-fold: firstly, we compare the full coherent amplitude of $\text{SU}(2)$ BF to its semi-classical approximation, in order to recover existing results and emphasize our argument\footnote{First comparisons for higher-valent building blocks were also recently presented in \cite{Allen:2022unb}}; for the Lorentzian EPRL model, convergence with the asymptotic approximation was studied in great detail in \cite{Dona:2019dkf,Gozzini:2021kbt}. Secondly, for both types of models, we parametrize slight deviations of boundary data away from the critical points and demonstrate the exponential suppression of the vertex amplitude as representations are scaled up uniformly. All numerical results for the $\text{SU}(2)$ BF model were obtained from a numerical code written in the \verb|Julia| programming language, by contracting a $\text{SU}(2)$ $\{15j\}$-symbol against the overlap of coherent and spin network intertwiners. The simulations were performed on the Ara Cluster at FSU Jena. For the Lorentzian EPRL model, we used the package \verb|sl2cfoam-next| \cite{Gozzini:2021kbt} and performed the calculations on a consumer level laptop.

\subsubsection{Brief recap of numerical tests of semi-classical formula}

\begin{figure}
    \centering
    \includegraphics[width=0.5\textwidth]{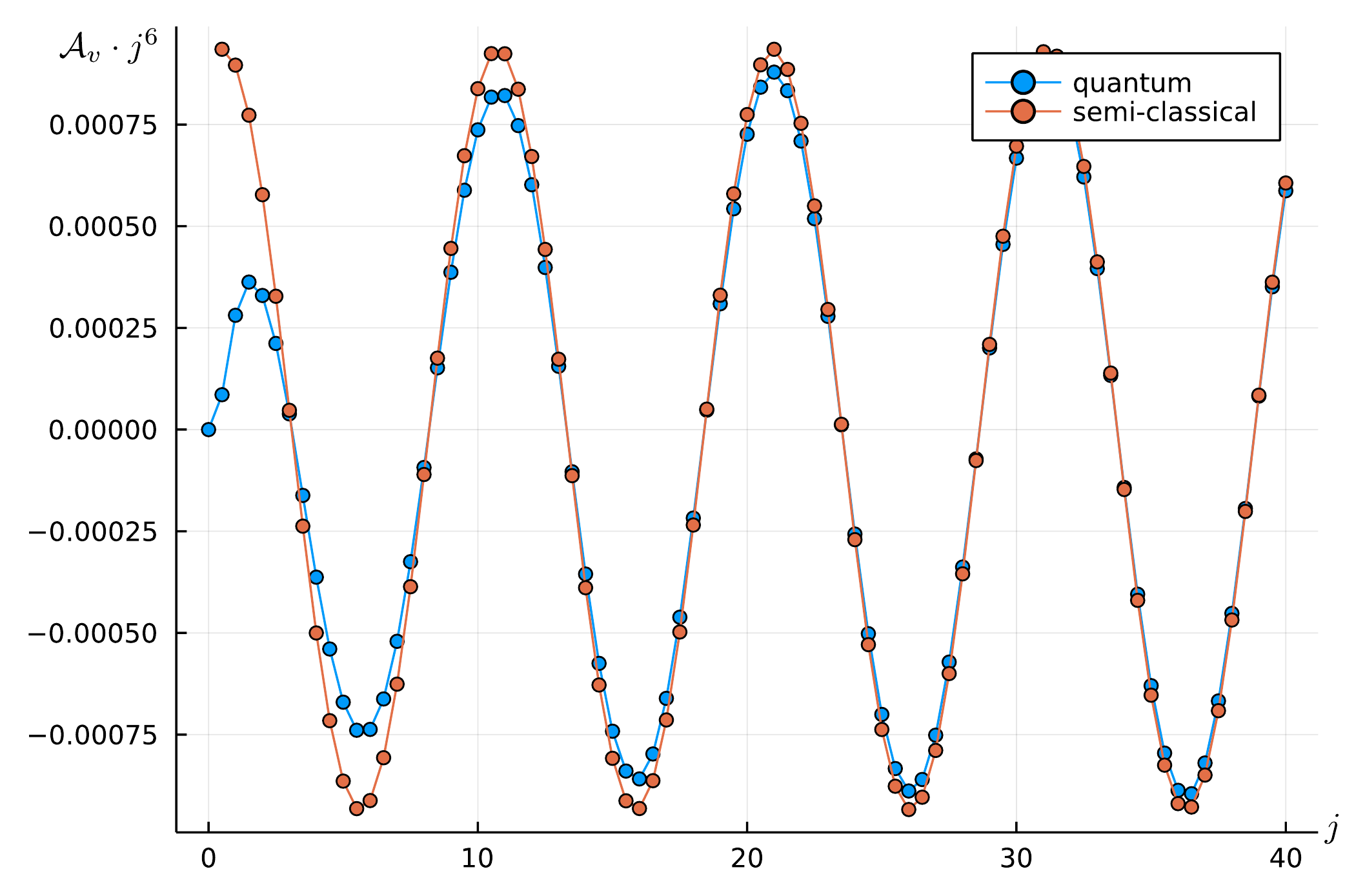}
    \caption{Comparison of the full coherent amplitude and its semi-classical approximation for equilateral boundary data up to $j=40$.}
    \label{fig:semi_compare}
\end{figure}

Before we explore the exponential suppression away from critical points, let us begin with a consistency check of our new algorithm to compute the coherent $\text{SU}(2)$ vertex amplitude. We compare the (rescaled) coherent amplitude to its semi-classical approximation for equilateral boundary data  with spins up to $j=40$ in fig. \ref{fig:semi_compare}, where we have multiplied both amplitudes by  $j^6$ to compensate the polynomial suppression of the vertex amplitude. The agreement rapidly improves as we increase all spins, and both amplitudes are only distinguishable at the maxima of oscillations. These results are in agreement with the findings in \cite{Dona:2017dvf}.
We have additionally studied isosceles 4-simplices as boundary data, the results for which we omit since they also agree with previous reports. Again, for the case of the Lorentzian EPRL model we refer to \cite{Gozzini:2021kbt}.

\subsubsection{Examples of suppression}

At the level of a single vertex amplitude, little is known about an analytical formula \textit{away} from the critical points. 
Interestingly, such an analsys was done for bulk representations in spin-foams consisting of multiple vertices in the context of the so-called ``flatness problem'' \cite{Bonzom:2009hw,Magliaro:2011qm,Hellmann:2012kz,Hellmann:2013gva,Han:2013hna,Oliveira:2017osu,Engle:2020ffj}. 
In \cite{Han:2021kll} the authors show in case of the $\Delta_3$-triangulation (where three 4-simplices share one bulk triangle and all edges are part of the boundary) that for large but finite representations there are non-negligible contributions from configurations away from critical points, interpreted as complex critical points. In contrast to real critical points, which enforce a vanishing deficit angle in the associated triangulation, the complex ones allow for small but non-vanishing angles and contribute significantly to the path integral in an appropriate regime. This is in line with numerical results in \cite{Dona:2020tvv,Gozzini:2021kbt}, where exponential suppression was shown for boundary data with a non-vanishing deficit angle in the bulk. We stress however that these complex critical points are sub-leading relative to the actual critical configurations, and they must eventually become negligible as the representations are further increased.

In order to explore the amplitudes around critical points, deviations from criticality are understood as follows: starting from a critical configuration, e.g. an equilateral Euclidean 4-simplex or an isosceles Lorentzian 4-simplex, we modify the boundary data in two different ways. The first variant introduces a violation of the closure condition in the normal vectors associated to a tetrahedron. These violations get more and more severe, leading to a stronger exponential suppression. In the second case we demand closure for all tetrahedra, but change their dihedral angles away from the critical point\footnote{To be precise, non-closing data also violate non-matching; thus we expect non-closing data to show more suppression, but both cases cannot be readily compared.}. Thus the shapes of triangles of glued tetrahedra do not match, resulting in an exponential suppression. Again, we expect a more severe violation to result in a stronger exponential suppression compared to the critical point. The parameters used for non-closing data are collected in table \ref{tab:non-closing}, and the ones used for non-matching data can be found in table \ref{tab:non-matching}. 

\begin{table}[]
    \centering
    \begin{tabular}{|c|c|c|c|c|}
    \hline
        Case & $\vec{h}_1$ & $\vec{h}_2$ & $\vec{h}_3$ & $\vec{h}_4$ \\
        \hline
        equilateral & $(0,0,1)$ & $(0.0, 0.9428, -0.3333)$ & $(0.8165, -0.4714, -0.3333)$ & $(-0.8165, -0.4714, -0.3333)$ \\
        \hline
        non-closing 1 & $(0,0.1411,0.99)$ & $(0.0, 0.9428, -0.3333)$ & $(0.8165, -0.4714, -0.3333)$ & $(-0.8165, -0.4714, -0.3333)$\\
        \hline
        non-closing 2 & $(0,0.4359,0.9)$ & $(0.0, 0.9428, -0.3333)$ & $(0.8165, -0.4714, -0.3333)$ & $(-0.8165, -0.4714, -0.3333)$\\
        \hline
        non-closing 3 & $(0,0.6,0.8)$ & $(0.0, 0.9428, -0.3333)$ & $(0.8165, -0.4714, -0.3333)$ & $(-0.8165, -0.4714, -0.3333)$ \\
        \hline
    \end{tabular}
    \caption{parametrization of non-closing boundary data with respect to an equilateral tetrahedron.}
    \label{tab:non-closing}
\end{table}

\begin{table}[]
    \centering
    \begin{tabular}{|c|c|c|}
    \hline
        Case & $\Phi_1$ & $\Phi_2$ \\
        \hline
        equilateral & $-\frac{1}{3}$ & $-\frac{1}{3}$ \\
        \hline
        non-matching 1 & $-\frac{1}{3}$ & $-\frac{1}{3} + 0.01$\\
        \hline
        non-matching 2 & $-\frac{1}{3}$ & $-\frac{1}{3} + 0.1$ \\
        \hline
        non-matching 3 & $-\frac{1}{3} + 0.1$ & $-\frac{1}{3} + 0.1$  \\
        \hline
    \end{tabular}
    \caption{parametrization of non-matching boundary data with respect to an equilateral tetrahedron. A coherent tetrahedron is parametrised by the areas of its four triangles, here all equilateral, and two dihedral angles, as defined in appendix \ref{AppendixS}. }
    \label{tab:non-matching}
\end{table}

\subsubsubsection{In $\text{SU}(2)$ BF theory}



\begin{figure}
    \centering
    \includegraphics[width=0.45 \textwidth]{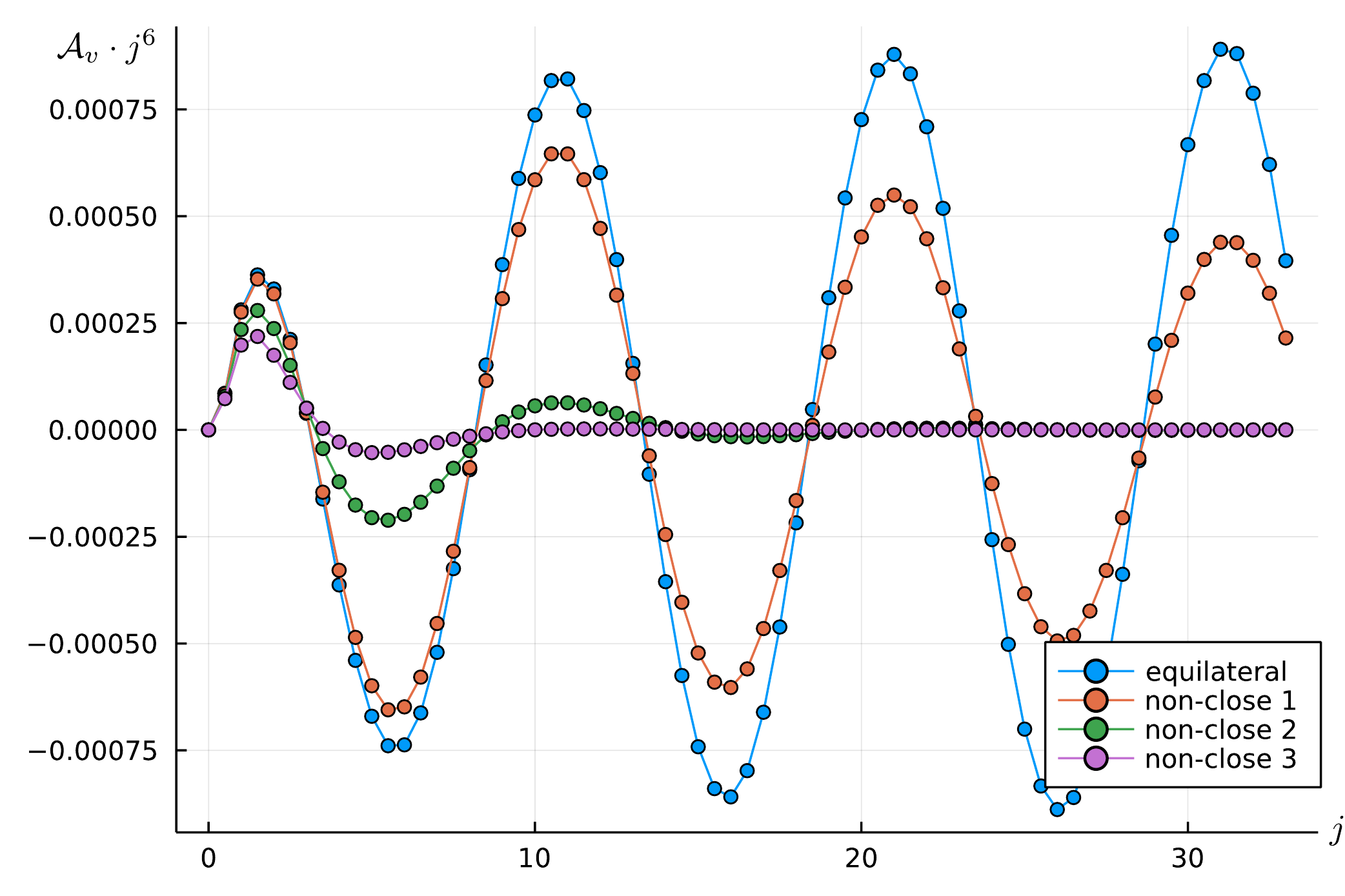} \; \;
    \includegraphics[width=0.45 \textwidth]{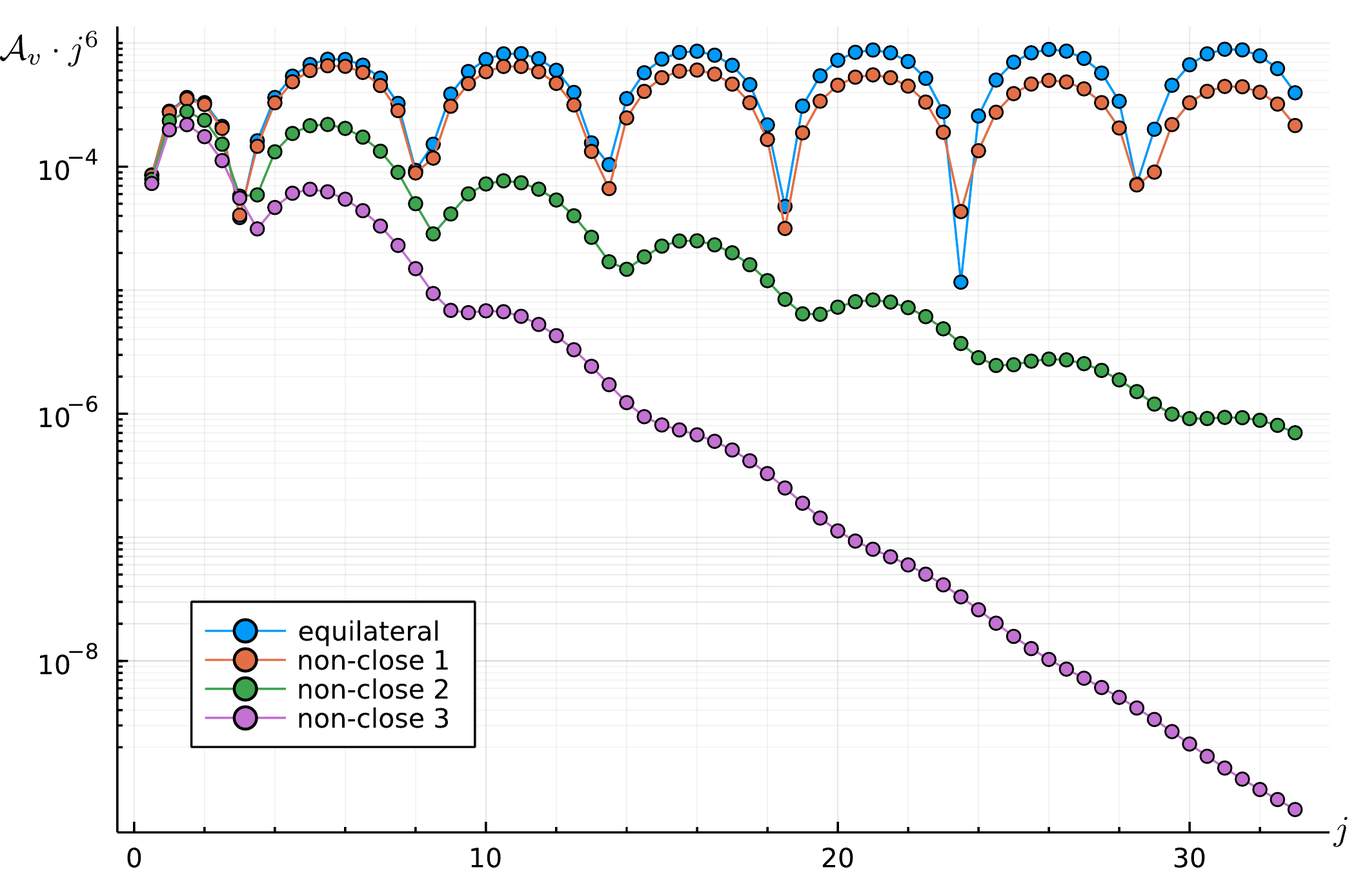}
    \caption{Comparison of coherent $\text{SU}(2)$ BF theory vertex amplitudes for equilateral and non-closing boundary data (close to equilateral).}
    \label{fig:non_close_all_su2}
\end{figure}

\begin{figure}
    \centering
    \includegraphics[width= 0.45 \textwidth]{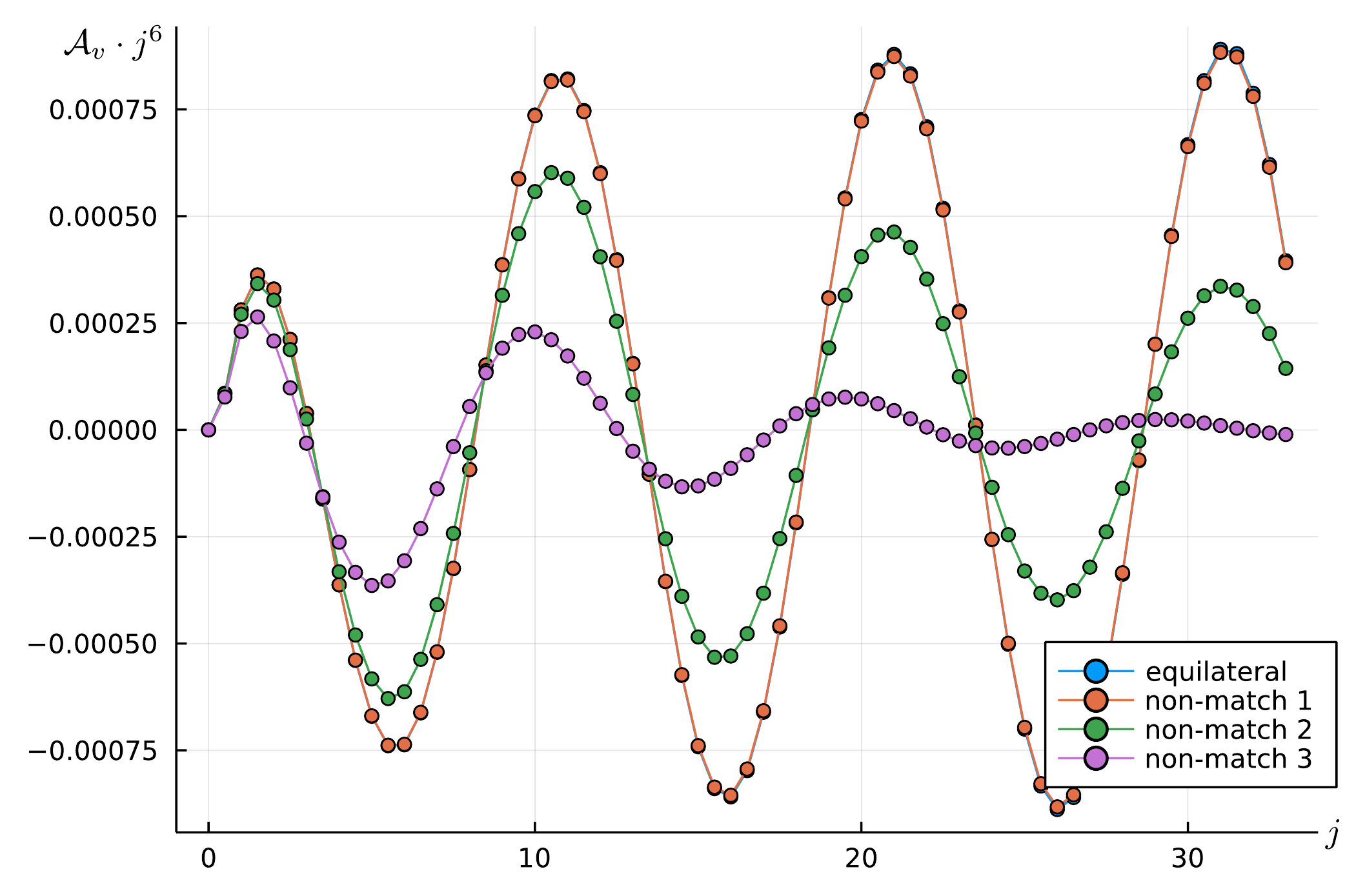} \; \;
    \includegraphics[width= 0.45 \textwidth]{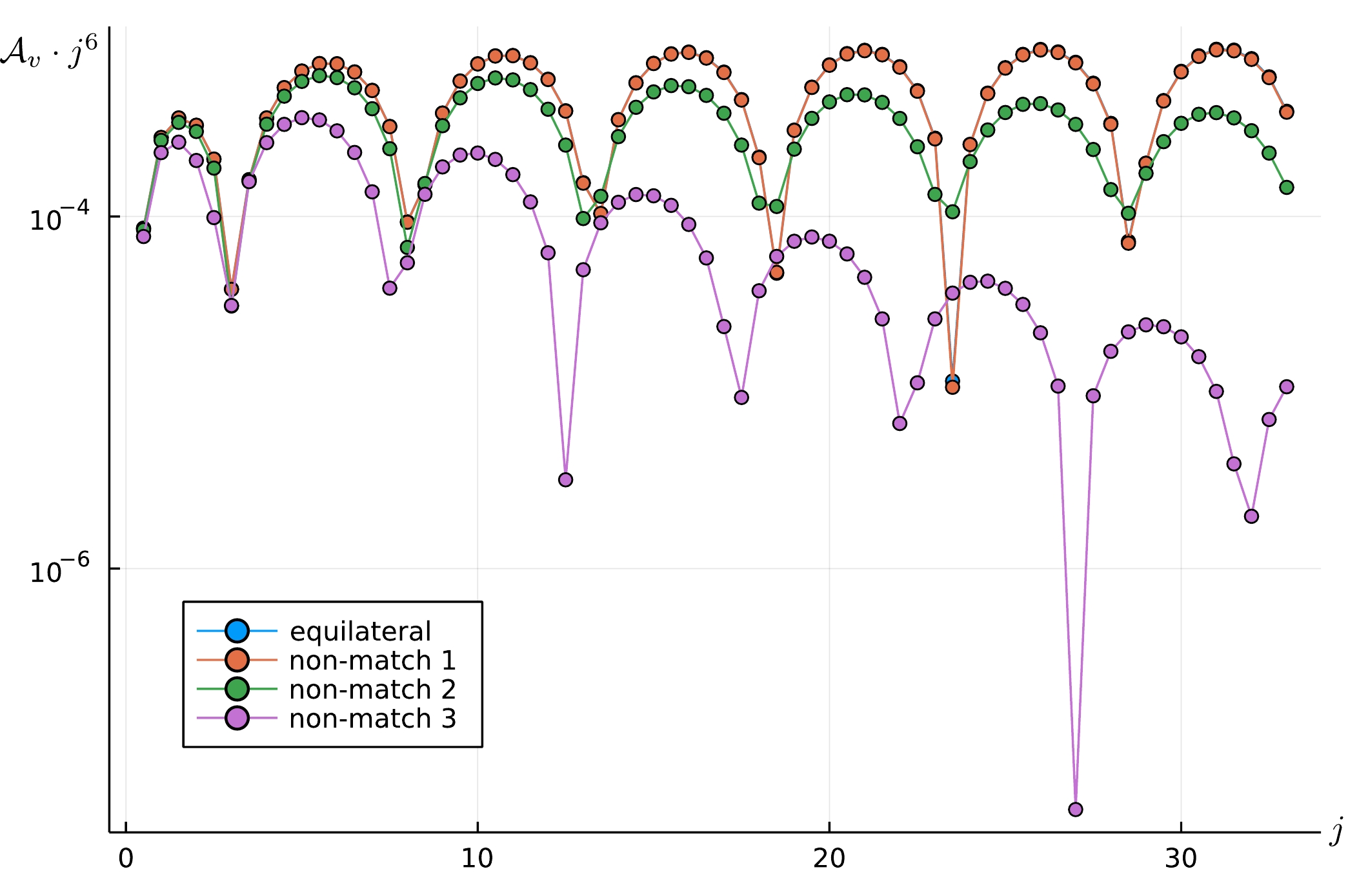}
    \caption{Comparison of coherent $\text{SU}(2)$ BF theory vertex amplitudes for equilateral and non-matching boundary data (close to equilateral).}
    \label{fig:non_match_all_su2}
\end{figure}

As a first example, consider an equilateral Euclidean 4-simplex, in which all tetrahedra deviate from equilateral tetrahedra, either by non-closure or non-matching. While the deviations are rather small, we expect a significant decline of the amplitude, since all five tetrahedra deviate. The results for non-closing tetrahedra are shown in fig. \ref{fig:non_close_all_su2}, and those for non-matching ones can be found in fig. \ref{fig:non_match_all_su2}. As expected, the exponential suppression is clear for non-closing data: the two larger deviations suffer a very strong suppression, such that the amplitudes are roughly an order of magnitude smaller compared to the equilateral case already at $j=10$. Moreover, the oscillations cease in these cases. For the smallest non-closing deviation studied, however, we see a much weaker exponential suppression compared to the rescaled critical amplitude and persisting oscillations, which will probably only seize at even larger spins.

For non-matching data we observe a similar behaviour, but overall with less exponential suppression. Indeed, for the smallest deviation studied, barely any decay is visible even though we deviate all tetrahedra and study the amplitude up to spins $j > 30$. These numerical results thus suggest that for a single vertex amplitude and large but finite representations we can always find a (sufficiently) small region around the critical points in which the exponential suppression is (still) negligible, which is in line with the results found in \cite{Han:2021kll} for the bulk deficit angle of the $\Delta_3$ triangulation. As we increase the spins the region around the critical point shrinks, since stronger deviations get more and more suppressed. Note that in particular the cases we studied in fig. \ref{fig:non_close_all_su2} are still close to equilateral tetrahedra, and the exponential suppression for strongly violating data is more severe. 

The final case we consider is an unusual one. In \cite{Dona:2019dkf}, the authors study the $\text{SL}(2,\mathbb{C})$ EPRL coherent vertex amplitude for a Lorentzian isosceles 4-simplex, whose tetrahedra are all space-like. The boundary data corresponding to this 4-simplex is a critical point for Lorentzian EPRL, and below we will recover some of the authors' results. Before doing so, we consider the same boundary data in the context of $\text{SU}(2)$ BF theory, where it does not correspond to a critical point and is exponentially suppressed. Still, this data is interesting for a different reason: a key ingredient in efficiently computing the $\text{SL}(2,\mathbb{C})$ EPRL vertex amplitude is to express it as a sum of $\text{SU}(2)$ $\{15j\}$ symbols for auxiliary labels contracted with matrices encoding the action of boosts \cite{Speziale:2016axj}. The infinite sum over the auxiliary labels starts at the label $j$ of the boundary spins. Therefore, a $\text{SU}(2)$ $\{15j\}$ symbol evaluated for Lorentzian boundary data is related to the first term appearing in the expansion the full $\text{SL}(2,\mathbb{C})$ vertex amplitude.

\begin{figure}
    \centering
    \includegraphics[width = 0.45 \textwidth]{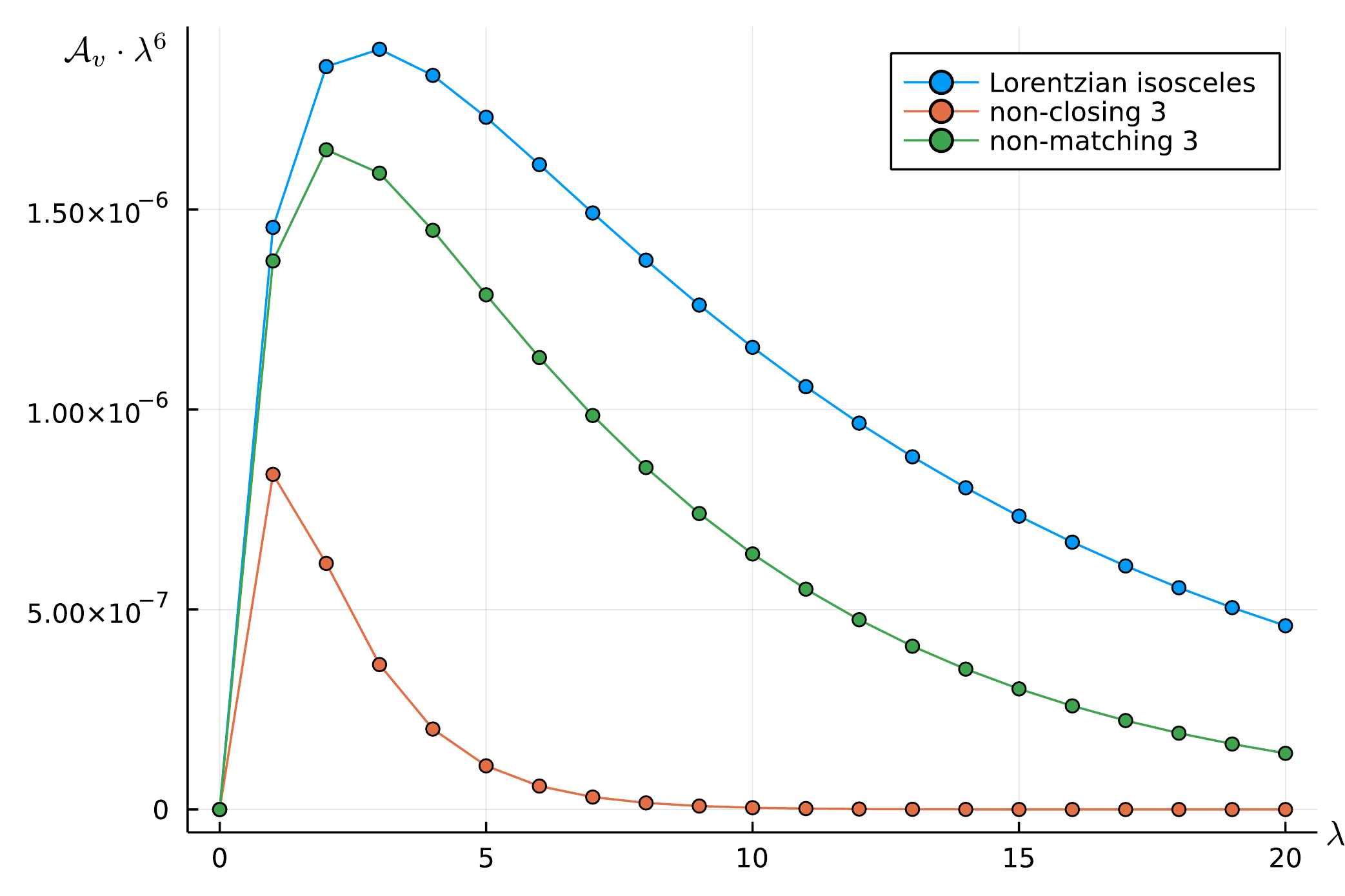} \; \;
    \includegraphics[width = 0.45 \textwidth]{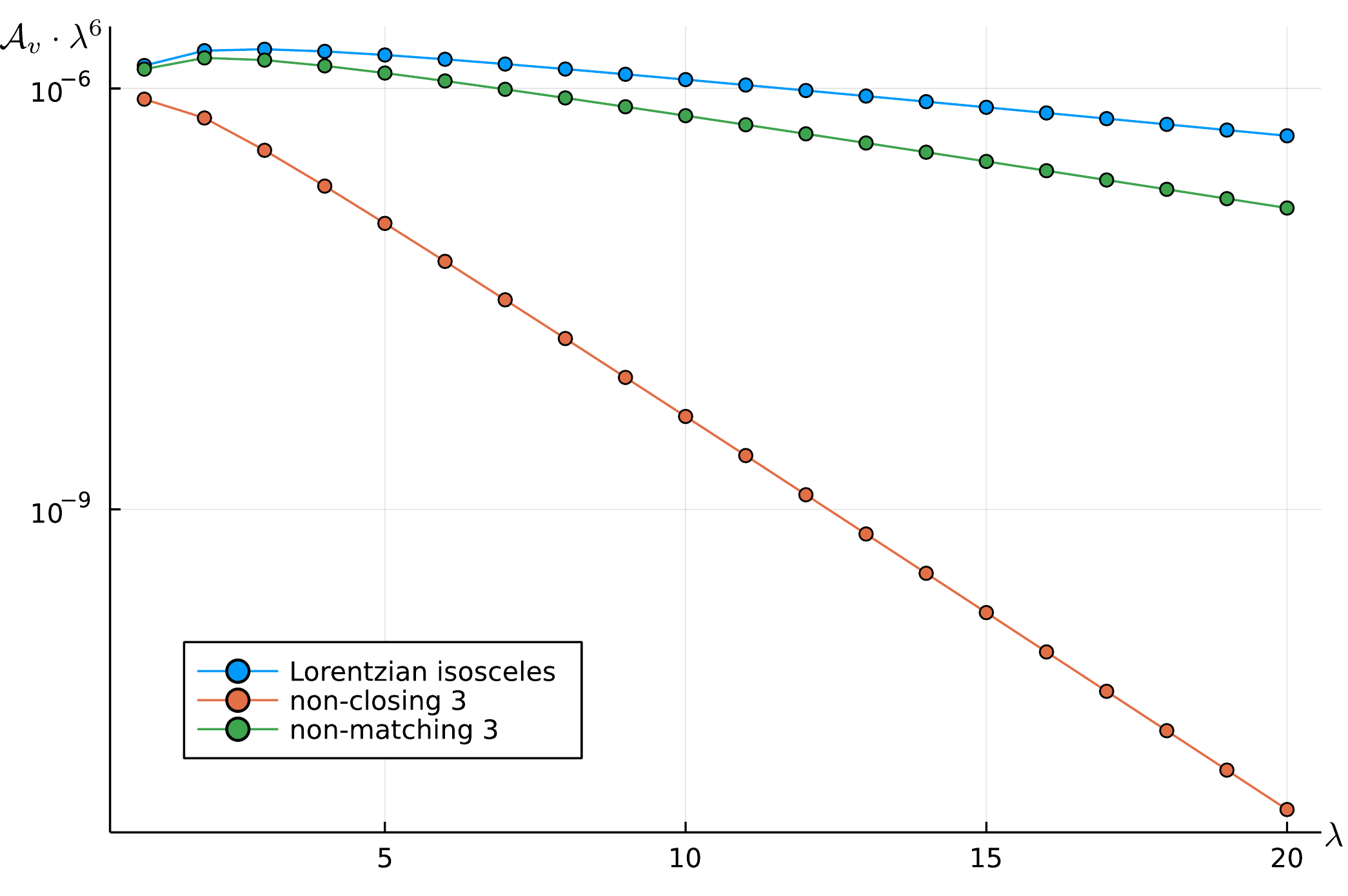}
    \caption{Comparison of $\text{SU}{2}$ BF theory vertex amplitudes for boundary data corresponding to an isosceles Lorentzian 4-simplex and non-closing / non-matching deviations away from it on a single tetrahedron.}
    \label{fig:Lorentzian_SU2}
\end{figure}

In fig. \ref{fig:Lorentzian_SU2} we plot the coherent $\text{SU}(2)$ vertex amplitude evaluated for the isosceles Lorentzian 4-simplex, labeled by a set of ten spins $\{5,5,5,5,2,2,2,2,2,2\}$. The first four spins correspond to an equilateral tetrahedron, while the remaining spins label four isosceles tetrahedra with spins $\{5,2,2,2\}$. We have additionally considered two configurations, in which we replace the equilateral tetrahedron either by the largest non-closing or non-matching configurations used before. Since none of these configurations correspond to a critical point of $\text{SU}(2)$ BF theory, none of them oscillate and they all decay exponentially. Still this Lorentzian 4-simplex is singled out even in $\text{SU}(2)$ BF theory; deviating away from it results in an even stronger exponential decay, in particular for the non-closing configuration. From this we can draw several conclusions. The first term in the infinite sum over auxiliary $\text{SU}(2)$ spins of the $\text{SL}(2,\mathbb{C})$ EPRL vertex amplitudes contributes significantly. While it decays exponentially, its suppression is less than one might have naively expected. However, it is also clear that this single term is not sufficient to recover the oscillatory nature of Lorentzian EPRL vertex amplitudes, as observed in \cite{Dona:2019dkf}, and additional terms are necessary to get the correct polynomial scaling behavior. It would be interesting to better understand how the full $\text{SL}(2,\mathbb{C})$ amplitude emerges as a sum over $\text{SU}(2)$ $\{15 j\}$ symbols, e.g. to learn whether the infinite sum can be truncated with little error.

\subsubsubsection{In Lorentzian EPRL}

We study the coherent vertex amplitudes for both an equilateral Euclidean 4-simplex and the aforementioned isosceles Lorentzian 4-simplex in the $\text{SL}(2,\mathbb{C})$ EPRL model using the \verb|sl2cfoam-next| package. We ran the code on a consumer laptop, such that we were not able to probe large values of $\lambda$ or shells $s$ (encoding the truncation of the infinite sum over virtual spins), but we believe the domain we considered is sufficient to clearly show the exponential suppression away from the critical points of the vertex amplitude. For both sets of boundary data we modify only a single tetrahedron by the same deviations as above. Both simulations were run for a Barbero-Immirzi parameter of $\gamma = 0.5$. Regarding the shell number, we chose $s=1$ for the Euclidean simplex and $s=4$ for the Lorentzian one. Due to the similarity of results, we only plot the non-closing deviations; qualitatively, the non-matching cases are similar to the previous $\text{SU}(2)$ ones and simply show less exponential decay.

\begin{figure}[ht!]
    \centering
    \includegraphics[width = 0.45 \textwidth]{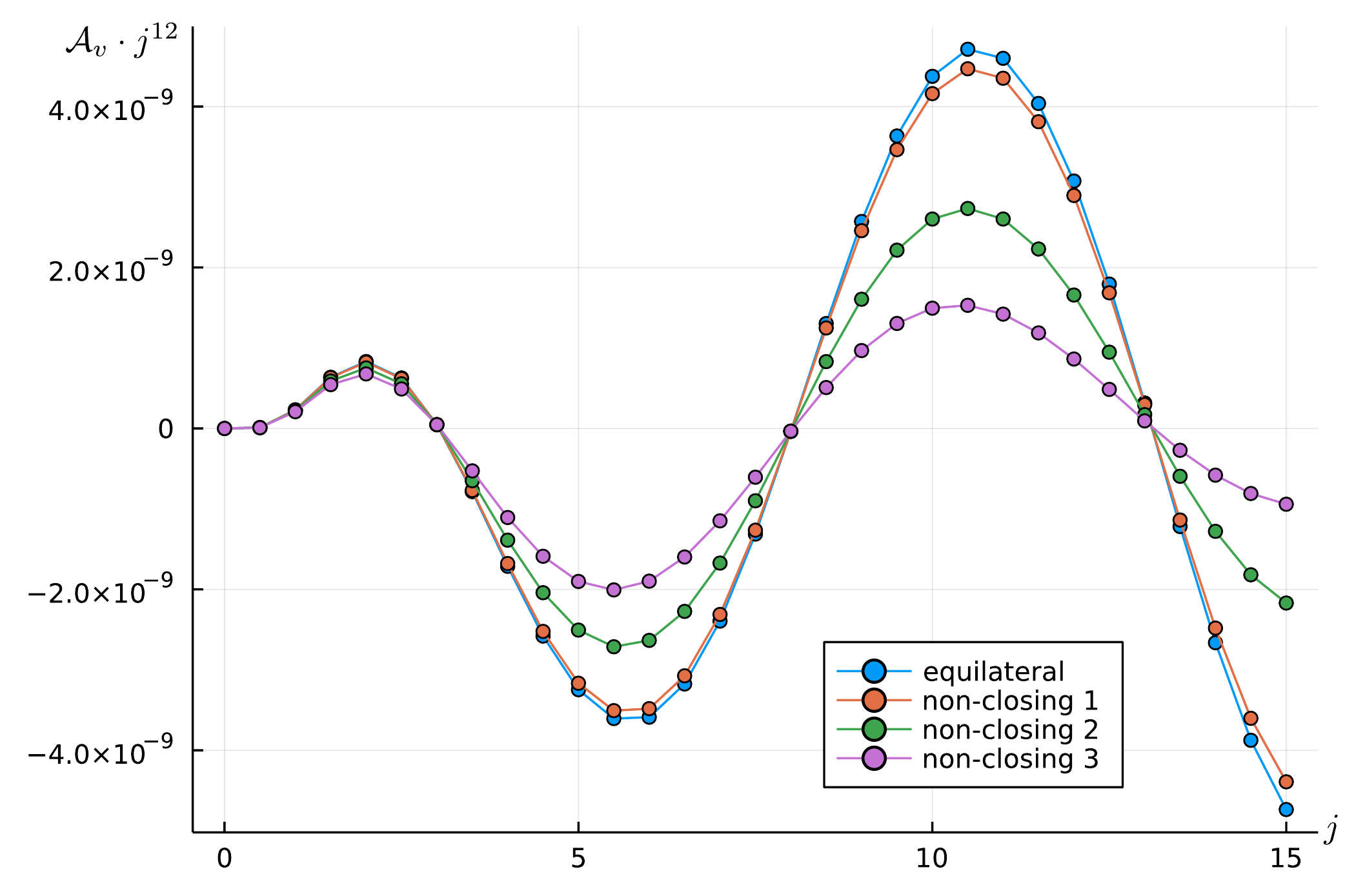} \; \;
    \includegraphics[width = 0.45 \textwidth]{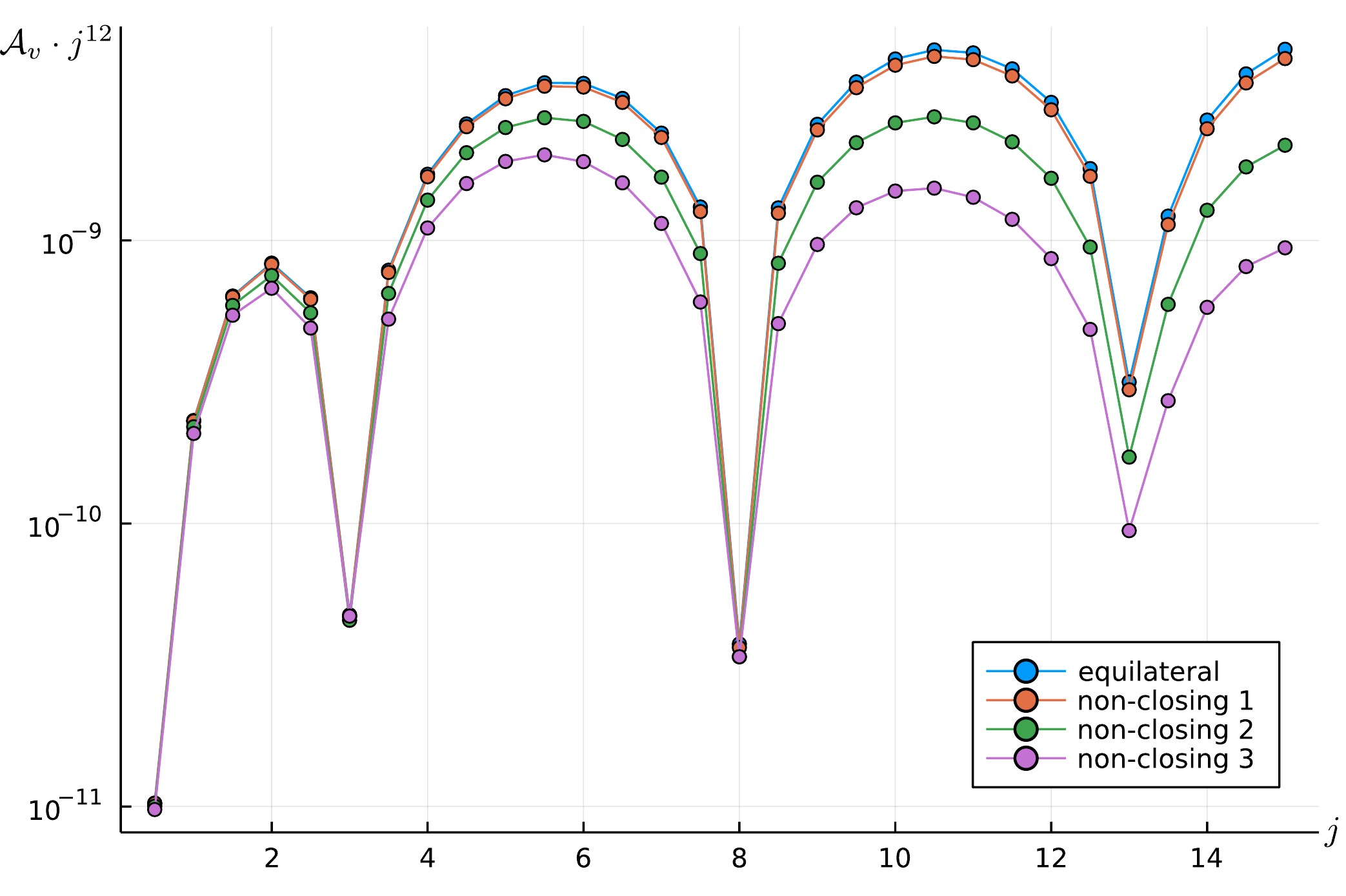}
    \caption{Comparison of $\text{SL}(2,\mathbb{C})$ vertex amplitudes for boundary data corresponding to an equilateral Euclidean 4-simplex and non-closing deviations of a single tetrahedron away from those data. The Immirzi parameter is chosen to be $\gamma = 0.5$, the number of shells is $s=1$.}
    \label{fig:sl2c_euclidean}
\end{figure}

\begin{figure}[ht!]
    \centering
    \includegraphics[width = 0.45 \textwidth]{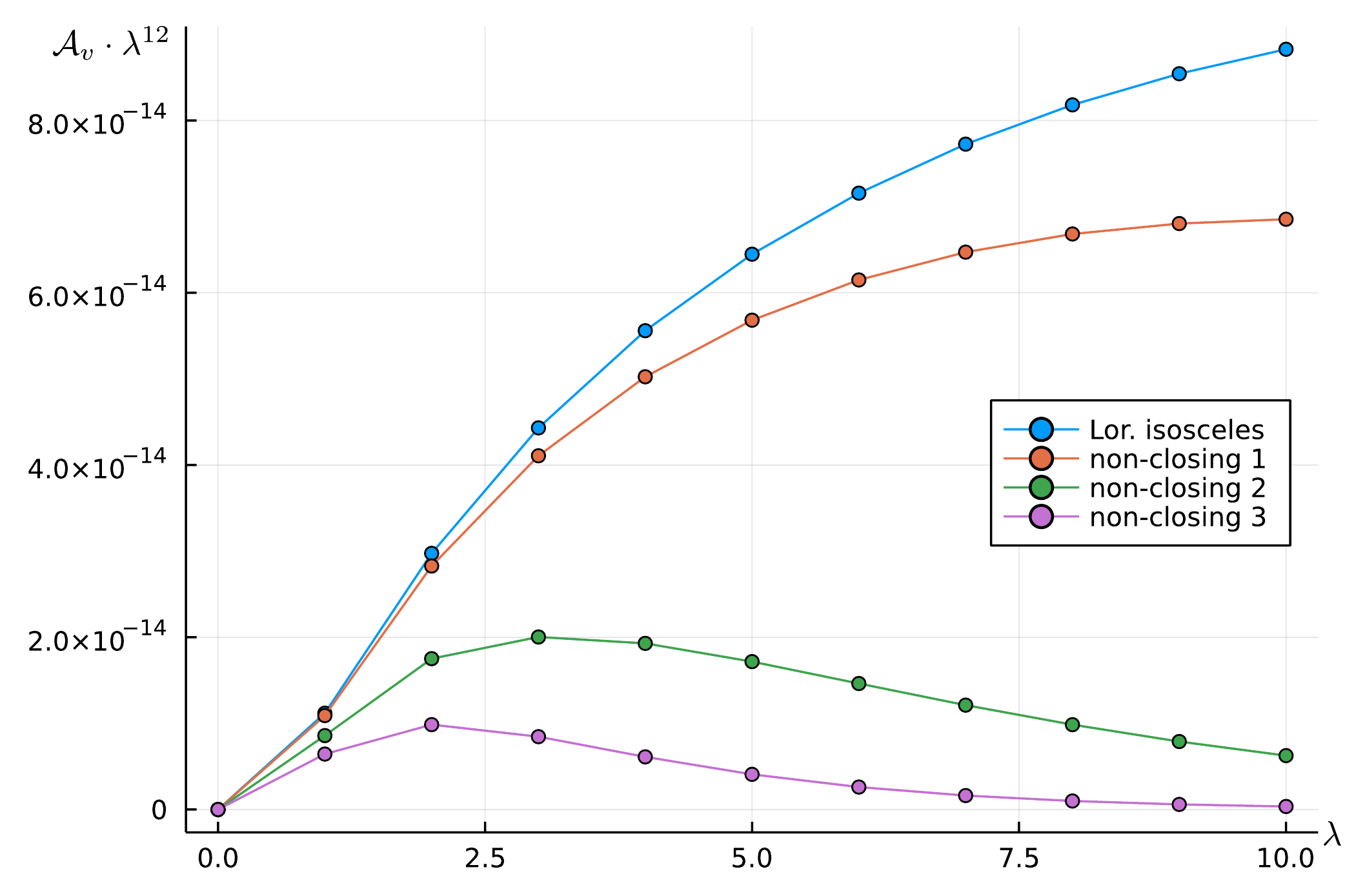} \; \;
    \includegraphics[width = 0.45 \textwidth]{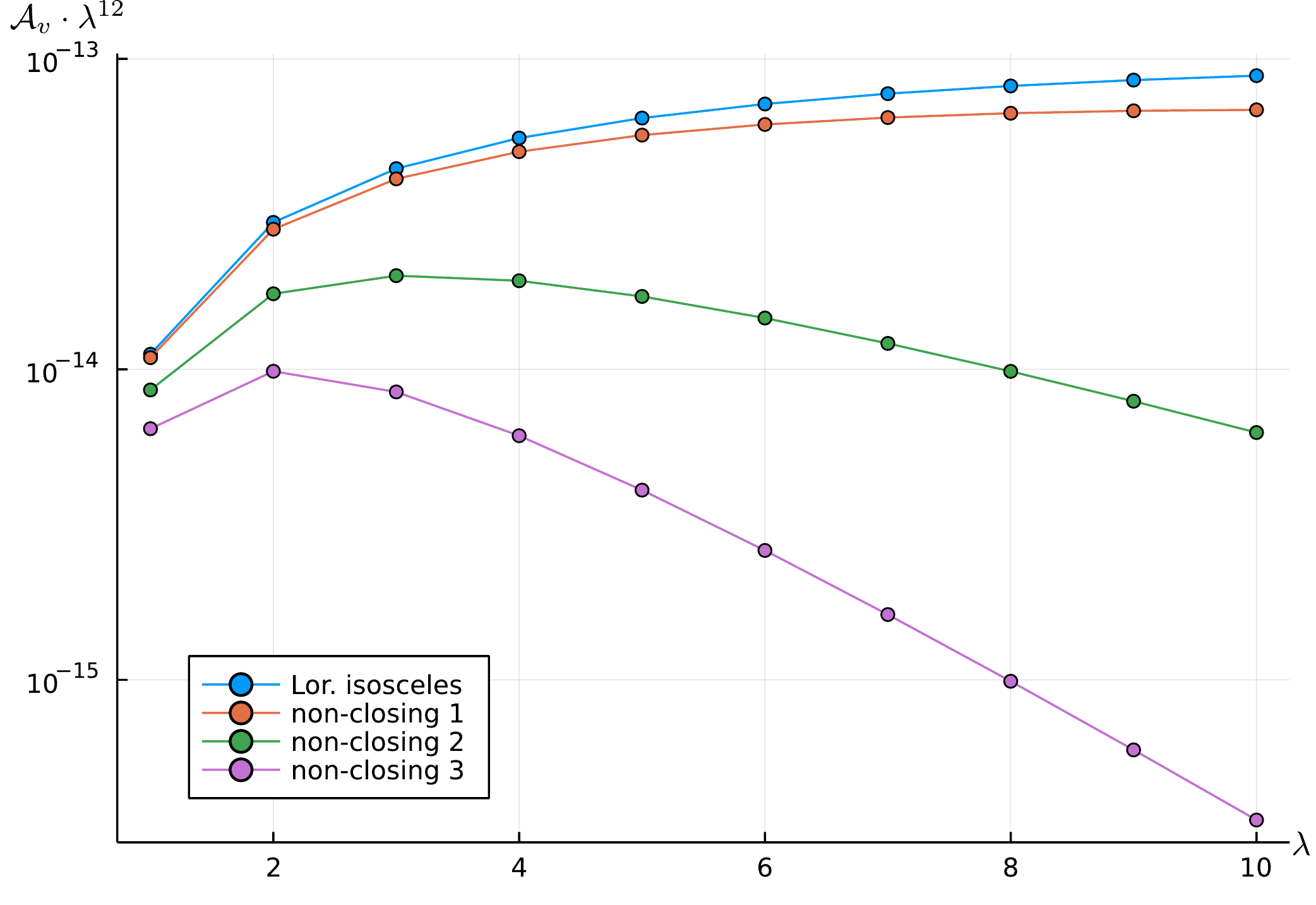} \; \;
    \caption{Comparison of $\text{SL}(2,\mathbb{C})$ vertex amplitudes for boundary data corresponding to Lorentzian isosceles 4-simplex and non-closing deviations of one tetrahedron away from this critical point. The Immirzi parameter is chosen to be $\gamma = 0.5$, the number of shells is $s=4$.}
    \label{fig:sl2c_lorentzian}
\end{figure}

In figure \ref{fig:sl2c_euclidean} we plot the coherent vertex amplitude for an equilateral Euclidean 4-simplex, where we can recognize an exponential decay away from the critical point at already fairly small representations. For the smallest non-closing deviation, increasing $\lambda$ further would be necessary to capture more half periods and clearly see the exponential suppression. Note that, since we only deviate a single tetrahedron rather than all five of them, the suppression is less accentuated than in the $\text{SU}(2)$ case.

In figure \ref{fig:sl2c_lorentzian} we plot the amplitude for boundary data corresponding to an isosceles Lorentzian 4-simplex, now also for non-closing boundary data for a single tetrahedron. We find a far stronger suppression compared to the Euclidean case, yet this is due to the different boundary data. At $\lambda=10$ we are computing the vertex amplitude for representations $(\rho,n)=(50,20)$ (see section \ref{gluesl} for the necessary representation theory), which is significantly larger than $10$ for the same $\lambda$ in the equilateral case. Again, for the smallest non-closing deviation, this small $\lambda$ is not sufficient to observe an exponential decay. Still, these results confirm again our expectation that for increasing spins all vertex amplitudes that significantly contribute to the path integral are located in a small but non-singular region around the critical points.

All the above results support the heuristic expectation that spin-foam models should possess a regime in which only small regions around the critical points of each vertex amplitude contribute to the path integral. At the same time, it is clear that more thorough investigations are necessary to make qualitative statements more quantitatively precise, e.g. by identifying for which representations this statement is valid. We leave this for future research.

\section{Gluing constraints} \label{gluing_constraints}


In this section we shall formally introduce and define the \emph{gluing constraints} for the SU(2) BF and Lorentzian EPRL-FK spin-foam models. As mentioned in section \ref{hybrid_alg}, the gluing constraints appear through the process of fully disentangling the coherent vertex amplitudes assigned to vertices of a 2-complex via insertions of identities expressed in terms of the coherent states. We shall consider the case where the 2-complex is dual to a triangulation by simplices; the vertex amplitudes are thus associated to 4-simplices of the dual triangulation. 

In the coherent state representation, the partition function associated to a 2-complex can be expressed as a product of the vertex amplitudes after doubling the gauge group integration and inserting resolutions of the identity on the representation spaces of that group, expressed in terms of coherent states. The partition function would then take the schematic form 
\be
{\cal Z} = \sum_{\{ \chi_t\}} \int \prod_{(\sigma,\sigma')}\prod_{t} d_{\chi_t}\, \d h_t \, \scalebox{0.55}{\input{vertex_left.tikz} }\,\, h_t  \, \, \scalebox{0.55}{\input{vertex_right.tikz} }\,.
\ee
As one may see, the resulting vertex amplitudes are not all independent, as the same coherent data associated to a common edge is shared between a pair of vertices. There is however a straightforward way to gain better control over the mutual dependence of the vertices. We can isolate the vertex amplitudes by inserting at the common edge an extra identity \eqref{Idcoh} in terms of coherent states,
\be \label{cohAmp2}
{\cal Z} = \sum_{\{ \chi_t\}} \int \prod_{(\sigma,\sigma')}\prod_{t} d_{\chi_t}^2\, \d h_t \, \d k_t \, \scalebox{0.55}{\input{vertex_left.tikz} }\,\, h_t \, \scalebox{0.55}{\input{gluing0.tikz} }\, k_t \, \, \scalebox{0.55}{\input{vertex_right.tikz} }
\ee
such that now every vertex amplitude carries its own coherent data at the boundary. In this manner we get for every (bulk) edge shared by a pair of vertices a term which is represented by the four lines in-between the amplitudes in \eqref{cohAmp2}. It is a tensor product of inner products between coherent states associated to the edges belonging to the pair of vertices. We refer to the extra terms as gluing constraints as they specify how pairs of vertices glue or fit together.


As it stands, the gluing constraints do not share any of the symmetry properties of the vertex amplitude. Since these are ultimately derived from the symmetries of the Haar measure, we may remedy this by proposing a definition of the gluing constraints which involves an integration over the gauge group of the theory. We chose to do so, and define the gluing constraints to generally be given by
\be\label{glconst}
G_\tau= \int_{\cal G} \d\mu(g) \, \bigotimes_{t=1}^4 \, \langle {J \,\triangleright}  \chi_t,  h_t | D^{\chi_t}( g_\tau) | \chi_t,  k_t \rangle  = h_t \, \scalebox{0.8}{\input{gluing.tikz} }\, k_t \;\; . 
\ee
We would like to remark that the particular symmetry properties of $G_\tau$ depend on what one considers to be its domain of definition. Indeed, if one takes the boundary data of the constraints to be general group elements $h_t, k_t \in \text{SU}(2)$, then the constraints are \textit{invariant} under the action of $g,g' \in \text{SU}(2)$ by virtue of the bi-invariance of the Haar measure, i.e. $G_\tau(g h_t, g' k_t) = G_\tau(h_t, k_t)$. On the other hand, if one considers the boundary data to be given by coherent states (which, we remind the reader, are constructed from a reference state as $\ket{\chi, h}=h \ket{\text{ref}_\chi},\, D^\chi(h) \in \text{SU}(2)/\text{U}(1)$), then the gluing constraints are only invariant up to a phase factor (a property which extends to the coherent vertex amplitude), and this has been termed \textit{covariance} in the literature \cite{Dona:2017dvf} for the case of the vertex amplitude. The reason is that the group action on a coherent state results in a second coherent state up to a phase, as $\text{SU}(2)/\text{U}(1)$ is not closed under multiplication. Thus it holds only that $|G_\tau(g \triangleright \vec{h}_t,g' \triangleright \vec{k}_t)|=|G_\tau(\vec{h}_t,\vec{k}_t)|$. In order to avoid having to deal with multiplicative phase factors we will, for the rest of the paper, take $G_\tau=G_\tau(h_t, k_t)$.

Although including a group integration in the definition of the constraints can be argued to increase its computational complexity, it has the advantage of allowing the constraints to more closely resemble the structure of a vertex amplitude. The gluing constraints may then serve as a testing ground for properties of the actual vertex (we shall see in section \ref{asympt} that these objects share the same qualitative behaviour in the asymptotic limit), and we further speculate that the gauge invariance afforded by the group integration may be potentially useful in the construction of the hybrid algorithm. 

Before moving on towards the explicit definition of the gluing constraints for different classes of models, we would like to point out that such objects have already made an appearance in the literature to some extent. The gluing constraint of equation \eqref{glconst} involves two coherent intertwiners (not necessarily the same) associated to every bulk edge. In the special case when the two coherent intertwiners are the same, the gluing constraint becomes the norm of that coherent intertwiner. The asymptotics of the norm of coherent intertwiners has been studied in \cite{Livine:2007vk} for ${\cal G} = \rm SU(2)$.  The gluing constraints appear also in \cite{Dona:2020xzv} as a special case of general invariants of $\rm SL(2,\mathbb C)$, and the authors have studied some of its asymptotic properties. In the remainder of this paper we complement these previous results by studying the constraints in detail for different spin-foam models.

\subsection{Gluing constraints for SU(2) BF theory}

We shall first consider the gluing constraints for the $\rm SU(2)$ BF spin-foam model with compact $\rm SU(2)$ as its gauge group. 
The boundary data associated to the gluing constraint induces $\rm SU(2)$ coherent states, which in turn serve as a basis for the construction of several other coherent states appearing in spin-foam models. 

As usual, we label the unitary and irreducible representations of $\rm SU(2)$ by spins $j \in \frac{\mathbb{N}}{2}$. The canonical basis states are taken to be $|j,m\rangle$, for integers $m=-j,\cdots,j$. An SU(2) coherent state is defined by a group action on the highest or lowest weight states, i.e. 
\be
\ket{j, h} =  D^j(h) \ket{j,j}\,, \quad |j, h ] :=  D^j(h) \ket{j,-j}\, ,
\ee
where $D^j$ denotes a Wigner representation matrix.
Although it is usual to define such coherent states by restricting to the quotient space $k \in \text{SU}(2)/\text{U}(1)$, we will allow for any $k \in \text{SU}(2)$ according to the discussion of the previous section. We further introduce a complex structure on $\mathbb{R}^4$, inducing the antilinear map
\begin{equation}
    \begin{gathered}
    J: \; \mathbb{C}^2 \rightarrow \mathbb{C}^2 \\
    v \mapsto -i \sigma_2 \overline{v}\,,
    \end{gathered}
\end{equation}
which can be used to invert and conjugate-transpose any $\text{GL}(2, \mathbb{C})$ matrix $g$ by $JgJ^{-1} = \det g (g^\dagger)^{-1}$, and which can be canonically extended to $\mathbb{C}^{2n}$. Due to the standard Clebsh-Gordan isomorphism $\ket{j, j} \simeq \ket{\frac{1}{2}, \frac{1}{2}}^{2j}$, it moreover holds that $J^2=(-1)^{2j}$. Finally, note that the complex structure $J$ has an important geometric meaning, which justifies its use in the spin-foam models: while a coherent state $\ket{j, h}$ is associated to the vector $\vec{h}$, the state $J \ket{j, h}$ is associated to its symmetric vector $- \vec{h}$.

Having reviewed the necessary representation theory, the gluing constraint for $\rm SU(2)$ associated to an edge dual to a tetrahedron $\tau$ with four faces is defined by 
\ba \label{gcsu2}
G_{\tau}^{\rm SU(2)} &:=& \int_{\rm SU(2)} \d \mu(g) \, \bigotimes_{i=1}^4 \, \langle J \triangleright j_i,{ h}_i\, | D^{j_i}(g) \, | j_i, { k}_i \rangle =  h_i \,\, \scalemath{0.8}{\input{gluing.tikz} }\,\,  k_i .
\ea
$D^{j_i}(g)$ is the Wigner matrix associated to $g$ and $\d \mu(g)$ is the SU(2) bi-invariant Haar measure\footnote{The  definition of the gluing constraint can be easily generalized to any $n$-valent node associated to a polygon with $n$ faces.}.  
Recalling the isomorphism $\ket{j ,j} \simeq \ket{\frac{1}{2}, \frac{1}{2}}^{2j}$, the constraint may also be formulated as
\ba \label{gcsu2-p}
G_{\tau}^{\rm SU(2)} = \int_{\rm SU(2)} \d \mu(g) \, \prod_{i=1}^4 \, \langle  J{ h}_i  | g | { k}_i \rangle^{2j_i},
\ea
where $g$ is now in its defining representation. This gives an integral expression for the gluing constraint.\footnote{One can also consider using the coherent states $| h_i]$. } The function $G_{\tau}^{\rm SU(2)}$ can also be expressed as a sum through the Peter-Weyl theorem for compact groups. The SU(2) integrals can be decomposed in terms of an intertwiner basis, which for a four-valent node is given by
\be
\int  \d \mu(g) \, \bigotimes_{i=1}^4 \, D^{j_i}(g) = \sum_\iota d_\iota \, | j_i , \iota \, \rangle \langle j_i , \iota |
\ee
where $d_\iota = 2 \iota + 1$ are dimension factors and the orthogonal basis states  
\be\label{wig4j}
| j_i , \iota \rangle = \sum_{m_i} \begin{pmatrix} j_1 & j_2 & j_3 & j_4 \\ m_1 & m_2 & m_3 & m_4 \end{pmatrix}^{(\iota)} |j_1,m_1 \rangle  \cdots  |j_4 ,m_4 \rangle
\ee
are expressed in terms of the Wigner $4jm$ symbols. Another formulation of the constraint is thus given by a sum over intertwiner labels 
\ba \label{gsu2sum}
G_{\tau}^{\rm SU(2)} 
=  \sum_{\iota} d_{\iota}   \, \overline{c}_\iota( j_i,  h_i) \, c_\iota( j_i, k_i) = \sum_\iota  d_{\iota}  \,\,  \input{gluing2.tikz} \,,
\ea
where we have defined $c_{\iota}( j_i, k_i) :=  \langle j_i , \iota | j_i, k_i \rangle$ as the coherent $4j$ symbol (the coherent states expressed in the  basis of the invariant subspace). The integral expression \eqref{gcsu2-p} is usually useful for performing critical point analysis whiles the summation expression \eqref{gsu2sum} is useful for numerical computations, e.g. when computing the vertex amplitude for coherent boundary data \cite{Dona:2017dvf} as we did in section \ref{hybrid_algo}. 

\subsection{Gluing constraints for Euclidean EPRL/FK models} 


Next, we consider the gluing constraints for the Euclidean EPRL and FK spin-foam models, for which the associated group is $\rm Spin(4)$. The $\rm Spin(4)$ group is well known as the spin covering of  $\,\rm SO(4) $, and under the classification of simple Lie groups via Dynkin diagrams it is isomorphic to $\rm SU(2) \times SU(2)$. The group isomorphism leads to an isomorphism between the irreducible representations $D^{(j_+ \, , \,j_-)} \simeq D^{j_+} \otimes D^{j_-} $ labelled by spins. For the definition of the EPRL and FK spin-foam models one also requires an additional constant $\gamma$, known as the Barbero-Immirzi parameter. In these models the intertwiners of $\rm Spin(4)$ are derived from those of $\rm SU(2)$ using an injection (see \cite{Barrett2009a} for more details) constructed from a Clebsch-Gordan intertwining map $C_{j}^{j_+ \, , \,j_-}$, 
which maps the $\rm SU(2)$ representation $D^j$ into the highest spin $j = j_+ + j_-$ and lowest spin $j = j_+ - j_-$ subspace of $D^{(j_+ \, , \,j_-)}$ for $\gamma <1$ and $\gamma >1$, respectively. The spins $j_\pm$ are related to $\gamma$ by
\be
j_{\pm} = \frac{ j}{2}|1\pm \gamma|\, .
\ee
Starting from $\rm SU(2)$ coherent states $|j_i, k_i \rangle $ one can construct coherent states for $\rm Spin(4)$ using $C_{j}^{j_+ \, , \,j_-}$, and we define the gluing constraints associated to a four-valent node to be 
\ba \label{gcsp4}
G_{\tau}^{\rm Spin(4)} &:=& \int_{\rm Spin(4)} \d \mu(g) \bigotimes_{i=1}^4 \, \langle J \triangleright j_i, h_i  \, \circ \,C_{j_i}^{j_i^+ \, \, j_i^-} \,  |  D^{j_+}(g^+) \, D^{j_-}(g^-) |\,  C_{j_i}^{j_i^+ \, \, j_i^-} \circ j_i, k_i \rangle \,,
\ea
having denoted $\d \mu(g) = \d \mu(g^+) \d \mu(g^-)$. The cases for both models are considered separately.
\begin{itemize}
    \item Euclidean EPRL model
    
    For $\gamma <1$, the Clebsch-Gordan map injects into the highest spin subspace  $j_i = j_i^+ + j_i^-$. The symmetrizers of $j_i^+$ and $j_i^-$ can be absorbed into the symmetrizer of highest spin $j_i$, so that the tensor product property allows the gluing constraint to be split into a product as
\ba\label{gcsp4p-split}
G^{{\rm EPRL}}_{\gamma<1} &=& \int_{\rm Spin(4)} \d \mu(g) \prod_{i=1}^4 \, \langle J \triangleright j^+_i, { h}_i  | g^+ |j^+_i, { k}_i \rangle^{2j_i^+}  \, \langle  J \triangleright j_i^-, { h}_i  | g^- | j_i^-,{ k}_i \rangle^{2j_i^-} \,.
\ea
Hence we simply get that for $\gamma <1$ the gluing constraint satisfies $G^{{\rm EPRL}}_{\gamma<1} =G_{\tau+}^{\rm SU(2)} \times G_{\tau-}^{\rm SU(2)}$. 
    
In the case for $\gamma >1$, the Clebsch-Gordan map now injects into the lowest spin subspace  $j_i = j_i^+ - j_i^-$. The symmetrizers of $j_i$ and $j_i^-$ can be absorbed into that of $j_i^+$. Following the construction in \cite{Barrett2009a}, the gluing constraints can be defined by insertions of identity over $\rm SU(2)$ coherent states.    
    \item FK model

For the FK spin-foam model \cite{Freidel:2007py}, SO(4) coherent states are given in terms of SU(2) coherent states through $|j^+, h^+ \rangle \otimes |j^-, h^- \rangle$ and $|j^+, h^+ \rangle \otimes \overline{|j^-, h^- \rangle}$ for $\gamma<1$ and $\gamma >1$, respectively, where $h^+ = h^-$. For $\gamma <1$, this leads to the same gluing constraints \eqref{gcsp4p-split} as for the $\gamma <1$ EPRL model. 
If one takes $\gamma>1$, the FK gluing constraints take the form
\ba\label{gcsp4p}
G^{\rm FK}_{\gamma>1} &=& \int_{\rm Spin(4)} \d \mu (g)\prod_{i=1}^4 \, \langle  { h}_i J | g^+ | { k}_i \rangle^{2j_i^+}  \, \overline{\langle  { h}_i J | g^- | { k}_i \rangle^{2j_i^-} }.
\ea

\end{itemize}

\subsection{Gluing constraints for Lorentzian EPRL, FK models} 
\label{gluesl}

The underlying gauge group for the Lorentzian EPRL, FK spin-foam models is  $\sl2c$, the double covering of the Lorentz group $\rm SO(3,1)$.
The representations of $\sl2c$ are constructed on the space $\mathcal{D}_{(n_1, n_2)}$ of homogeneous functions of two complex variables \cite{ruhl1970the,Gelfand1963}, i.e. functions $F: \mathbb{C}^2 \rightarrow \mathbb{C}$ satisfying 
\begin{equation}
    F(\lambda z_1, \lambda z_2) = \lambda^{n_1-1} \bar\lambda^{\bar n_2 -1} F(z_1,z_2), \q  \lambda \in \mathbb C \,,
\end{equation}
such that the representation $D^{(n_1, n_2)}$ acts on functions through the usual transposed matrix multiplication in $\mathbb{C}^2$,
\begin{equation}
    \begin{gathered}
    D^{(n_1,n_2)}:\;  \sl2c \rightarrow \text{Aut}(\mathcal{D}_{(n_1, n_2)}) \\
    D^{(n_1, n_2)}(g) F(z) = F(g^Tz)\,, \quad z\in \mathbb{C}^2\,.
    \end{gathered}
\end{equation}
The relevant representations for spin-foam models are those contained in the so-called principal series, characterized by the restriction $n_1=\bar n_2$. It is usual to redefine $n_1=(-n+i\rho)/2$ and  $n_2=(n+i\rho)/2$ with $n \in \mathbb Z, \, \rho \in \mathbb R$, and collect these variables in the label $\chi=(n, \rho)$. Such principal series representations are irreducible, and they are unitary under the inner product
\begin{equation}
\begin{gathered}
    \braket{F_1, \, F_2}=\int_{\rm \mathbb{C}P} \omega\, \bar F_1(z), F_2(z)\,, \\
    \omega= \frac{i}{2} (z_1\d z_2- z_2 \d z_1 ) \wedge (\bar z_1\d \bar z_2- \bar z_2 \d \bar z_1 )\,,
\end{gathered}
\end{equation}
where $\rm \mathbb{C}P$ denotes that the integral is to be computed over a section of the bundle $\mathbb{C}^{2*} \rightarrow \rm \mathbb{C}P$, and the result is independent of the choice of section. We further remark that there exists an intertwining isomorphism (defined up to normalization, possibly depending on $\chi$),
\begin{equation}
    \begin{gathered}
    \mathcal{A}: \; \mathcal{D}_\chi \rightarrow  \mathcal{D}_{-\chi} \\
    \mathcal{A} D^ {-\chi}(g) = D^\chi(g) \mathcal{A}\,,
    \end{gathered}
\end{equation}
such that we may restrict our attention to $n, \rho \geq 0$. This intertwiner can be used to construct a bilinear form $(\cdot, \cdot)$ on $\mathcal{D}_\chi$,
\begin{equation}
    (F_2, \, F_2):=\braket{\mathcal{J} F_1, \, F_2}\,,
\end{equation}
having defined $\mathcal{J}F=\overline{\mathcal{A}F}$. This map plays a similar geometrical role as the complex structure $J: \, \mathbb{C}^2 \rightarrow \mathbb{C}^2$ of the previous section. Finally, there is a canonical orthonormal basis for $\mathcal{D}^\chi$ given by the functions
\begin{equation} 
    F^\chi_{j,m}(z)=\sqrt{\frac{2j+1}{\pi} } \, \| z\|^{2(i \rho/2 -1-j)} D^j_{n/2,m}(g(z))\,,
\end{equation}
for $D^j_{ab}(g)$ a unitary irreducible representation of $\text{SU}(2)$, and 
\begin{equation}
g(z) = \begin{pmatrix}
z_1 & -\bar z_2 \\ z_2 & \bar z_1
\end{pmatrix} \in \text{SU}(2)\,, \q  |z_1|^2 + |z_2|^2 =1 \,.
\end{equation}

As before, coherent states are constructed from the action of $\text{SU}(2)$ on a reference state. Restricting the representation labels to the ones relevant to the model, $\chi=(2\gamma j, 2j)$, a general coherent state $\ket{g,\, j}:= D^ \chi(g) F^\chi_{j,j}$ thus reads
\begin{equation}\label{s2cohc}
    \ket{g,\, j}= \sqrt{2j+1}\, ||z||^{2j(i\gamma -1)-2} \braket{z,  \overline{g} +}^{2j}\,, \quad g \in \text{SU}(2)\,,
\end{equation}
\begin{equation}
    \mathcal{J} \ket{g,\, j}= \mathcal{N}\, \sqrt{2j+1}\, ||z||^{2j(i\gamma -1)-2} \braket{z, \overline{g} -}^{2j}\,, \quad g \in \text{SU}(2)\,,
\end{equation}
where $\ket{\pm}$ form the canonical basis of $\mathbb{C}^2$, and $\mathcal{N}$ is a multiplicative factor depending on the choice of normalization of $\mathcal{A}$. The usual choice is to take $\mathcal{N}=(-1)^{2j}e^{-i \mathrm{arctan} \gamma},\, \gamma>0$, a convention adopted in \cite{Barrett2010}. With these objects in place, the gluing constraints for the Lorentzian EPRL model read\footnote{Note that the Haar measure satisfies $\d \mu(g)=\d \mu(g^T)$.}
\begin{equation}
\label{sl2glue2}
    G_\tau^{\text{SL}(2,\mathbb{C})}(\overline{h}_i,\overline{k}_i)=\int\dif \mu(g) \prod_{i=1}^4 \int_{\mathbb{C}P} \omega'(z_i)\,\frac{\braket{g z_i, g z_i}^{j_i(i \gamma-1)-1}}{ \braket{z_i,z_i}^{j_i(i\gamma+1)+1}}  \braket{g z_i, k_i}^{2j_i}\braket{Jh_i, z_i}^{2j_i}\,,
\end{equation}
with $\dif \mu(g)$ the Haar measure on $\sl2c$; since the Haar measure is only defined up to a constant for locally compact groups, its normalization will later be discussed in \ref{asysl}. We have also included a number of numerical factors in the measure $\omega'$, defined as
\begin{equation}
    \omega'(z)=\overline{\mathcal{N}} \,\frac{2j+1}{\pi}\, \omega(z)\,.
\end{equation}


    


\section{Asymptotic gluing constraints away from critical points} \label{asymptotics_gluing}

As discussed before, the gluing constraints allow one to study the asymptotics of a given spin-foam model by a priori decoupling the critical configurations at each vertex, so that the well-known results concerning 1-vertex asymptotics may be applied individually. The role of the gluing constraints would then be to restrict the dominant configurations at each neighbouring vertex 
so as to enforce consistent gluing among simplices. Heuristically \cite{Asante:2020qpa} this is expected to be implemented via a localized (presumably Gaussian-like) distribution over the boundary data. We dedicate this section to the asymptotic study of the gluing constraints for the most ubiquitous spin-foam models, and we shall show that they indeed reduce to Gaussian functions, peaked at gluing tetrahedra.  

\subsection{Hörmander's asymptotic theorem }

Since we are interested in studying how the constraints enforce gluing in the asymptotic regime, we must approximate their defining integral not only at its critical point, but also in a surrounding neighbourhood. To this end we shall make thorough use of Hörmander's theorem \cite{Hormander2003} (Theorem 7.7.12) on the asymptotic evaluation of integrals subject to free parameters, which we reproduce here for the reader's convenience.
\begin{Theorem}[Hörmander]
\label{hormander}
Let $S(x,y)$ be smooth and complex-valued in a neighbourhood $K$ of $(0,0) \in \mathbb{R}^{n+m}$, such that $\Im{S} \geq 0$, $\Im{S}(0,0)=0$, $S'_x(0,0)=0$ and $\det{S''_{xx}}(0,0) \neq 0$. Consider furthermore $u \in \mathcal{C}_0^\infty(K)$. Then
\begin{equation}
\label{asympt}
    \int \dif x \; u(x,y) e^{i \lambda S(x,y)} = \left(\frac{2 \pi i}{\lambda}\right)^{n/2} \frac{u^0(y) e^{i \lambda S^0(y)}}{\sqrt{\left(\det{S''_{xx}}\right)^0(y)}}  + \mathcal{O}\left(\lambda^{-n/2-1}\right)\,,
\end{equation}
where the superscript $f^0(y)$ denotes an $x$-independent residue in the residue class of $f(x,y)\, \mathrm{mod} \,\mathcal{I}$, for $\mathcal{I}$ the ideal generated by the partial derivatives $\partial_{x^i} S$.
\end{Theorem}
To be clear, one says that $r$ is in the residue class $[a]_b$ if $a=r\, \mathrm{mod} \, b$. Thus the possible $f^0(y)$ are defined by expansions of the type
\begin{equation}
\label{Malgrange}
    f(x,y)=f^0(y) + \frac{\partial S}{\partial x^i}(x,y) q^i(x,y)\,,
\end{equation}
for $n$ smooth functions $q^i(x,y)$. $f^0$ is not unique, but another choice of representative will induce a correction of the same order as the error term in equation \eqref{asympt}. That $f(x,y)$ can be brought into this form near the origin is a consequence of the Malgrange preparation theorem \cite{Hormander2003}, which can be thought of as a division theorem with remainder $f^0(y)$. As it stands, however, the explicit form of the coefficients in equation \eqref{Malgrange} are difficult to obtain. We remedy this by resorting to a second theorem of \cite{Hormander2003}.
\begin{Theorem}
\label{x-X}
Let $b_j(x,y), \; j+1,...,n$, be smooth and complex-valued in a neighbourhood $K$ of $(0,0) \in \mathbb{R}^{n+m}$, such that $b_j(0,0)=0$ and $\det \partial_{x^i} b_j \neq 0$. Then
\begin{equation}
    \mathcal{I}(b_1,...,b_n)=\mathcal{I}(x_1-X_1(y),..., x_n-X_n(y))
\end{equation}
for some $X_j \in \mathcal{C}^\infty$ vanishing at the origin. 
\end{Theorem}
Making use of the Malgrange preparation theorem \eqref{Malgrange} $N$ times, and then applying theorem \ref{x-X}, equation \eqref{Malgrange} may be written as
\begin{equation}
\label{MalgrangeN}
    f(x,y)= \sum_{|\alpha| < N} f^{\alpha}(y) \left(x-X(y) \right)^\alpha \mathrm{mod}\,\mathcal{I}^N\,,
\end{equation}
for some smooth $f^\alpha(y)$ near the origin and $\mathcal{I}^N=\{\sum_{|\alpha|=N} s^\alpha \left(x-X(y) \right)^\alpha\, |\, s^\alpha \in \mathcal{C}^\infty(K)  \}$. Note that $\alpha=(\alpha_1,...,\alpha_n)$ stands for a multi-index set, and $|\alpha|=\sum_j \alpha_j$. Note furthermore that the $|\alpha|=1$ term is an element of $\mathcal{I}^N$, and thus it can be omitted from equation \eqref{MalgrangeN}. This $x$-polynomial representation of $f(x,y)$ will be useful in what follows, as it allows a direct comparison with the function's Taylor series. 

\subsection{Asymptotics of SU$(2)$ gluing constraints}
\label{AsySU2}

The SU$(2)$ case is the simplest among the ones we will study, so we shall start by deriving its asymptotic behaviour first. As a function of the boundary data for a certain choice of spins, the gluing constraint of equation \eqref{gcsu2} may be brought into the form,
\begin{equation}
\label{su2glue}
    G_{j_i}(h_i,k_i)=\int_{\text{SU}(2)} \dif \mu(g)\, \exp\left\{\sum_{i=1}^4 \ln \braket{J\, \triangleright \, h_i ,j_i\,|\, D^{j_i}( g )\, |\, k_i ,j_i}\right\}\,,
\end{equation}
where $\ket{k , j}$ stands for a coherent state $D^j(k) \ket{j, j}$ in the Hilbert space $\mathcal{H}^{j}$ associated to the unitary and irreducible $\text{SU}(2)$ representation of spin $j$.

\subsubsection{Malgrange expansion}
In order to identify the residue in equation \eqref{su2glue}, we first approximate the exponent (which we will refer to as the action) by a second order Taylor series using coordinates $g^I$, $I=1,2,3$,
\begin{equation}
\label{taylor}
    S(g,y)=S(g_c,y)+ \partial_I S(g_c,y) (g-g_c)^I + \frac{1}{2} \partial_{IJ}^2 S(g_c, y) (g-g_c)^I (g-g_c)^J + \mathcal{O}(g^3)\,,
\end{equation}
where we let $y$ stand collectively for the boundary data and $g_c$ is a critical point of $S(g,y)$ at some particular configuration $y_c$ of the boundary, i.e.  $\partial_I S(g_c,y_c)=0$. The $N=3$ Malgrange expansion \eqref{MalgrangeN} of the same function reads
\begin{equation}
    S(g,y)=S^0(y) + S^2_{IJ}(y)(g-X(y))^I(g-X(y))^J \; \mathrm{mod}\, \mathcal{I}^3\,,
\end{equation}
and, after matching monomials in $g$ and disregarding elements in $\mathcal{I}^3$, one finds
\begin{equation}
\label{general_as}
    \begin{cases}
            S^2_{IJ}(y)=\frac{1}{2} H_{IJ}\,, \\
            S^2_{IJ}(y)X^I(y)=- \frac{1}{2}\left(\partial_J S(g_c,y) - H_{IJ} g_c^I\right)\,, \\
            S^0(y)=S(g_c,y)-\frac{1}{2}\partial^I S(g_c,y)\, \partial^J S(g_c,y)\,  H^{-1}_{IJ}\,,
    \end{cases}
\end{equation}
where $H_{IJ}=\partial_{IJ}^2 S(g_c,y)$ stands for the Hessian matrix. These equations uniquely specify a representative $S^0$ in the residue class $[S]_\mathcal{I}$. 

\subsubsection{Haar measure and coordinates in SU$(2)$}

Performing the Haar integral in equation $\eqref{su2glue}$, as well as computing the derivatives appearing in \eqref{general_as}, requires one to choose coordinates $\phi_i: U_i \in G \rightarrow \mathbb{R}^3$ on the group manifold $G=\text{SU}(2)$. However, since we are interested in making use of theorem \ref{hormander} rather than analytically evaluating the integral, we can substantially simplify the discussion by implicitly picking useful coordinates and explicitly specifying only the values of the derivatives in those coordinates. 

With a slight abuse of notation (we omit the dependence on $\phi_i$), let $\dif g$ be the exterior derivative of the preimage of a chart,
\begin{equation}
    \dif g  : \mathbb{R}^3 \rightarrow T_{g} G \, \,.
\end{equation}
The differential and right-multiplication $R_g$ for matrix groups satisfies $\dif R_g^{-1} X= X g^{-1}$ for $X \in TG, \, g \in G$, and hence the map $\dif g g^{-1}$ must take values in the Lie algebra $\mathfrak{g}\simeq T_eG$. One may think of this object as a 1-form\footnote{This  amounts to a coordinate representation of the usual Maurer-Cartan 1-form.} in $\mathbb{R}^3$ with values in $\mathfrak{g}$, such that it admits an expansion in terms of $\sigma_I$ (Pauli matrices) generators
\begin{equation}
\label{maurercartan}
\dif g g^{-1}=\frac{i}{2} \sigma_J \Omega^{J} \,, \quad \Omega^J \in T^*\mathbb{R}^3 \simeq \mathbb{R}^3\,.
\end{equation}
Generally, then one sees that coordinate derivatives of $g\in G$ may always be written as $\partial_I g = \frac{i}{2} \Omega\indices{^J_I} \sigma_J g$, for $\Omega\indices{^J_I}$ a matrix of coefficients dependant on the choice of charts $\phi_i$. A particular simple choice of coordinates is that in which the matrix of coefficients reduces to the identity,
\begin{equation}
    \partial_I g = \frac{i}{2} \sigma_I g\,,
\end{equation}
and this is the choice we make for coordinates $g^I$ on SU$(2)$ throughout our analysis. 

Besides derivative terms, the only object in theorem \ref{hormander} which depends on the choice of charts is the Haar measure. But this too can be identified without explicitly defining the map $g=g(g^I)$. Indeed, note that the 1-form of equation \eqref{maurercartan} is right-invariant. We may thus construct a measure on $G$ by taking the trace of its third exterior power, which is bi-invariant by virtue of the cyclicity property of the trace operator,
\begin{equation}
\label{maurertrick}
    \dif \mu(g)=N \tr\left[(\dif g g^{-1})^{\wedge 3}\right] = N \frac{3}{2}  \Omega^1\wedge \Omega^2 \wedge \Omega^3\,.
\end{equation}
Given that for compact groups the normalized bi-invariant Haar measure is unique, one can determine $N$ by computing \eqref{maurertrick} in both the $g^I$ coordinates and some other standard coordinates for which the measure is known, e.g. Euler angle coordinates \cite{ruhl1970the}. Doing so fixes $N=\frac{2}{3}(4\pi)^{-2}$, and the normalized measure for SU$(2)$ in adapted coordinates reads
\begin{equation}
    \dif \mu(g) = (4\pi)^{-2} \dif g^1 \wedge \dif g^2 \wedge \dif g^3\,.
\end{equation}

\subsubsection{Asymptotic formula}

We may now proceed with the derivation of the asympotic expansion. Note that, under the usual Clebsch-Gordan isomorphism, one has the identification $\ket{j , j}=\ket{\frac{1}{2} ,  \frac{1}{2}}^{2j}$. The action of equation \eqref{su2glue} may be rewritten as
\begin{equation}
    S(g, y)=\sum_{i=1}^4 2j_i \ln \braket{J h_i|\, g\, |\, k_i}\,,
\end{equation}
where $\ket{k}=k\ket{\frac{1}{2}\,, \frac{1}{2}}$ and $\ket{Jh}=h \ket{\frac{1}{2}\,, -\frac{1}{2}}$. A critical point is characterized by a vanishing $\partial_I S$ derivative, which reads
\begin{align} 
\label{clo}
\partial_I S(g,y)&=i \sum_{i=1}^4  j_i \frac{\braket{Jh_i\,|\, \sigma_I g \, |\, k_i}}{\braket{Jh_i\,|\, g \,|\, k_i}} \nonumber \\
&= i \sum_{i=1}^4 j_i \frac{\left[\pi(g)\vec{k}_i -\vec{h}_i-i\pi(g)\vec{k}_i \times \vec{h}_i\right]}{1-\pi(g)\vec{k}_i \cdot \vec{h}_i}^{(I)}\,,
\end{align}
where we repeatedly used the spin homomorphism restricted to $\text{SU}(2) \rightarrow \text{SO}(3)$,
\begin{equation}
    \begin{gathered}
    \pi:    \quad \text{SL}(2,\mathbb{C}) \rightarrow \text{SO}^+(3,1) \\
     g \mapsto \pi(g) \text{ s.t. }   g \sigma_\mu g^\dagger = \pi(g)\indices{^\nu_\mu}\sigma_\nu\,,
    \end{gathered}
\end{equation}
and all vectors $\vec{k}_i, \vec{h}_i$ are defined as $\vec{k}=\pi(k)\hat{e}_3$. On the other hand, the real part of the action satisfies
\begin{equation}
    \mathrm{Max}\,\Re S  \leq \sum_{i=1}^4 2j_i \ln \, \mathrm{Max}\,|\braket{J h_i|\, g\, |\, k_i}| = \sum_{i=1}^4 2j_i \ln 1\,,
\end{equation}
so that $\Re S\leq 0$ (as required by theorem \ref{asympt}) and $\Re S=0$ is attained at
\begin{equation}
\label{real}
    g\ket{k_i}=e^{i \phi_i} \ket{J h_i}\,, \quad \phi_i \in [0,2\pi)\,.
\end{equation}
Under this condition, equation \eqref{clo} reduces to the expected closure relation for both sets $\{h_i\},\, \{k_i\}$ of boundary data. By making use of the complex structure map $J$, we may characterize $g$ through the eigensystem
\begin{equation}
\label{eigen}
    \begin{cases}
         g\ket{k_i}=e^{i \phi_i} \ket{J h_i} \\
         g\ket{J k_i}=-e^{- i \phi_i} \ket{ h_i}
    \end{cases}
    \Rightarrow g=e^{-i \phi_i \vec{\sigma} \cdot \vec{h}_i} h_i (-i \sigma_2)  k_i^\dagger\,, \; \forall i\,,
\end{equation}
and, using the spin homomorphism map, one can show
\begin{equation}
\label{rotation}
    \pi(g) \vec{k}_i = - \vec{h}_i\,,
\end{equation}
i.e. there are critical points if the boundary vectors can be rotated into each other. 


Suppose now that equation \eqref{eigen} admits at least one solution, labeled by $\{\hat{g}, \hat{\phi_i}\}$. Then it must be the case that, for any other solution $\{g, \phi_i\}, $
\begin{equation}
    e^{-i ( \phi_i-\hat{\phi_i}) \vec{\sigma} \cdot \vec{h}_i} = e^{-i ( \phi_j-\hat{\phi_j}) \vec{\sigma} \cdot \vec{h}_j}  \,, \quad i \neq j\,.
\end{equation}
Making the simplifying assumption that the boundary vectors are not all colinear, the previous set of equations implies $\phi_i=\hat{\phi}_i \vee \phi_i=\hat{\phi}_i+\pi$ mod $2\pi$. For non-colinear boundary data, then, the two critical point configurations are given by
\begin{equation}
\label{crit}
    g_c= \pm \, e^{-i \hat{\phi}_i \vec{\sigma} \cdot \vec{h}_i} h_i (-i \sigma_2)  k_i^\dagger\,,
\end{equation}
and this holds for any $i=1,...,4$.

Although the critical configurations $g_c$ were determined for boundary data $h_i,k_i$, the gluing constraints are only defined up to a global $\text{SU}(2)$ gauge afforded by the bi-invariance of the Haar measure. The element $g_c$ in equation \eqref{crit} is thus actually gauge-dependent, and one is free to make the simplifying choice $g_c=\pm\mathds{1}$ by appropriately rotating the boundary. A straightforward computation then fixes the coefficients of the Malgrange expansion for non-colinear arbitrary data as follows:
\begin{equation}
    \begin{cases}
    H_{IJ}(y)=-\frac{1}{2}  j_i \left(\delta_{IJ} - V^i_I(y) V^i_J(y) \right)\,, \\
    \partial_I S(y)= i j_i V_I^i(y)\,, \\
    S^0(y)=S(\mathds{1},y)+\frac{1}{2}j_k j_l V_{I}^k(y) (H^{-1})^{IJ}(y)  V_J^l(y)\,, 
    \end{cases}
\end{equation}
having further defined 
\begin{equation}
     \vec{V}_i(y)= \frac{\vec{k}_i -\vec{h}_i -i\vec{k}_i \times \vec{h}_i}{1-\vec{k}_i \cdot \vec{h}_i}\,.
\end{equation}
Finally, we remark that, since we considered a series expansion of $S(g,y)$ to second order in equation \eqref{taylor}, the argument of the square root in theorem \ref{asympt} approximates to $(\det H)^0(y)= (\det H )(\mathds{1},y)$. The resulting asymptotic expansion thus reads
\begin{equation}\label{asympsu2}
    G_{j_i}(h_i,k_i)\simeq  \frac{(1+(-1)^{2 \sum_i j_i})\prod_i {\braket{Jh_i\,|\,k_i}^{2j_i}}}{\sqrt{32 \pi}\sqrt{-   \det H}}\exp\left\{ \frac{1}{2}  j_k j_l V_{I}^k (H^{-1})^{IJ}  V_J^l\right\}\,,
\end{equation}
where the prefactor $(- 32\pi)^{-1/2}= (4\pi)^{-2} \cdot (-2\pi )^{3/2}$ is obtained from the normalization of the Haar measure and from the numerical factor of theorem \ref{asympt}, respectively.

\subsection{Asymtpotics of Lorentzian EPRL gluing constraints}
\label{asysl}

For the Lorentzian EPRL model, recall the gluing constraints from equation \eqref{sl2glue2}, 
\begin{equation}
    G_{j_i}(\overline{h}_i,\overline{k}_i)=\int \dif \mu(g) \prod_{i=1}^4 \int_{\mathbb{C}P} \omega'(z_i)\,\frac{\braket{g z_i, g z_i}^{j_i(i \gamma-1)-1}}{ \braket{z_i,z_i}^{j_i(i\gamma+1)+1}}  \braket{g z_i, k_i}^{2j_i}\braket{Jh_i, z_i}^{2j_i}\,.
\end{equation}
Having defined our object of interest, the arguments laid down in section \ref{AsySU2} apply with little modification. We start by rewriting the constraints in a form adapted to asymptotic analysis,
\begin{equation}
\label{sl2action}
    S(g,z,y)=\sum_{i=1}^4 2j_i \left(\ln \braket{g z_i, k_i} \braket{Jh_i, z_i} + \ln \frac{\braket{gz_i, g z_i}^{\frac{i\gamma-1}{2}}} {\braket{z_i, z_i}^{\frac{i \gamma+1}{2}}} \right)\,,
\end{equation}
\begin{equation}
    u(g,z)=\prod_{i=1}^4 \frac{ \dif \mu(g) \omega'(z_i)}{\braket{z_i, z_i} \braket{gz_i, g z_i}}\,,
\end{equation}
\begin{equation}
     G_{j_i}(h_i,k_i)=\int_{\text{SL}(2,\mathbb{C})} \int_{\mathbb{C}P} u(g,z) e^{S(g,z,y)}\,,
\end{equation}
with a slight abuse of notation in the definition of $u(g,z)$; we still refer by $y$ to the collection of external parameters $h_i, k_i$. Once more we see from \eqref{sl2action} that $\Re{S} \leq 0$, and the maximum is attained at
\begin{equation}
\label{lambdachi}
    \ket{g z_i}=\lambda_i \ket{k_i}\,, \quad \ket{z_i}=\chi_i \ket{Jh_i}\,, \quad \lambda_i,\chi_i \in \mathbb{C}\,,
\end{equation}
moreover implying
\begin{equation}
\label{lambdachi2}
    \ket{Jh_i}=\frac{\lambda_i}{\chi_i} g^{-1} \ket{k_i}\,.
\end{equation}
We may straightforwardly characterize the critical points of $S(g,z,y)$ through the first derivatives in the spinor and group variables; as before, we pick local coordinates $g^I, \, I=1,...,6$ for the special linear group such that $\partial_I g = \frac{i}{2} \Sigma_I g$, having denoted the generators of the algebra by $\Sigma=(\vec{\sigma}, i \vec{\sigma})$. One then finds
\begin{equation}
    \partial_I S=\sum_{i=1}^4 i j_i \left[\frac{i\gamma-1}{2} \frac{\braket{gz_i,  (\Sigma_I-{\Sigma}^\dagger_I) g z_i}}{\braket{gz_i, gz_i}} - \frac{\braket{gz_i,  \Sigma^\dagger_I k_i}}{\braket{gz_i, k_i}}\right]\,,
\end{equation}
\begin{equation}
    \partial_{z_i^a}S=2j_i \left[\frac{\braket{Jh_i, a}}{\braket{Jh_i, z_i}} +\frac{i\gamma-1}{2} \frac{\braket{gz_i, ga}}{\braket{gz_i,gz_i}} - \frac{i\gamma+1}{2} \frac{\braket{z_i, a}}{\braket{z_i, z_i}} \right]\,,
\end{equation}
for $z_i^a$ the $a$th component $\braket{a,z_i}$. We remark that, due to the conjugation property of Wirtinger derivatives, if $\Re S=0$ then $ \partial_{z_i^a}S=0 \Leftrightarrow  \partial_{\overline{z}_i^a}S=0$. Under equation \eqref{lambdachi}, a vanishing gradient $\partial S=0$ reduces to
\begin{equation}
\label{critical}
    \sum_{i=1}^4 j_i \vec{k}_i =0\,, \quad \ket{Jh_i}=\frac{\overline{\chi}_i}{\overline{\lambda}_i} g^\dagger \ket{k_i}\,,
\end{equation}
and, as expected, one identifies a closure condition in the first equation above. 
\subsubsection{Rotation of the boundary data}
Equations \eqref{lambdachi2} and \eqref{critical} admit a similar treatment as that of the previous section. A general element $g \in \text{SL}(2,\mathbb{C})$ can be polar-decomposed in terms of a pure boost $b=e^{\vec{\beta} \cdot \vec{\sigma}}$ and a unitary $a \in \text{SU}(2)$ as $g=ba$. One may then combine those equations into the eigenvalue condition
\begin{equation}
    b \ket{k_i} = \left|\frac{\lambda_i}{\chi_i}\right| \ket{k_i}\,,
\end{equation}
from where it follows
\begin{equation}
\label{eigensystemlor}
    \begin{cases}
         b\ket{k_i}= \left|\frac{\lambda_i}{\chi_i}\right| \ket{k_i} \\
         b\ket{J k_i}= \left|\frac{\chi_i}{\lambda_i}\right| \ket{J k_i}
    \end{cases}
    \Rightarrow b=e^{\ln\left|\frac{\lambda_i}{\chi_i}\right| \vec{k}_i \cdot \vec{\sigma}}\,, \; \forall i\,.
\end{equation}
If one assumes that the boundary data corresponds to non-colinear vectors, then, through \eqref{eigensystemlor}, equations \eqref{lambdachi2} and \eqref{critical} imply
\begin{equation}
   g \ket{Jh_i}=\frac{\lambda_i}{\chi_i} \ket{k_i}\, \quad g\in \text{SU}(2)\,, \quad \left|\frac{\lambda_i}{\chi_i}\right|=1\,,
\end{equation}
a result entirely analogous to the critical point condition \eqref{real} of the $\text{SU}(2)$ model, and thus the discussion there can be immediately carried over. Given that $\text{SU}(2)$ is a subgroup of $\text{SL}(2,\mathbb{C})$, and the Haar measure induces an $\text{SL}(2,\mathbb{C})$ symmetry at the level of the boundary data, we may again pick a convenient gauge such that $g_c=\pm \mathds{1}$. 
\subsubsection{Haar measure normalization}
The Haar measure for locally-compact groups is only unique up to a multiplicative constant. While this constant is entirely conventional, it is still useful to derive an explicit expression for the Haar measure in adapted coordinates for a given choice of convention. Noting that $\mathfrak{sl}(2,\mathbb{C})$ is a complex Lie algebra, one may generally write locally
\begin{equation}
    \dif g g^{-1}=\frac{i}{2}\Omega_{I}^{\; J} \sigma_J \dif x^I\,, \quad \Omega_{I}^{\; J} \in \mathrm{GL}(3,\mathbb{C})\,,
\end{equation}
for some complex coordinate chart $x^I$. The Haar measure for $\sl2c$, up to a factor $N$, thus reads
\begin{align}
\label{haarslgen}
    \dif \mu(g)&=N \tr\left[(\dif g g^{-1})^{\wedge 3}\wedge (\dif g g^{-1})^{\dagger \wedge 3}\right] \nonumber \\
&= N \frac{ 2\cdot 3!^2 \cdot 6^2}{2^6} |\det \Omega |^2 \dif x^1\wedge \dif \overline{x}^1 \wedge ... \wedge \dif x^3\wedge \dif \overline{x}^3\,.
\end{align}
An often-used choice of measure in the spin-foam literature \cite{Dona:2019dkf,Gozzini:2021kbt} is the one of Rühl, which can be e.g. formulated in the parametrization (Appendix of \cite{ruhl1970the})
\begin{equation}
\label{ruhl}
    g=\begin{pmatrix}
    a_{11} & a_{12} \\
    a_{21} & a_{22}
    \end{pmatrix}\,, \quad \dif \mu(g)_R=\frac{(2\pi)^{-4}}{|a_{22}|^2} \left(\frac{i}{2}\right)^3\, \dif a_{12}\wedge \dif \overline{a}_{12} \wedge \dif a_{21}\wedge \dif \overline{a}_{21} \wedge \dif a_{22}\wedge \dif \overline{a}_{22}\,.
\end{equation}
The normalization factor of equation \eqref{haarslgen} can be made to agree with the convention of Rühl by direct comparison. Letting $x^I$ now stand for the parametrization of equation \eqref{ruhl} one finds $|\det \Omega |^2=2^4 |a_{22}|^{-2}$, and hence, requiring $\dif \mu(g)=\dif \mu(g)_R$,
\begin{equation}
    N_R=\frac{-i\cdot (2 \pi)^{-4}}{3!^2\cdot 6^2 \cdot 2^2 }.
\end{equation}
Having fixed $N_R$, the Haar measure in the real adapted coordinates $g^I$ mentioned above, i.e. coordinates such that $\partial_I g=\frac{i}{2}  \Sigma_I g$, 
takes the simple form
\begin{equation}
  \dif \mu(g)=\frac{1}{(4 \pi)^4} \dif g^1\wedge ... \wedge \dif g^6\,.
\end{equation}
\subsubsection{Hessian matrix}
As was previously mentioned, the $z$-integral of equation \eqref{sl2glue2} requires a choice of section in $\mathbb{C}^{2*}$, which has up to now remained unspecified. To simplify calculations, we shall fix that section before taking derivatives. Since the set $\{k_i, Jk_i\}$ spans $\mathbb{C}^2$, we are free to pick
\begin{equation}
    \ket{z_i}=\ket{k_i}+\beta \ket{J k_i}\,,
\end{equation}
where $\beta \in \mathbb{C}$; this choice considerably simplifies the discussion.  We remark that this is indeed a global section of the bundle $\mathbb{C}^{2*}\rightarrow \rm \mathbb{C}P$, since - under a judicious choice of the range of $\beta$ - it crosses every line through the origin only once. Such a choice further restricts the complex parameters to $\lambda_{i}=1$, and equation \eqref{lambdachi} determines $\beta_c=0$. We now list below all relevant derivatives to the asymptotic analysis (note that we include only the symmetric part of all second derivatives), denoting by $S_c$ an evaluation at the critical point $S(g_c,\beta_c,y)$:

\begin{equation}
     \partial_{\overline{\beta}_i}S_c=0\,, \quad \partial_{\beta_i}S_c=-2 \Theta_i \,, \quad \partial_I S_c^R=-i \Gamma_I\,, \quad \partial_I S_c^B=-i \gamma \Gamma_I\,, 
\end{equation}
\begin{equation}
\partial_{\overline{\beta}_i}^2 S_c=0\,, \quad \partial_{\beta_i}^2 S_c = -2\frac{\Theta_i^2}{j_i}\,, \quad \partial_{\beta_i \overline{\beta}_i}^2S_c=-2j_i\,,
\end{equation}
\begin{equation}
    \partial_{I \beta_i}^2S_c^R=0\,, \quad  \partial_{I \beta_i}^2S_c^B= j_i(1-i\gamma)\kappa_{iI}\,, \quad  \partial_{I \overline{\beta}_i}^2S_c^R=-ij_i \overline{\kappa}_{iI}\,, \quad  \partial_{I \overline{\beta}_i}^2S_c^B=-i \gamma j_i \overline{\kappa}_{iI}\,,
\end{equation}
\begin{equation}
    \partial_{IJ}^2S_c^{RR}=-\frac{1}{2} \sum_i j_i\left(\delta_{IJ}-k_{iI}k_{iJ} \right)\,, \quad  \partial_{IJ}^2S_c^{BB}=\frac{2i\gamma-1}{2} \sum_i j_i\left(\delta_{IJ}-k_{iI}k_{iJ}\right) \,,
\end{equation}
\begin{equation}
     \partial_{IJ}^2S_c^{BR}=\frac{i}{2} \sum_i j_i \left(\delta_{IJ}-k_{iI}k_{iJ} \right) \,, \quad   \partial_{IJ}^2S_c^{RB}=\frac{i}{2}\sum_i j_i\left(\delta_{IJ}-k_{iI}k_{iJ}\right) 
\end{equation}
where we defined $\kappa_{iI}=\braket{k_i, \sigma_I Jk_i}$ and $\Gamma_I=\sum_i j_i k_{iI}$, $\Theta_i=j_i\frac{1+\vec{h}_i \cdot \vec{k}_i}{\vec{h}_i \cdot \overline{\vec{\kappa}_i}}$ (note that $k_{iI}$ denotes the $I$-th component of $k_i$). We have also split the capital indices $I$ into a rotation and boost part (such that in the equations above $I=1,...,3$). The Hessian, which is a $14\times14$ symmetric matrix, has thus the schematic structure
\begin{equation}
     H(g_c,z_c,y)= \begin{tikzpicture}[baseline={-0.5ex},mymatrixenv]
        \matrix [mymatrix] (m)
        {
       H_{gg}^{RR} & H_{gg}^{RB} & H_{g\beta}^{R} & H_{g \overline{\beta}}^R \\
     &  H_{gg}^{BB} & H_{g\beta}^{B} & H_{g \overline{\beta}}^B\\
     & & H_{\beta \beta} & H_{\beta \overline{\beta}} \\
     & & & H_{\overline{\beta}\overline{\beta}} \\
        };
        \mymatrixbraceleft{1}{1}{$3$}
        \mymatrixbraceleft{2}{2}{$3$}
        \mymatrixbraceleft{3}{3}{$4$}
        \mymatrixbraceleft{4}{4}{$4$}
        \mymatrixbracebottom{1}{1}{$3$}
        \mymatrixbracebottom{2}{2}{$3$}
        \mymatrixbracebottom{3}{3}{$4$}
        \mymatrixbracebottom{4}{4}{$4$}
    \end{tikzpicture}\,.
\end{equation}
\subsubsection{Asymptotic formula}
Equations \eqref{general_as} immediately generalize to the present case by including derivatives with respect to both sets of integration variables. The prefactor term $u(g,z)$ can be shown to be constant for our choice of section at the critical points. Hence the final expression for the gluing constraints in the asymptotic regime reads
\begin{equation}\label{asympsl2c}
    G_{j_i}(\overline{h}_i,\overline{k}_i)\simeq \overline{\mathcal{N}}^4_{j_i} \frac{(1+(-1)^{2 \sum_i j_i}) \prod_i (2j_i+1)\braket{Jh_i\,|\,k_i}^{2j_i}}{32 \pi \sqrt{- \det H}}  \exp\left\{V_\alpha (H^{-1})^{\alpha \beta} V_\beta \right\}\,,
\end{equation}
where $V_\alpha$ is a 14-component vector formally defined as
\begin{equation}
    V=\left(-i \vec{\Gamma}, -i\gamma \vec{\Gamma}, -2\vec{\Theta}, \vec{0}_4  \right)\,,
\end{equation}
and $\overline{\mathcal{N}}^4_{j_i}$ is to be understood as a product of the phase $\mathcal{N}$ for every spin $j_i$.
The coefficient $(32\pi)^{-1}= (4\pi)^{-4} \cdot ( 2 \pi)^{-4} \cdot (2\pi )^{7}$ is obtained from the normalization of the Haar measure, from $u(g,z)$ and from the numerical factor of theorem \ref{asympt}, respectively. We remind the reader that $\mathcal{N}$ is a phase, as defined in section \ref{gluesl}.

\section{Numerical analysis} \label{numerical_analysis}


The gluing constraints defined in \eqref{glconst} are of a simple form compared to the actual vertex amplitude \eqref{vertamp}, and in the previous section we have derived their semi-classical approximation for general boundary data. Below we provide a numerical study of the gluing constraints and their properties, arguing they are well captured by their asymptotic formulae. We consider only the SU(2) BF and Lorentzian EPRL constraints, given in \eqref{gcsu2} and \eqref{sl2glue2} respectively. Note that the Lorentzian EPRL gluing constraint depends additionally on the Immirzi parameter $\gamma$.    

Our choice of parametrization for $\rm SU(2)$ is as follows:
\be \label{eulersu2}
g \equiv g(\phi,\theta,\psi) = e^{-i \phi\tfrac{ \sigma^3}{2}} \,  e^{-i \theta\tfrac{ \sigma^2}{2}}  \,e^{-i \psi\tfrac{ \sigma^3}{2}} , \q \q \, \, -\pi\leq \phi\leq \pi, \, \, -\tfrac\pi2\leq \theta \leq \tfrac\pi2 , \, \, -2\pi \leq \psi \leq 2\pi \,,
\ee 
with a corresponding Haar measure is $ \d \mu (g) = (4\pi)^{-2}\sin\theta\, \d\phi\, \d\theta\, \d \psi $. A coherent state $| j, k\rangle$ can be parametrized by a spin $j$ and two Euler angles, such that $k \equiv k(\phi,\theta) := k(\phi,\theta,-\phi)$. The group element $k$ maps to a unit vector $\vec k \in S^2$ directly through the Euler angles, and generally via the inner product $\vec{k}=\braket{k| \vec{\sigma} |k}$, as in section \ref{asymptotics_gluing}. A coherent intertwiner of a four-valent node can be described by the spins and angles associated to each leg. Coherent intertwiners associated to a classical tetrahedron satisfy the closure constraint $\mathcal{C}:=\sum_i^4 j_i \vec k_i = 0$, and hence not all the parameters are independent. We refer the reader to appendix \ref{AppendixS} for the choice of parametrization of coherent intertwiners used here. We term coherent intertwiners satisfying the closure condition \emph{closing intertwiners}, and otherwise we call them \emph{non-closing}. 

We consider as a first example a closing coherent intertwiner with all spins equal, i.e. $j_i = \lambda ,\,  i=1,\cdots, 4$.  Besides four spins, it is characterized by two dihedral angles $\{ \Phi_{1} , \Phi_{2} \}$ which parametrize the shape of the associated tetrahedron\footnote{The range of values of $\Phi_{1} , \Phi_{2}$ are restricted by simplex inequalities or generalized triangle inequalities.}.
The choice $\Phi_{1} = \Phi_{2} = \arccos\left(-\tfrac13 \right)$ yields an an equilateral tetrahedron. The unit-norm vectors orthogonal to the faces of such a tetrahderon can be determined from the spins and angles (see Appendix \ref{AppendixS}) up to a global rotation. Here we pick them to be
\be \label{nomequi}
\vec k_1 = \left(0,0,1 \right) ,\,\,\,\, \vec k_2 = \left(0,2\frac{\sqrt 2}{3}, -\frac13 \right), \,\,\,\,  \vec k_3 = \left(\sqrt{\frac23},-\frac{\sqrt 2}{3} , -\frac13 \right), \,\,\,\,  \vec k_4 = \left(-\sqrt{\frac23},-\frac{\sqrt 2}{3} , -\frac13 \right)  \, .
\ee
A second interesting case is the one of a closing intertwiner corresponding to a non-regular tetrahedron with equal areas, which can be obtained from the angles $\Phi_{1} = \arccos(-\frac13) ,\, \,\Phi_{2} = \frac\pi2$. Again, the corresponding normal vectors can be computed using the formula \eqref{nomdih} in Appendix \ref{AppendixS}.  The third and last example is that of a non-closing intertwiner with equal spins, obtained from the first example by changing one of the normal vectors. We rotate the polar angle $\theta_1$ of the first normal vector in \eqref{nomequi} by an angle $\delta\theta_1$. The new orthogonal vectors are thus
\be
\vec k_1' = (\sin(\delta\theta_1),0,\cos(\delta\theta_1)) \quad \vec k_i' = \vec k_i\,,\, i=2,3,4\,,
\ee
for which the closure constraint evaluates to
\begin{equation}
    {\cal C} = j_1(\vec k_1' - \vec k_1)=  j_1\left(\sin(\delta \theta_1),0,\cos(\delta\theta_1) -1\right)\,.
\end{equation} 
These three examples of boundary data, which we will use in the subsequent numerical analysis, are summarized in table \ref{tab:A}. In the next subsection, we shall study the gluing constraint between different pairings of these data.
\begin{table}[ht!]
\centering
\begin{tabular}{| c|l |c |}
 \hline
\,\, Intertwiner  \,\, & \hspace{2.1cm} parameters & feature \\ 
\hline
$ \kappa_A$ & \,\, $ j_i = \lambda, \,\, \Phi_{1} = \Phi_{2} = \arccos(-\tfrac13)$ & closing \\ 
\hline 
$ \kappa_B$ &\,\, $j_i = \lambda, \,\, \Phi_{1} = \arccos(-\frac13) ,\, \,\Phi_{2} = \frac\pi2$ & closing \\ 
\hline
$ \kappa_C$ &\,\, $j_i = \lambda, \,\, \vec k_1' =(\sin 1,0,\cos 1), \, \vec k_i' = \vec k_i ,i=2,3,4. \,\,\,  $ & \,\,  non-closing  \,\, \\ 
\hline
\end{tabular}
\caption{Examples of coherent intertwiners parametrized by spins and angles with all spins equal ($j_i = \lambda$).}
\label{tab:A}
\end{table}

\subsection{SU(2) gluing constraints}

Both the integral form \eqref{gcsu2-p} and the summation form \eqref{gsu2sum} of the constraints can be used efficiently for numerical analysis. We chose to use the integral representation,
\ba \label{gcsu2-p2}
G_{\tau}^{\rm SU(2)} = \int_{\rm SU(2)} \d \mu(g) \, \prod_{i=1}^4 \, \langle  { h}_i  | g | { k}_i \rangle^{2\lambda} \quad ,
\ea
repeated here for the reader's convenience. While the computational time for the above integration is independent of the spins $\lambda$, the integrand in the case of non-matching configurations with very large spins ($\lambda \sim 150$) becomes exceptionally small, such that the numerical integration turns less precise. 

\subsubsection{Scaling behaviour}

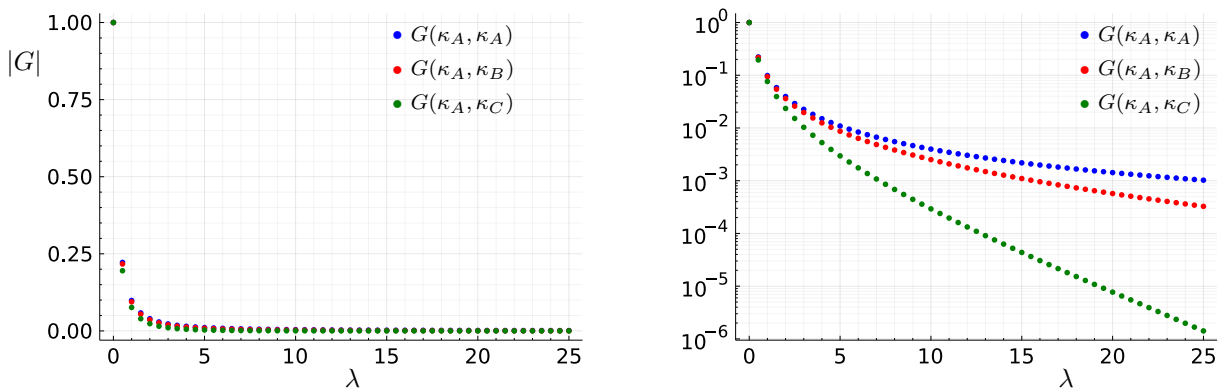
\begin{figure}[ht!]
\begin{picture}(500,145)
\put(10,5) { \includegraphics[scale=0.35]{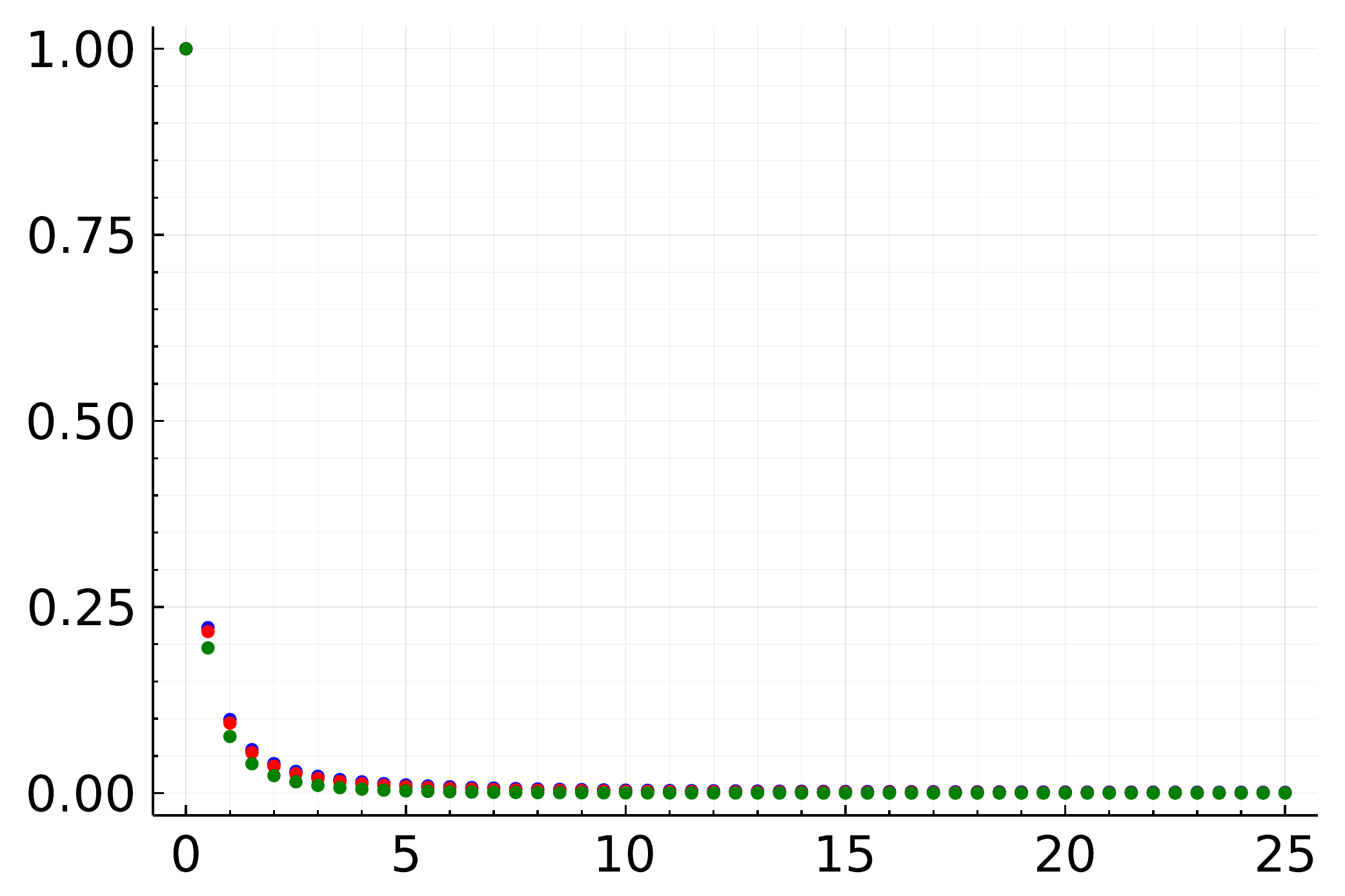} }
\put(250,5){ \includegraphics[scale=0.35]{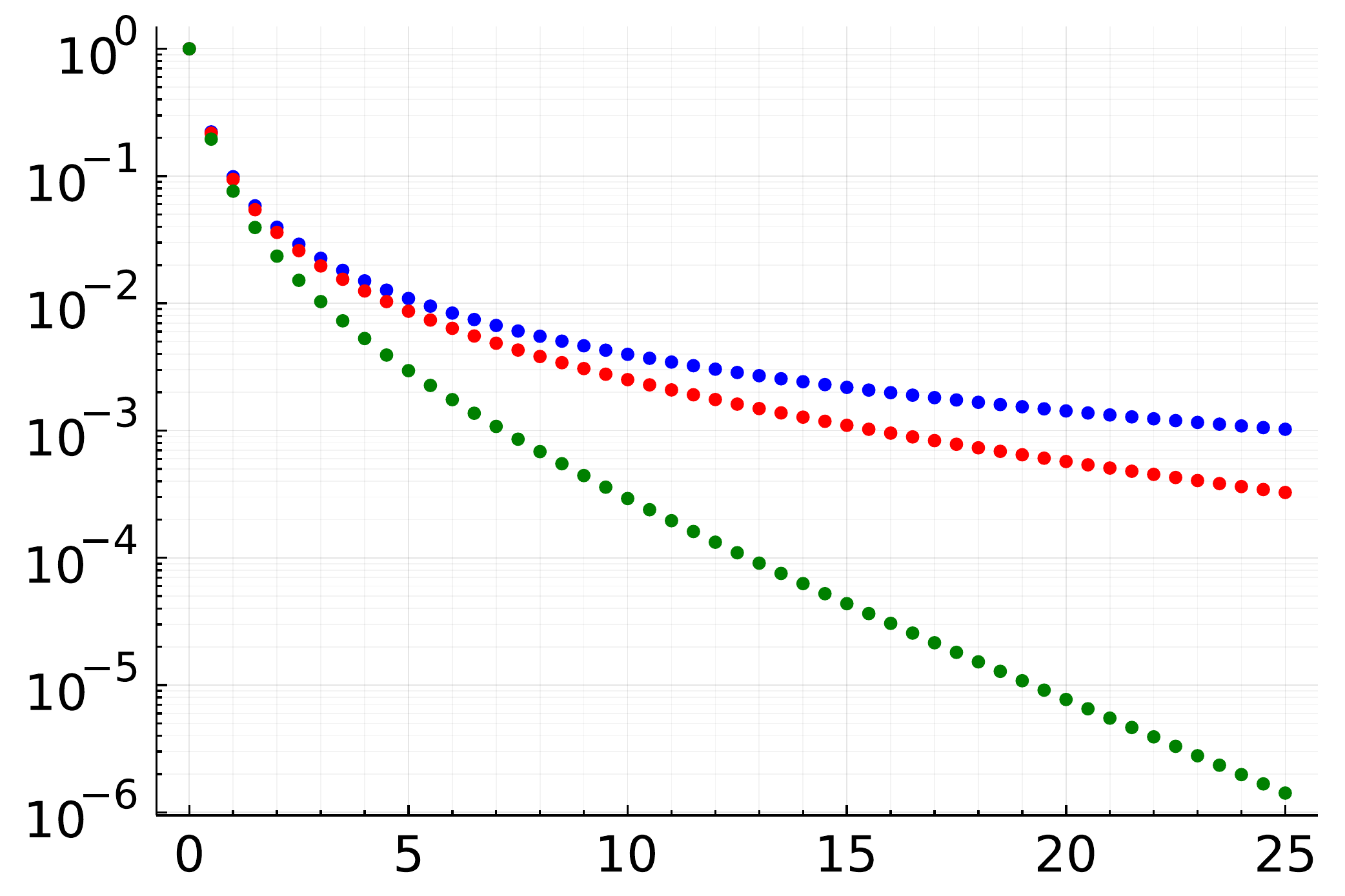} }

\put(2,120){$|G|$}
\put(130,0){$\lambda$}
\put(400,0){$\lambda$}

\put(150,133){\blue \circle*{3}} 
\put(155,131){\scriptsize $G(\kappa_A,\kappa_A)$}

\put(150,120){\red \circle*{3}}
\put(155,118){\scriptsize $G(\kappa_A,\kappa_B)$}

\put(150,107){\color{green!50!black} \circle*{3}}
\put(155,105){\scriptsize $G(\kappa_A,\kappa_C)$}

\put(410,133){\blue \circle*{3}} 
\put(415,131){\scriptsize $G(\kappa_A,\kappa_A)$}

\put(410,120){\red \circle*{3}}
\put(415,118){\scriptsize $G(\kappa_A,\kappa_B)$}

\put(410,107){\color{green!50!black} \circle*{3}}
\put(415,105){\scriptsize $G(\kappa_A,\kappa_C)$}

\end{picture}
\caption{A plot of the SU(2) gluing constraint as a function of spins. The left panel shows the absolute value $|G|$ for different pairings of boundary coherent intertwiners in table \ref{tab:A}. The right panel is the same plot with a log scale on the y-axis. } 
\label{fig:A}
\end{figure}

Figure \ref{fig:A} shows the absolute value of the gluing constraint as a function of spins, which generally decreases for higher spins $\lambda$. The log-scaled plot suggests a power-law decay in $\lambda$ for matching data, while non-matching and non-closing data both appear to exponentially decay. As mentioned previously, when two boundary intertwiners coincide the gluing constraint amounts to the norm of the intertwiner. It has been shown in \cite{Livine:2007vk} that the norm is exponentially suppressed unless the closure condition is satisfied. 
Here we find such a suppression also for the overlap between a closing intertwiner $\kappa_A$ and a non-closing intertwiner $\kappa_C$.
Moreover, if the two coherent intertwiners are both closing but not equal (not shape-matching), as for the pair $\kappa_A,\kappa_B$, the overlap is also suppressed. The suppression is slower compared to the example with a non-closing configuration. 




\begin{figure}[ht!]
\begin{picture}(500,145)
\put(24,5) { \includegraphics[scale=0.35]{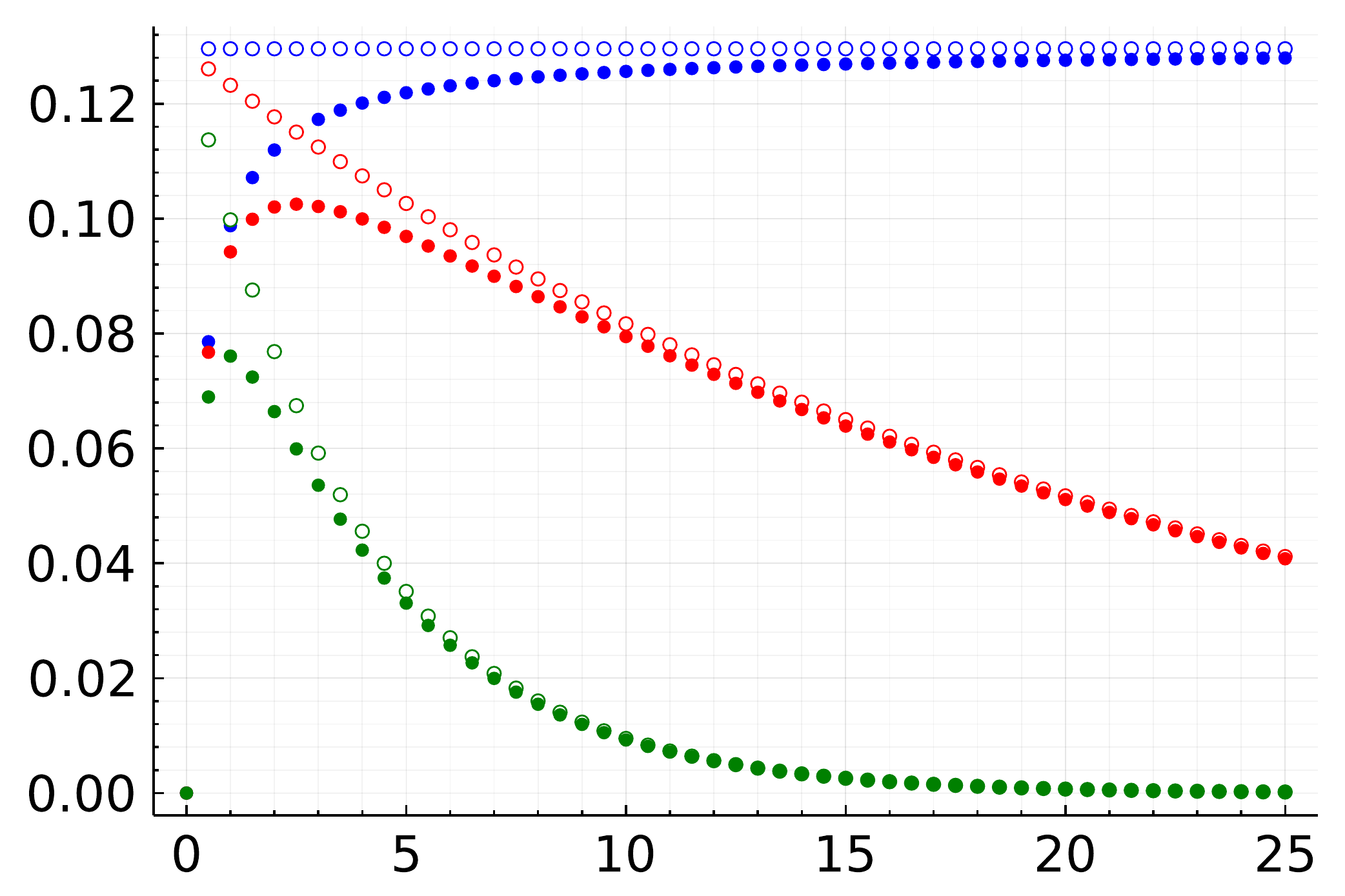} }
\put(250,5){ \includegraphics[scale=0.35]{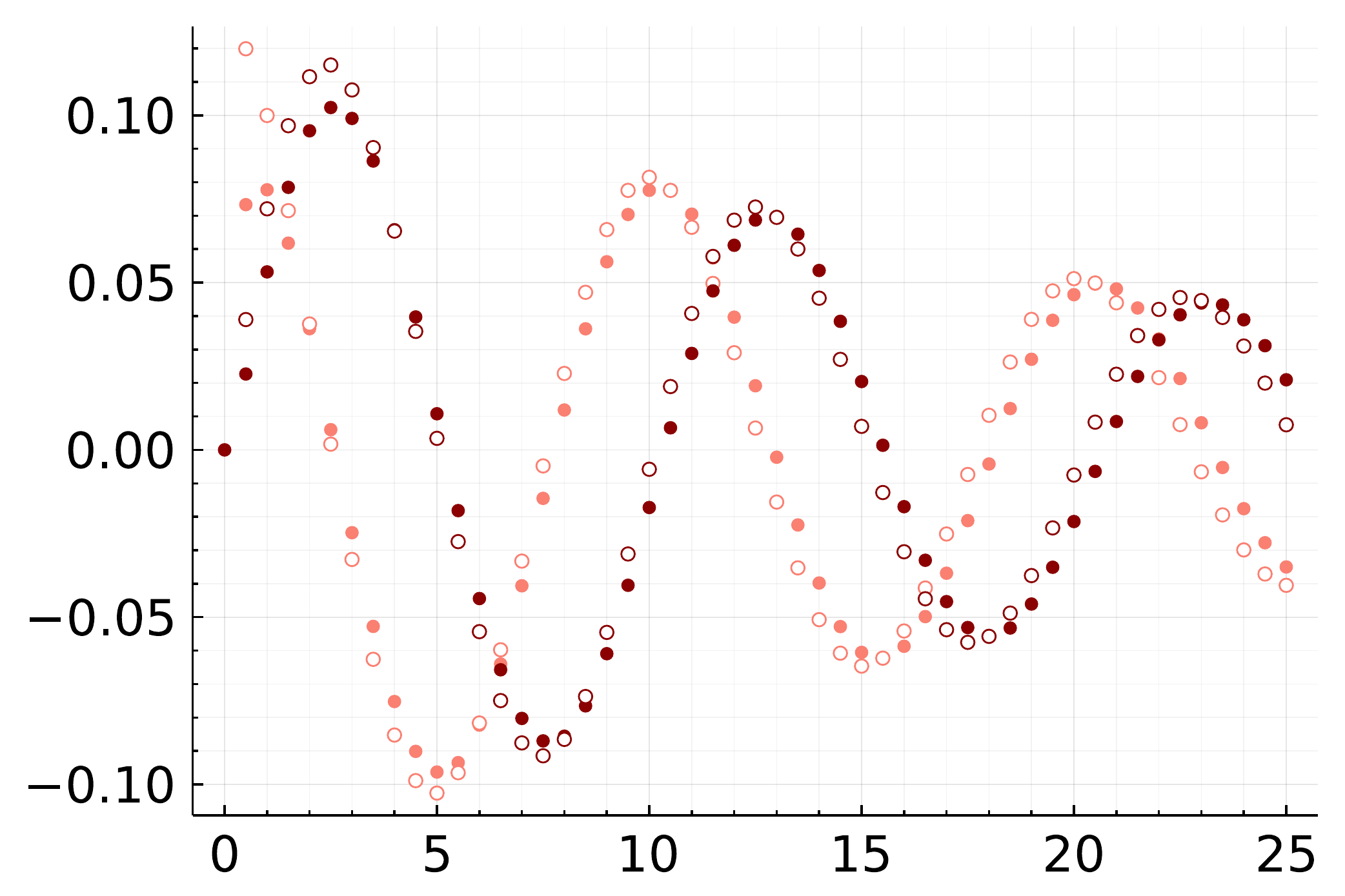} }

\put(2,120){$\lambda^\frac32 |G|$}
\put(240,120){$\lambda^\frac32 G$}

\put(130,0){$\lambda$}
\put(380,0){$\lambda$}

\put(160,118){\blue \circle*{3}}
\put(165,116){\scriptsize $G(\kappa_A,\kappa_A)$}

\put(160,105){\red \circle*{3}}
\put(165,103){\scriptsize $G(\kappa_A,\kappa_B)$}

\put(160,92){\color{green!50!black} \circle*{3}}
\put(165,90){\scriptsize $G(\kappa_A,\kappa_C)$}

\put(395,136){\color{red!50!white} \circle*{3}}
\put(400,134){\scriptsize $ {\rm Re}(G(\kappa_A,\kappa_B))$}
\put(395,124){\color{red!50!black} \circle*{3}}
\put(400,122){\scriptsize $ {\rm Im}(G(\kappa_A,\kappa_B))$}

\end{picture}
\caption{ A plot of SU(2) gluing constraints and their asymptotic formulas rescaled by $\lambda^{3/2}$ as function of half-integer spins. The left panel shows the absolute values of the gluing constraint for different boundary coherent data. The right panel shows the real and imaginary parts of the gluing constraint. The asymptotic data are plotted with empty circular points. The asymptotic formula closely approximates the gluing constraint already for spins $\lambda \sim 10$.}
\label{fig:B}
\end{figure}
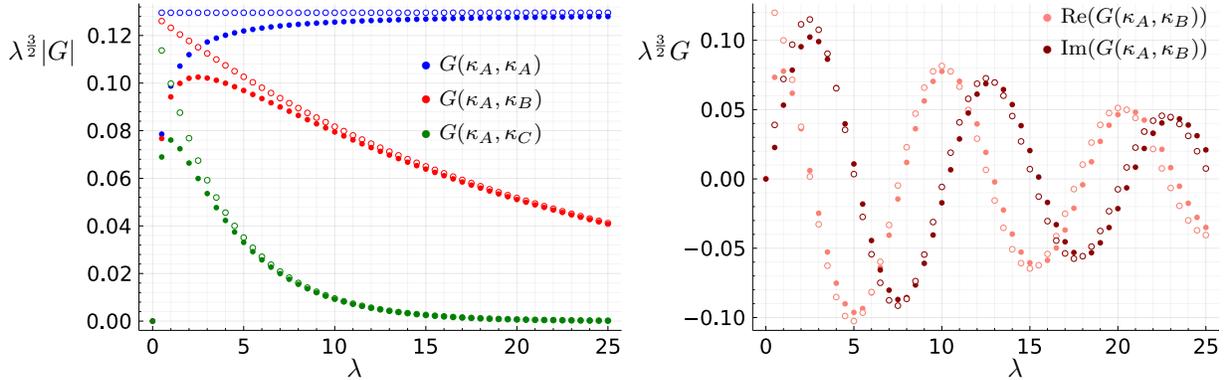

According to the analysis of section \ref{AsySU2}, the gluing constraints are supposed to have a power-law scaling $\sim \lambda^{-3/2}$ for non-collinear coherent states\footnote{We also checked that for colinear data the gluing constraints scale as $\sim \lambda^{-1}$.}. Accounting for this as in figure \ref{fig:B}, one again sees an exponential suppression for non-closing configurations and a slight suppression for non-shape matching configurations. We moreover find the asymptotic formula \eqref{asympsu2} matches the actual gluing constraint extremely well even for small spins.

\subsubsection{Closing but non-matching configurations}

Now consider the closing intertwiners $\kappa_A$ and $\kappa_{B}'$ as boundary states, where $\kappa_A$ is given in table \ref{tab:A} and $\kappa_B'$ is parametrized by  $\kappa_B'(\Phi_{2}) = \{j_i = \lambda, \arccos(-\tfrac13),\Phi_{2} \}$. By varying the angle $\Phi_{2}$ we can probe the behaviour of the constraints as a function of $\Phi_{2}$ for fixed spin $\lambda$. 
Since the boundary data satisfy closure, the gluing constraints probe how the shapes of two boundary tetrahedra fit together, i.e. how well they are shape-matched. We refer to $\Phi_{2}$ as a shape-matching parameter. 
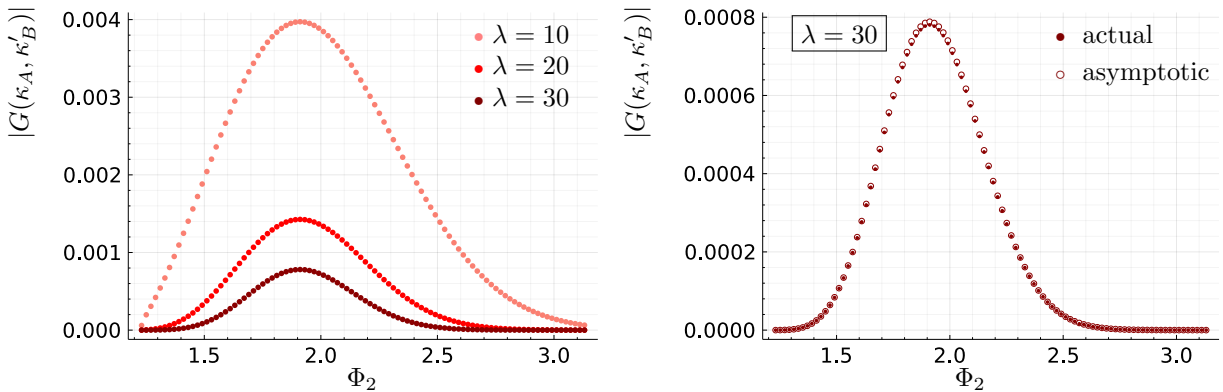
\begin{figure}[ht!]
\begin{picture}(500,145)
\put(15,5) { \includegraphics[scale=0.35]{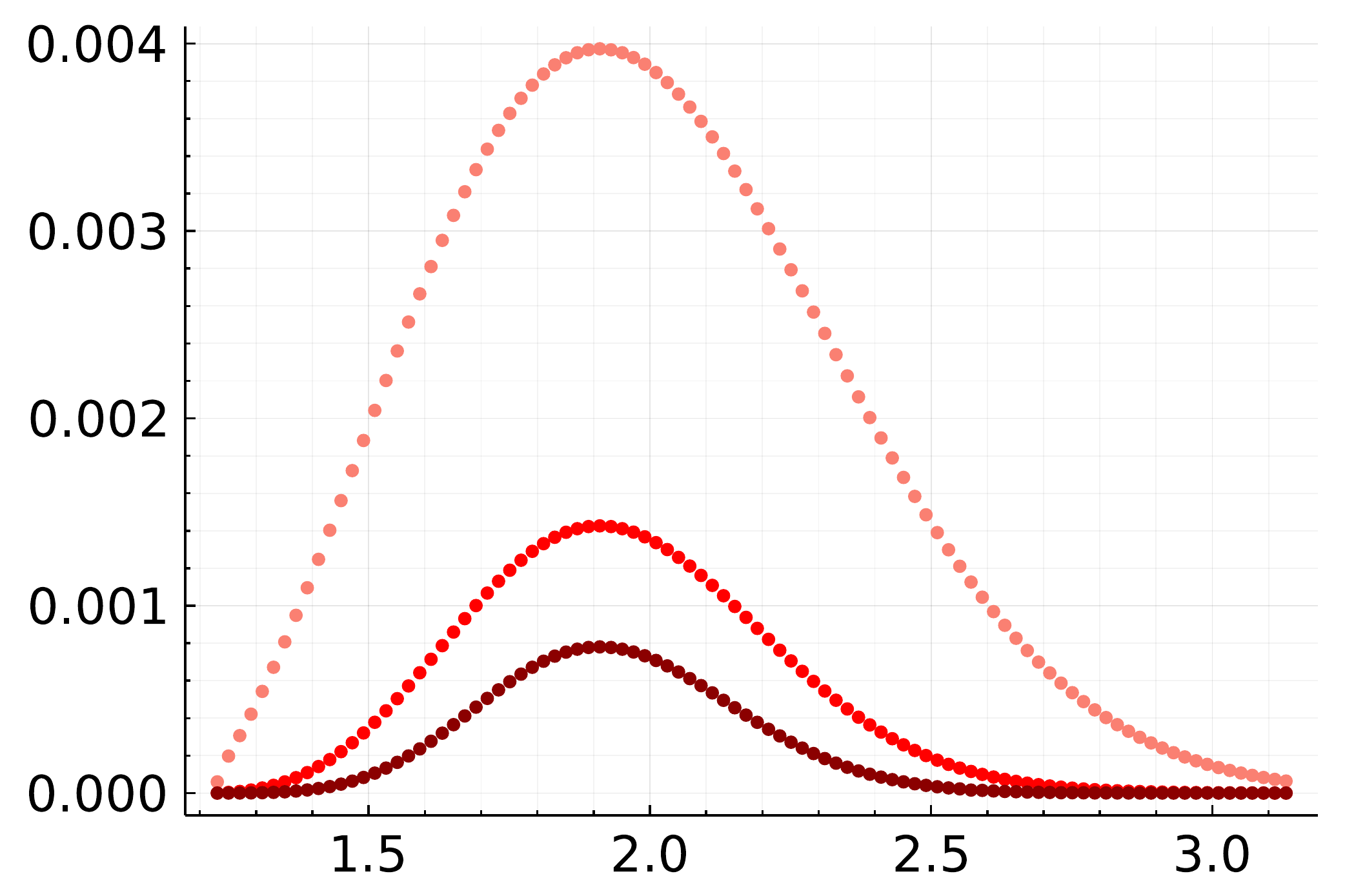} }
\put(250,5){ \includegraphics[scale=0.35]{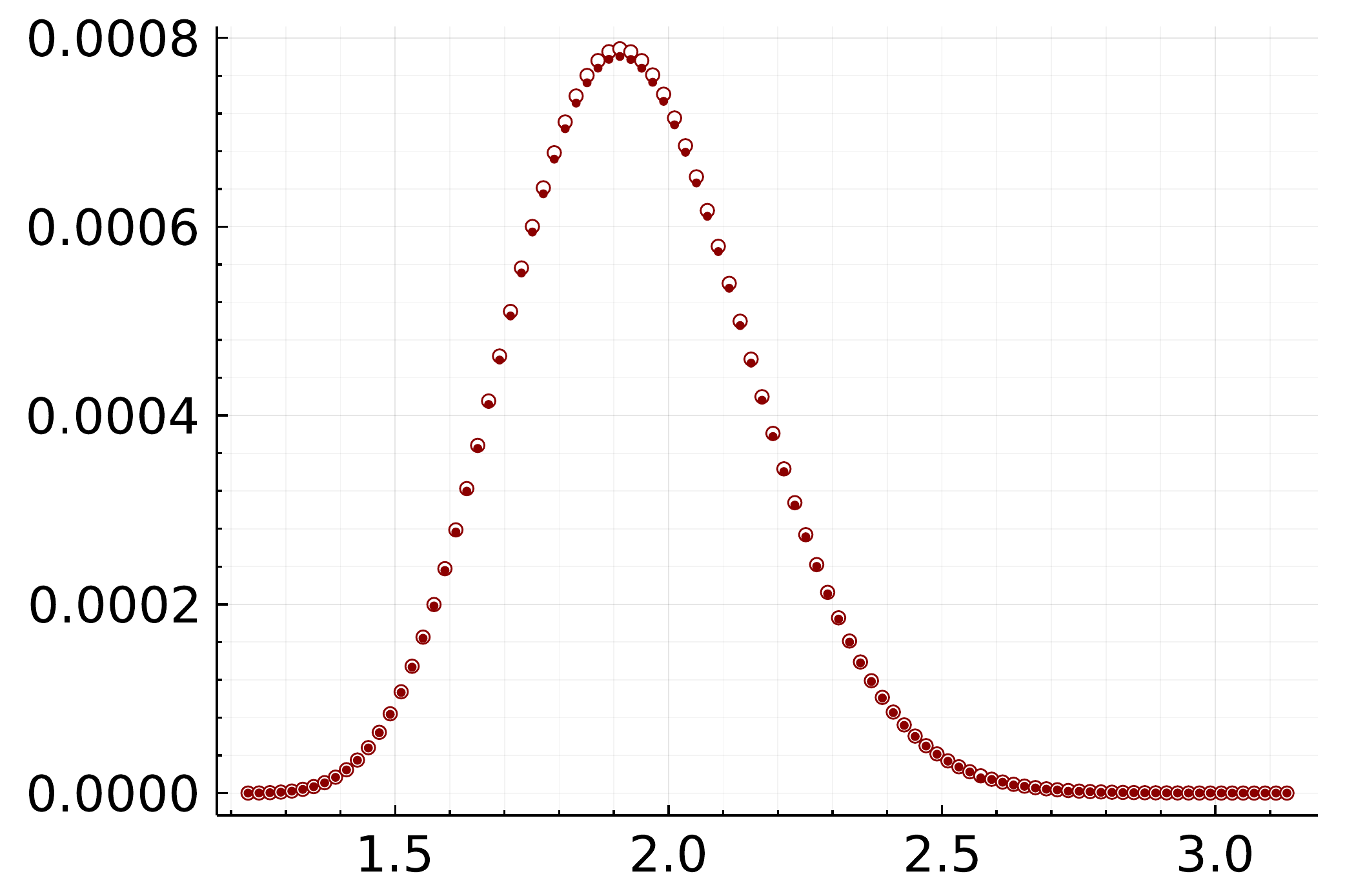} }

\put(2,90){\rotatebox{90}{ $ |G(\kappa_A,\kappa_B')| $} }

\put(235,90){\rotatebox{90}{ $ |G(\kappa_A,\kappa_B')| $} }

\put(130,0){$\Phi_2$}
\put(360,0){$\Phi_2$}

\put(180,132){\color{red!50!white} \circle*{3}}
\put(185,130){$\lambda = 10 $}

\put(180,120){\color{red} \circle*{3}}
\put(185,118){$\lambda = 20 $}

\put(180,108){\color{red!50!black} \circle*{3}}
\put(185,106){$\lambda = 30 $}

\put(400,132){\color{red!50!black} \circle*{3}}
\put(405,130){ actual }

\put(295,130){ $\boxed{\lambda=30}$ }


\put(398,116){\color{red!50!black} $\scalemath{0.8}{\circ}$}
\put(405,116){ asymptotic}

\end{picture}
\caption{SU$(2)$ gluing constraints for closing boundary data as a function of the shape-matching parameter at fixed spins. The plots show skewed Gaussian-like behaviours for large spins. The peaks correspond to shape-matching configurations at $\Phi_2 = \arccos(-\frac13)$. The right panel shows the asymptotic formula approximates almost exactly the amplitude.   }
\label{fig:C}
\end{figure}

In figure \ref{fig:C} we plot the constraints as functions of the shape-matching parameter. Notice that the constraints have broader tails than a Gaussian would, and their peaks are displaced from the center, being located at $\Phi_2 = \arccos(-\frac{1}{3}) $. 
It is precisely at this value of the shape-matching parameter that the boundary data satisfies the critical point equations of the usual asymptotic analysis found in the literature, corresponding to closed and shape-matched configurations. The right-hand side of figure \ref{fig:C} indicates that the refined asymptotic formula of equation \eqref{asympsu2} strongly reproduces the actual constraints even away from the critical point. 

\subsubsection{Non-closing and non-matching configurations}

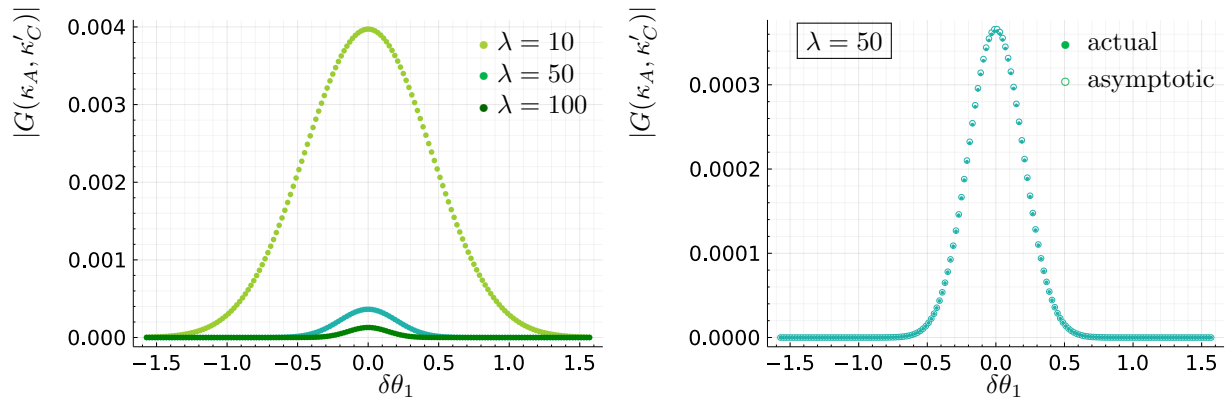
\begin{figure}[ht!]
\begin{picture}(500,145)
\put(15,5) { \includegraphics[scale=0.35]{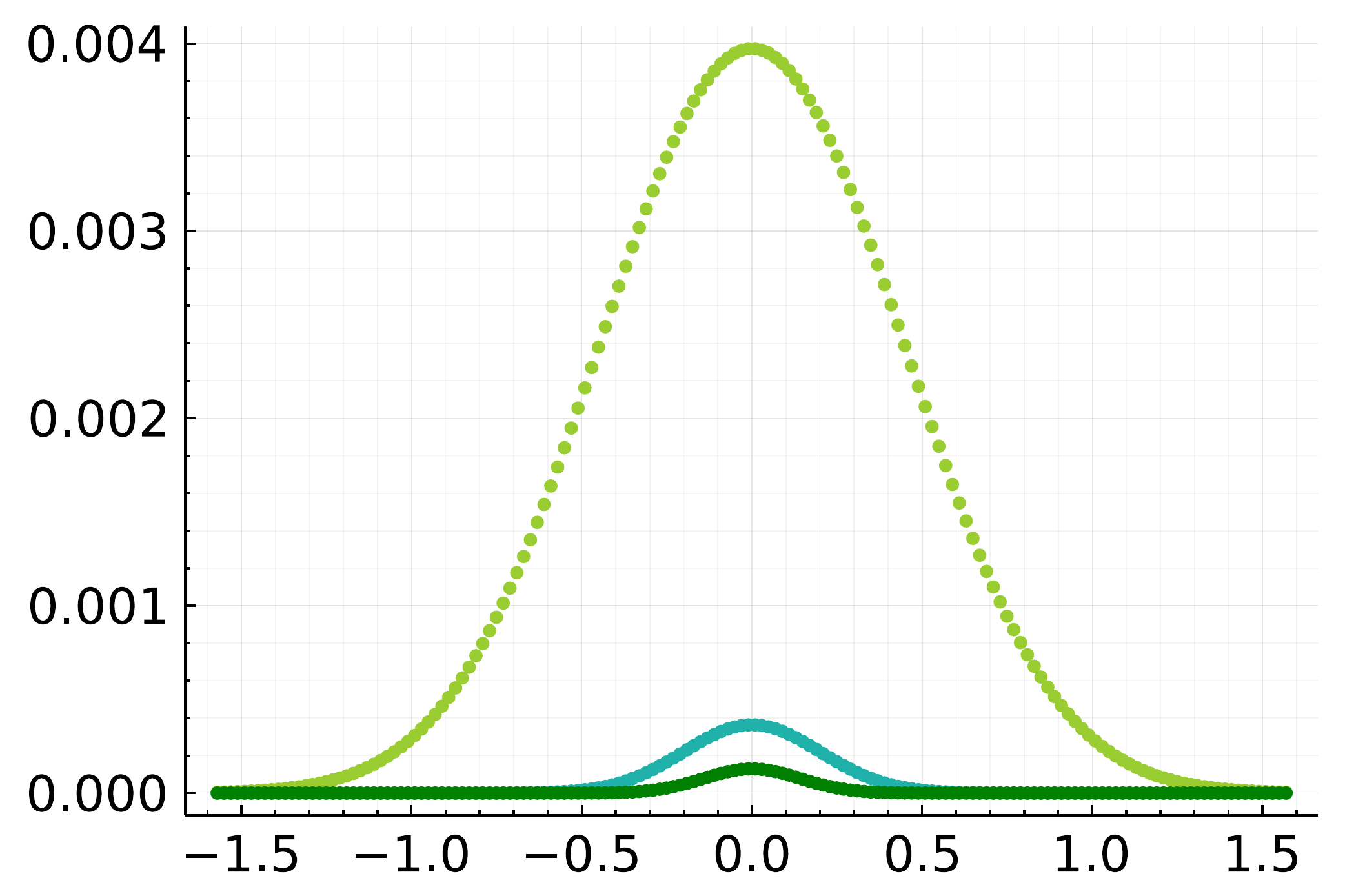} }
\put(250,5){ \includegraphics[scale=0.35]{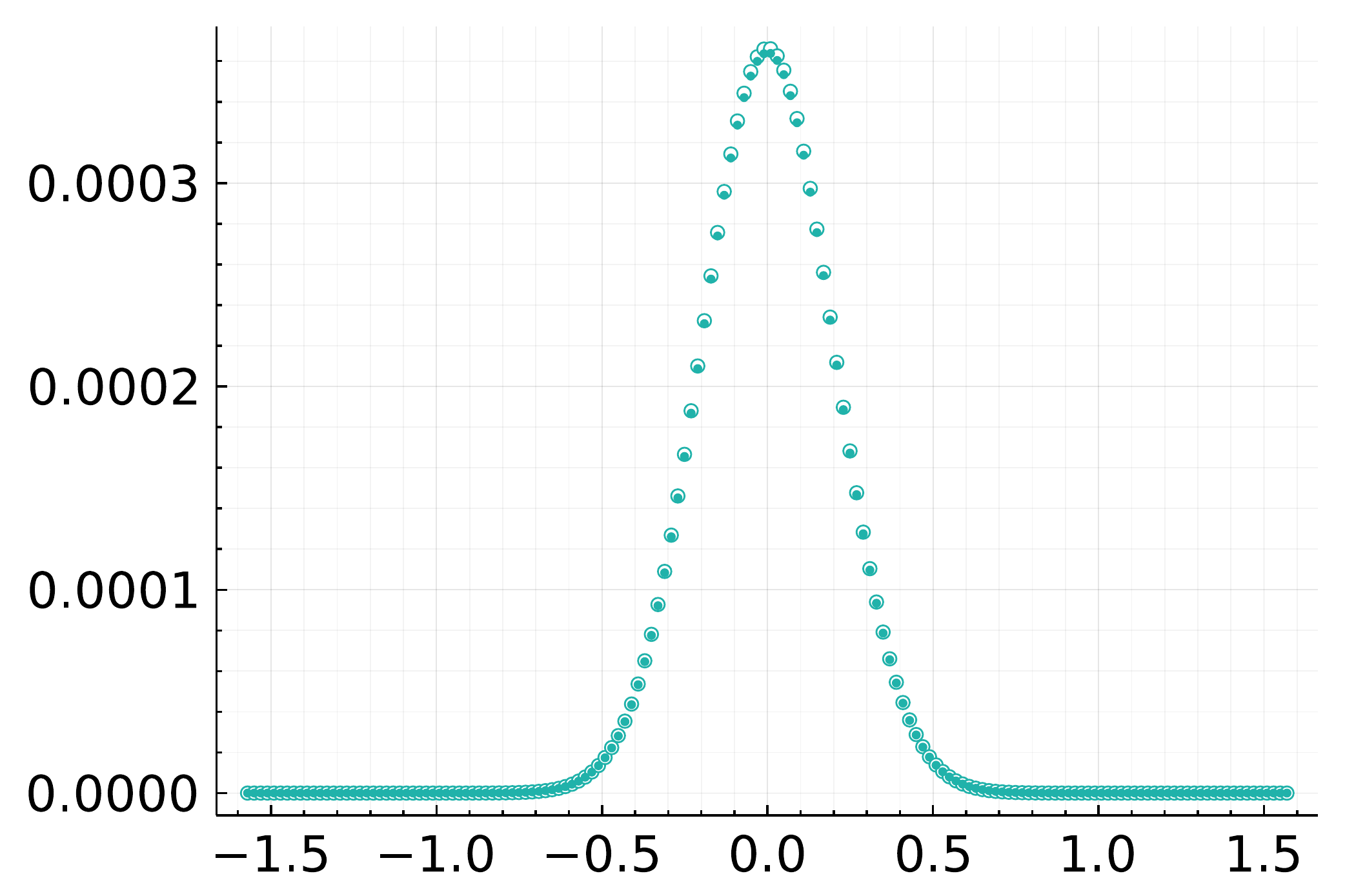} }

\put(2,93){\rotatebox{90}{ $ |G(\kappa_A,\kappa_C')| $} }

\put(235,93){\rotatebox{90}{ $ |G(\kappa_A,\kappa_C')| $} }

\put(140,0){$\delta \theta_{1}$}
\put(370,0){$\delta \theta_{1}$}

\put(180,132){\color{yellow!60!green} \circle*{3}}
\put(185,130){$\lambda = 10 $}

\put(180,120){\color{green!70!blue} \circle*{3}}
\put(185,118){$\lambda = 50 $}

\put(180,108){\color{green!40!black} \circle*{3}}
\put(185,106){$\lambda = 100 $}

\put(295,130){ $\boxed{\lambda=50}$ }

\put(398,116){\color{green!70!blue} $\scalemath{0.8}{\circ}$}
\put(405,116){ asymptotic}

\put(400,132){\color{green!70!blue} \circle*{3}}
\put(405,130){ actual }

\end{picture}
\caption{SU$(2)$ gluing constraints with boundary data $\kappa_A,\kappa'_C$ plotted as functions of angles (non-closing parameter) and for fixed spins. The right panel shows the Gaussian behaviour for both the actual amplitude and its asymptotic formula very well matched, peaked at $\delta\theta_1 = 0$.   }
\label{fig:D}
\end{figure}

Finally, we consider the constraints for non-closing configurations as a function of angles. 
To this end we make use of the intertwiner $\kappa_C$ described in table \ref{tab:A}, but allowing general variations $\delta\theta_1$, which we take to constitute a non-closing parameter for $\kappa_C'(\delta \theta_1)$. Figure \ref{fig:D} contains a plot of the overlap between the closing intertwiner $\kappa_A$ and the general one $\kappa_C'$ as a function of the non-closing parameter. The gluing constraints are Gaussian-shaped and their peak is located at $\delta \theta_1 = 0$, corresponding to a closing and shape-matched configuration $\kappa_A=\kappa_C'(0)$. Again, the asymptotic formula is in very good agreement with the numerical results on and away from the critical point. 

\subsection{$\sl2c$ gluing constraints}

We turn now to a study of the $\sl2c$ gluing constraints of equation \eqref{sl2glue2} and its asymptotic formula \eqref{asympsl2c}. The constraints themselves are defined via fourteen real improper integrations, coming from a six-dimensional integration over $\sl2c$ and two-dimensional real integrations over $\rm \mathbb C P^1$ for each of the four nodes. Since these integrals are known to be difficult to evaluate numerically, we chose to make use of \verb|sl2cfoam-next|\footnote{ The \emph{sl2cfoam-next} package uses a convention that fixes the phase so that it gives only real amplitudes. We therefore use it to compute only the absolute value of the gluing constraints.} developed in \cite{Gozzini:2021kbt} for the Lorentzian EPRL model. To do so, we formulated the constraints in terms of $\text{SU}(2)$ coherent intertwiners and boosted intertwiners, according to the proposal in \cite{Speziale:2016axj}. Diagrammatically, this is done through the set of equalities
\begin{equation}
 h_i\,\, \scalemath{0.8}{\input{gluing.tikz} }\,\, k_i \; =  \;  h_i\,\, \scalemath{0.8}{\input{cohboost1.tikz} }\,\, k_i\; =\; \sum_{\iota\, , \, \iota'}\; h_i\,\,\scalemath{0.8}{\input{cohboost2.tikz} }\,\, k_i\,,
\end{equation}
where the grey boxes represent $\text{SU}(2)$ integrations and the dotted box stands for a boost integration. Both the coherent intertwiners and the boosted intertwiner of the rightmost term can be natively computed in \verb|sl2cfoam-next|. In order to compare between the previous $\text{SU}(2)$ case and the current one, our numerical analysis uses the same boundary data as before. We moreover fix the Immirzi parameter to be $\gamma=0.123$. As we will show, the qualitative behaviour of the constraints for both models is approximately the same. 

\subsubsection{Scaling behaviour}

\begin{figure}[ht!]
\begin{picture}(500,145)
\put(20,5){ \includegraphics[scale=0.35]{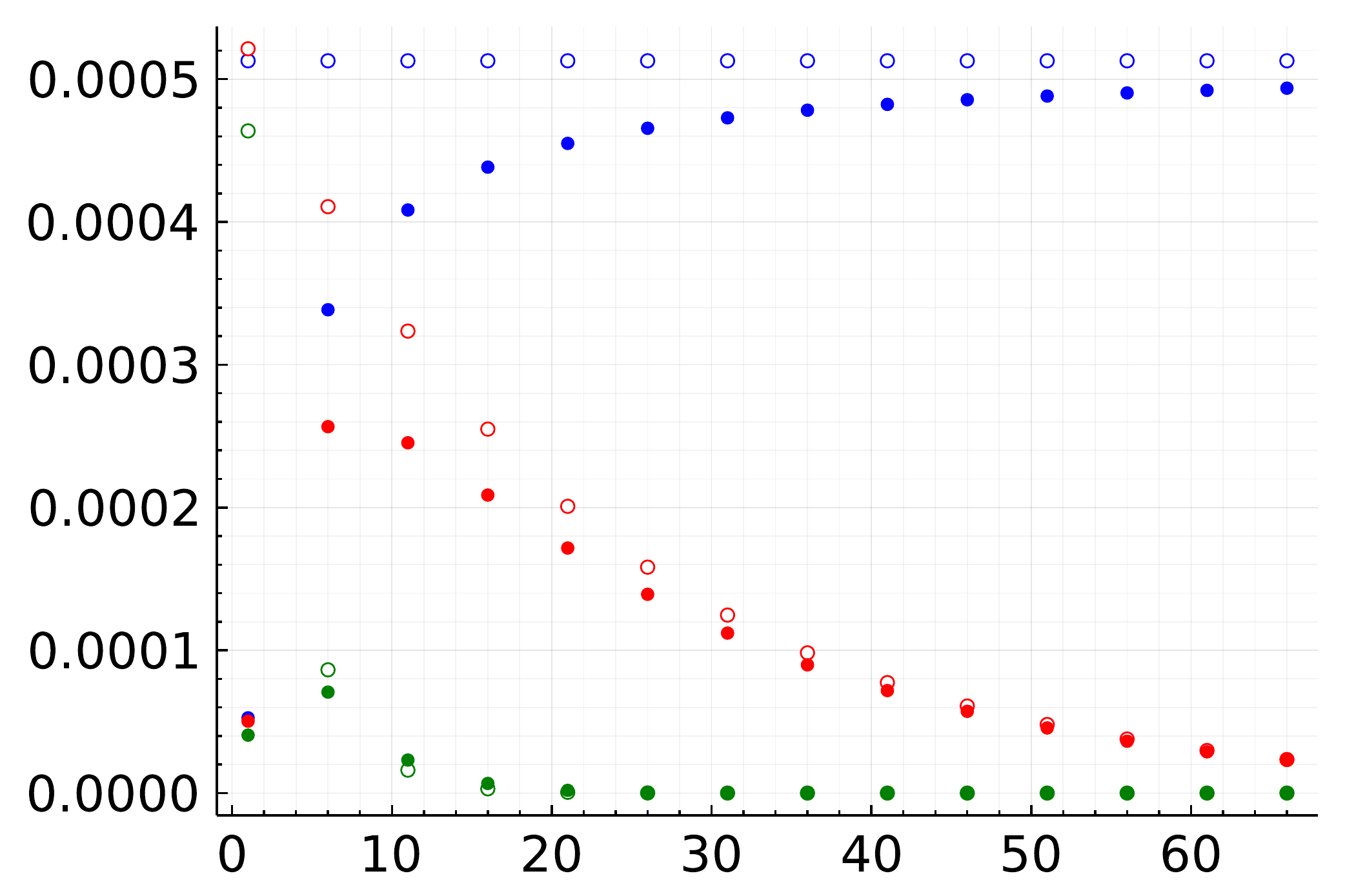} }
\put(250,5) { \includegraphics[scale=0.35]{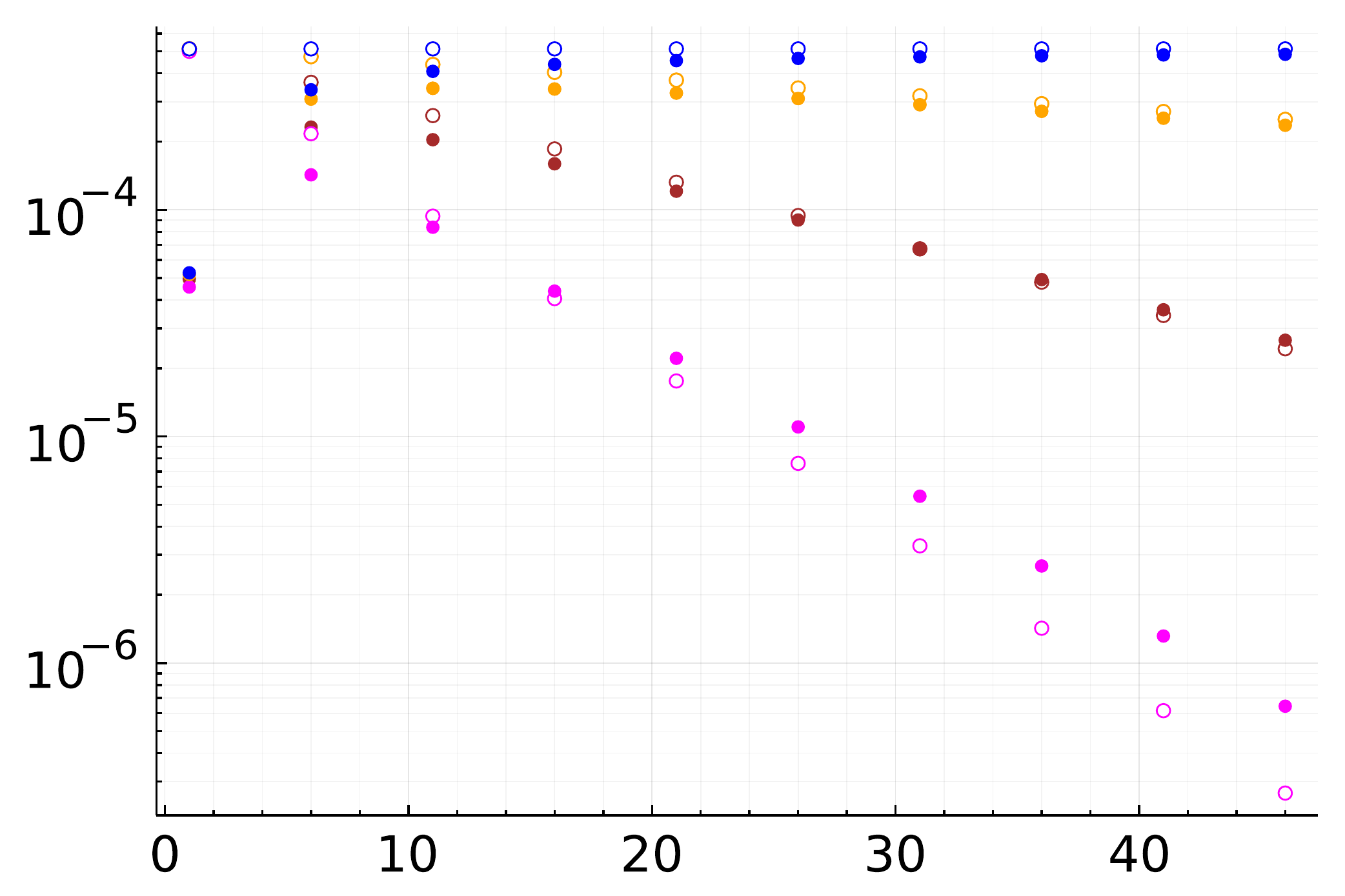} }

\put(130,0){$\lambda$}
\put(380,0){$\lambda$}

\put(0,130) {$\lambda'|G|$}

\put(160,108){\blue \circle*{3}}
\put(165,106){\scriptsize $G(\kappa_A,\kappa_A)$}

\put(160,95){\red \circle*{3}}
\put(165,93){\scriptsize $G(\kappa_A,\kappa_B)$}

\put(160,82){\color{green!50!black} \circle*{3}}
\put(165,80){\scriptsize $G(\kappa_A,\kappa_C)$}

\put(300,71){\blue \circle*{3}}
\put(305,69){\scriptsize $\delta \theta_1 = 0$}

\put(300,58){\color{orange} \circle*{3}}
\put(305,56){\scriptsize $\delta \theta_1 = 0.25$}

\put(300,45){\color{red!50!black} \circle*{3}}
\put(305,43){\scriptsize $\delta \theta_1 = 0.5$}

\put(300,32){\color{magenta!80!white} \circle*{3}}
\put(305,30){\scriptsize $\delta \theta_1 = 0.75$}
\end{picture}
\caption{The left panel contains a plot of $\sl2c$ gluing constraints for the data of table \ref{tab:A} scaled with $\lambda'$, for fixed $\gamma = 0.123$. It shows a power-law scaling for matching configurations and an exponential suppression otherwise. The right panel is a log-plot of the gluing constraints $ \lambda' \, |G(\kappa_A,\kappa_C)|$ and its asymptotic formula for different values of $\delta \theta_1$. The exponential suppression in the asymptotic formula is faster than the actual amplitude for sufficiently large deviations from the critical point.}
\label{fig:slC1}
\end{figure}
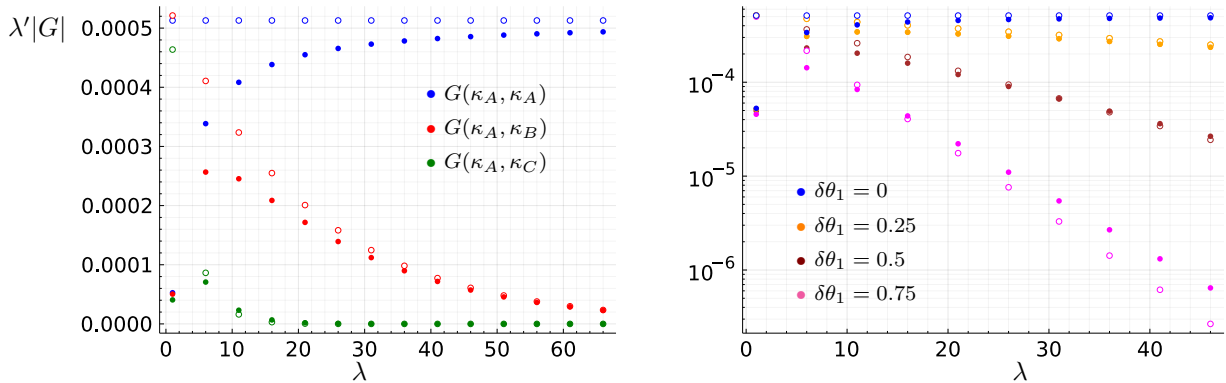

As per section \ref{asysl}, the gluing constraints are expected to scale with $~\lambda^{-3}$ at critical points, due to the fourteen-dimensional integration (contributing a $\lambda^{-7}$ scaling) and to the $(2\lambda+1)^4$ normalization factor in the inner products of coherent states \eqref{s2cohc}. Below, when ploting scaled gluing constraints, we will take the scale to be $\lambda'=\lambda^{-7}\times (2\lambda+1)^4$.
The left panel of figure \ref{fig:slC1} shows a power-law decay and exponential suppression for matching and non-matching boundary data, respectively. We find the qualitative behaviour of both the constraints and the asymptotic formula to be very similar to the $\text{SU}(2)$ case, although the convergence of both values seems to happen for much larger spins. 

The right panel in figure \ref{fig:slC1} is a log-plot of the gluing constraints for data of the type $\kappa_A, \kappa'_C$, i.e. for data which is displaced from criticality by some fixed amount $\delta \theta_1$. We observe a curious phenomenon where the asymptotic formula seems to slightly overestimate the suppression of the constraint, at a rate which is increasingly higher for larger deviations. We noticed these discrepancies become worse with larger values of the Immirzi parameter $\gamma$. After carefully checking our code, we believe that the source of this discrepancy is rooted in a numerical instability: the inverse of the Hessian matrix, which enters in the exponential function, depends substantially on very small changes of the matrix coefficients. We are therefore forced to take our results away from the critical points with a grain of salt, until a better understanding of these instabilities is achieved.




\subsubsection{Closing but non-matching configurations}

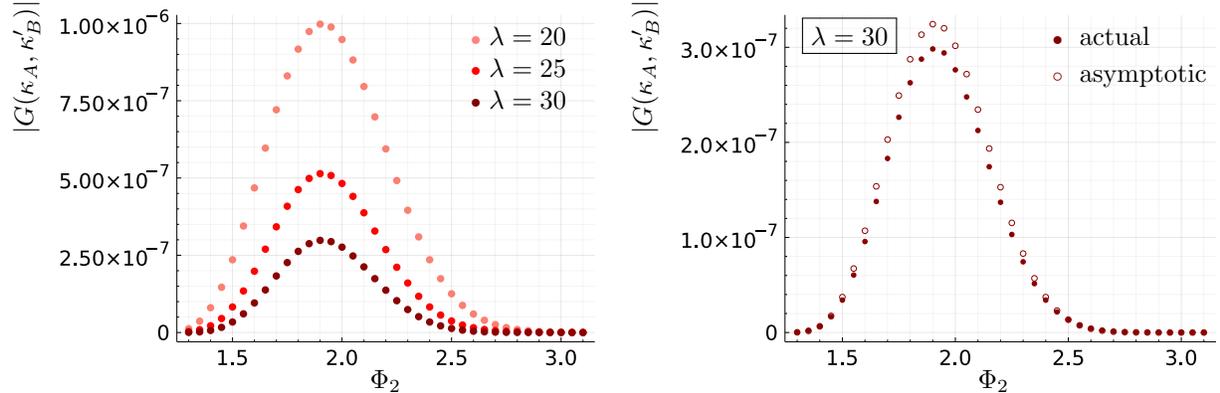
\begin{figure}[ht!]
\begin{picture}(500,148)
\put(15,5) { \includegraphics[scale=0.35]{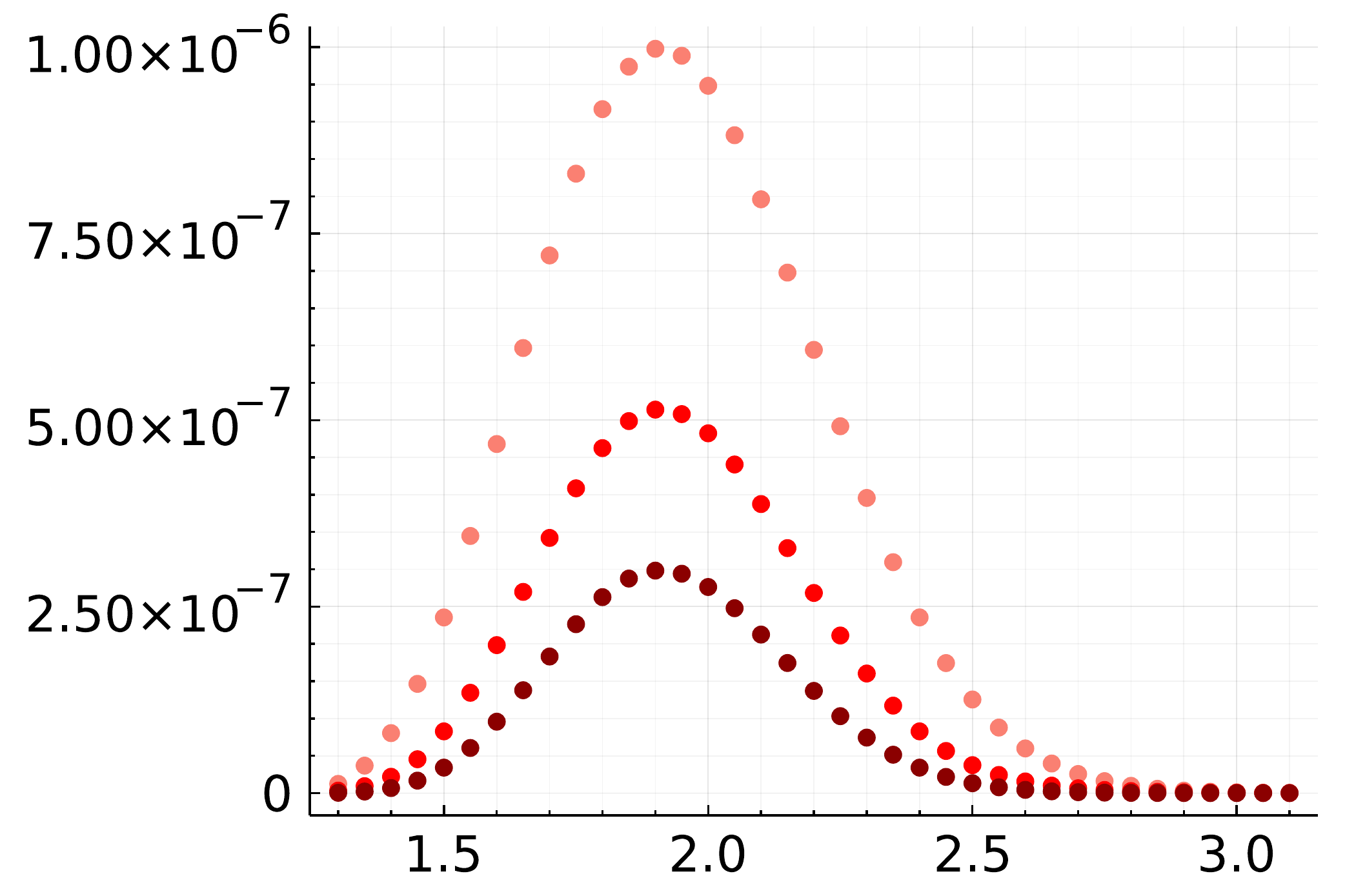} }
\put(250,5) { \includegraphics[scale=0.35]{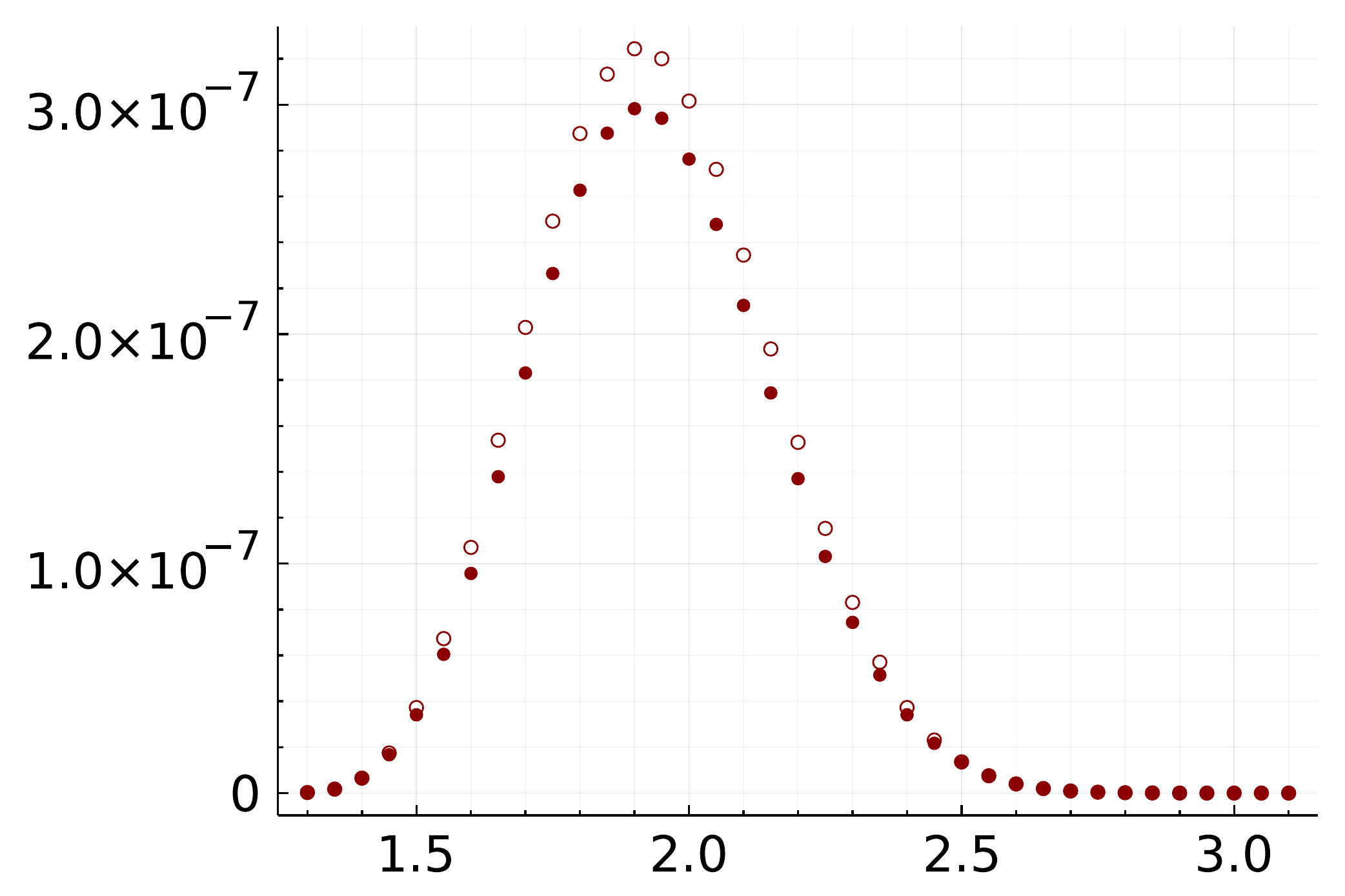} }

\put(5,93){\rotatebox{90}{ $ |G(\kappa_A,\kappa_B')| $} }

\put(239,93){\rotatebox{90}{ $ |G(\kappa_A,\kappa_B')| $} }

\put(180,132){\color{red!50!white} \circle*{3}}
\put(185,130){$\lambda = 20 $}

\put(180,120){\color{red} \circle*{3}}
\put(185,118){$\lambda = 25 $}

\put(180,108){\color{red!50!black} \circle*{3}}
\put(185,106){$\lambda = 30 $}

\put(300,130){ $\boxed{\lambda=30}$ }

\put(398,116){\color{red!50!black} $\scalemath{0.8}{\circ}$}
\put(405,116){ asymptotic}

\put(400,132){\color{red!50!black} \circle*{3}}
\put(405,130){ actual }

\put(140,0){$\Phi_2$}
\put(370,0){$\Phi_2$}

\end{picture}
\caption{Gluing constraints as functions of dihedral angles $\Phi_2$ show a skewed Gaussian-like behaviour. The peak is at matching configurations with $\Phi_2 = \arccos(-\tfrac13)$. The asymptotic formula matches the constraints well away from the critical point.  }
\label{fig:slC2}
\end{figure}

Similarly to the $\text{SU}(2)$ BF case, and according to the left panel of figure \ref{fig:slC2}, the gluing constraints as functions of the shape-matching parameter $\Phi_2$ are Gaussian-like but with broader tails. One finds moreover through the right panel that the asymptotic approximation reproduces the overall behaviour of the constraints, although, as previously mentioned, the convergence between both is worse at spin $\lambda=30$ compared to the BF model. 

\subsubsection{Non-closing and non-matching configurations}

For our last example of non-closing and non-matching configurations, we once more recover the general properties of $\text{SU}(2)$ BF gluing constraints. The asymptotic formula still matches the full amplitude reasonably well, and the highest absolute disparity is found at the peak. We have generally found this to be the case, and for the absolute error between the constraints and their asymptotic approximation to be smaller away from the critical point. 

\begin{figure}[ht!]
\begin{picture}(500,148)
\put(15,5) { \includegraphics[scale=0.35]{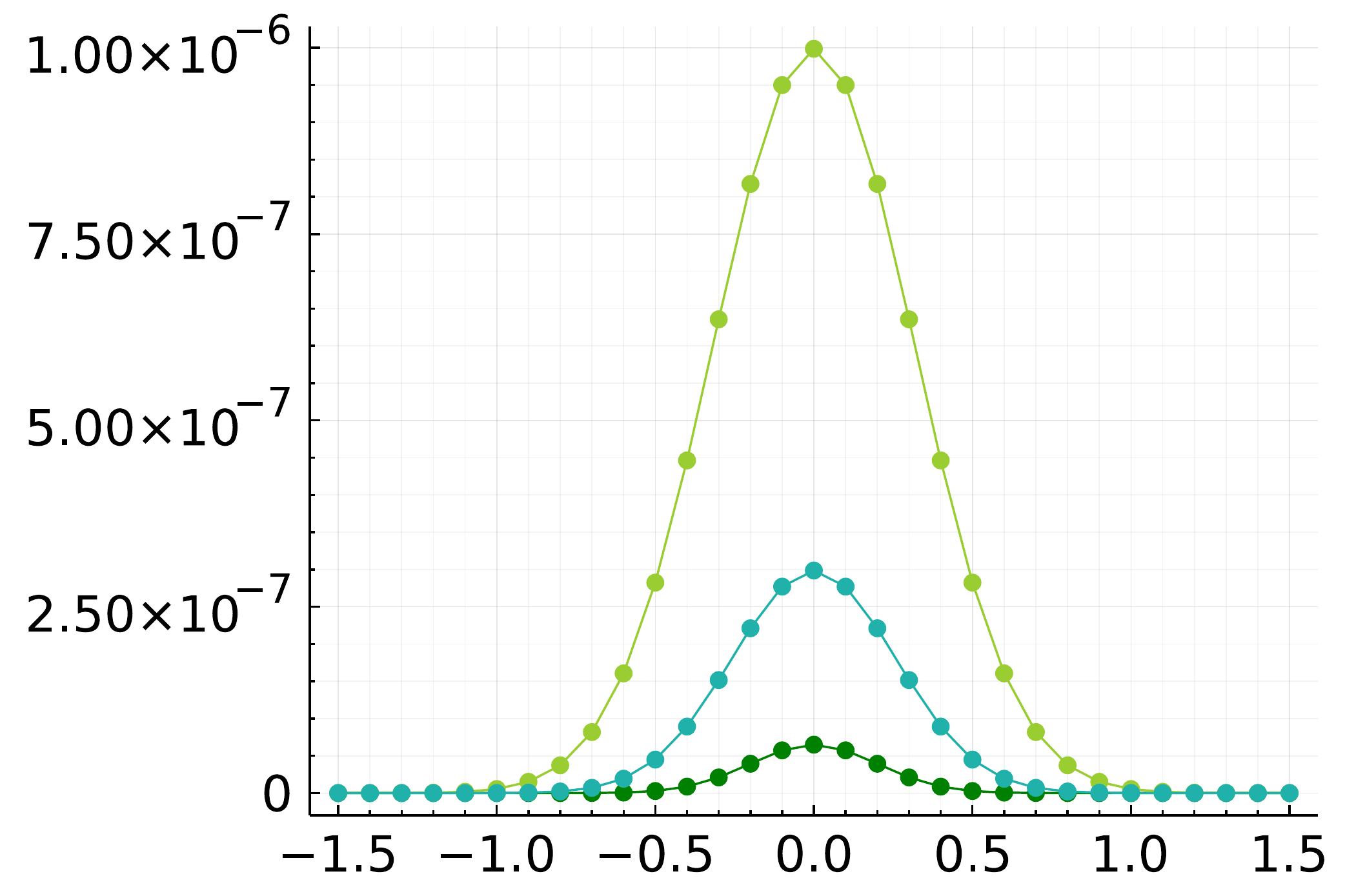} }
\put(250,5){ \includegraphics[scale=0.35]{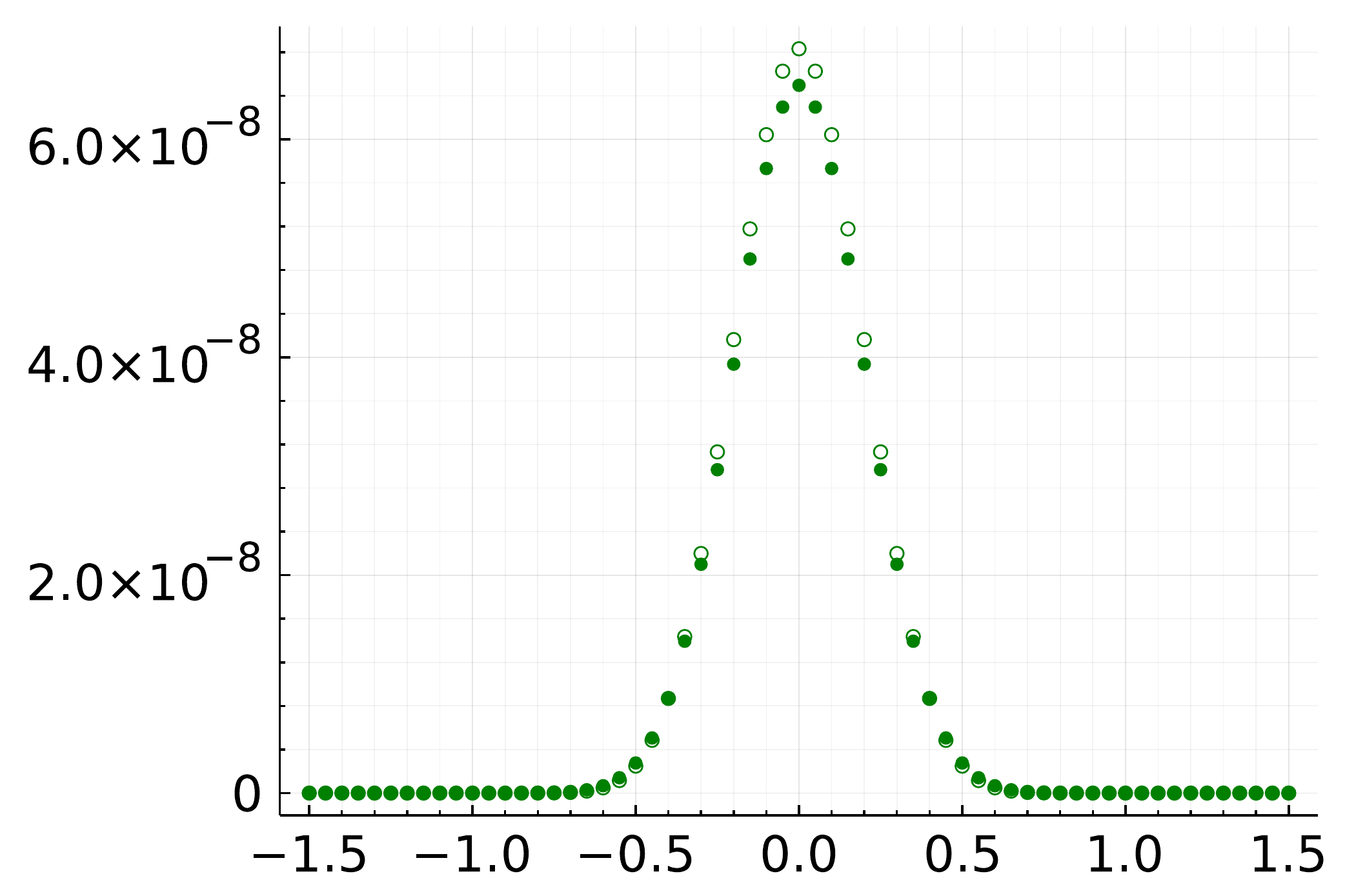} }

\put(2,93){\rotatebox{90}{ $ |G(\kappa_A,\kappa_C')| $} }

\put(235,93){\rotatebox{90}{ $ |G(\kappa_A,\kappa_C')| $} }

\put(140,0){$\delta \theta_{1}$}
\put(370,0){$\delta \theta_{1}$}
\thicklines
\put(180,132){\color{yellow!60!green} \circle*{3}}
\put(175,132){\color{yellow!60!green} \line(1,0){10}} 
\put(188,130){$\lambda = 20 $}

\put(180,120){\color{green!70!blue} \circle*{3}}
\put(175,120){\color{green!60!blue} \line(1,0){10}} 
\put(188,118){$\lambda = 30 $}

\put(175,108){\color{green!50!black} \line(1,0){10}}
\put(180,108){\color{green!50!black} \circle*{3}}
\put(188,106){$\lambda = 50 $}

\put(310,130){ $\boxed{\lambda=50}$ }

\put(398,116){\color{green!50!black} $\scalemath{0.8}{\circ}$}
\put(405,116){ asymptotic}

\put(400,132){\color{green!50!black} \circle*{3}}
\put(405,130){ actual }
\end{picture}
\caption{Gluing constraints as functions of the non-closing parameter $\delta \theta_1$. They are Gaussian with a larger standard deviation for large $\lambda$, and they are peaked at the critical $\delta \theta_1 = 0$. The asymptotic formula is well matched away from the peak.  }
\label{fig:slC3}
\end{figure}
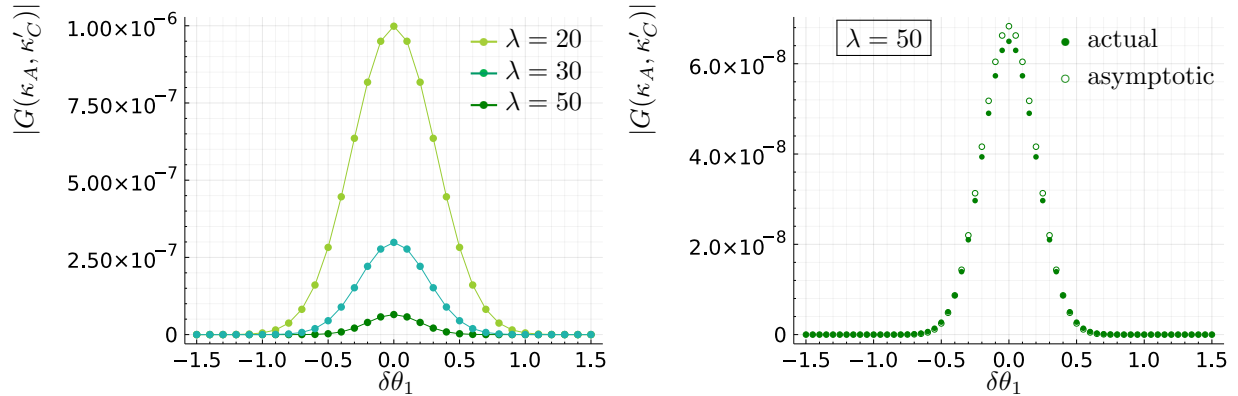

~~

\section{Discussion} \label{discussion}

In this article we have presented a new representation of the spin-foam partition function in terms of coherent states. By using resolutions of the identity we equipped each edge of the spin-foam with two sets of coherent data, one for each vertex. Each vertex thus carries its own independent set of data, and vertices are glued to their neighbours via gluing constraints. These constraints enforce shape matching weakly and are named after the gluing constraints defined in effective spin-foams \cite{Asante:2020qpa}. In this way, for large enough representations, one can apply the asymptotic formula of the vertex amplitude directly at each vertex, each of which is dominated only by its own critical points. From this we have conjectured a new quasi-geometric regime of spin-foams, which describes a superposition of semi-classical vertices glued in a non-matching way. Our future goal is to characterize this regime and use it to augment existing numerical algorithms by building a ``hybrid algorithm'', which switches to less costly semi-classical methods as soon as these provide a good approximation.

For now we focused mainly on the properties of gauge-invariant gluing constraints, which we defined for SU$(2)$ BF theory and the Lorentzian EPRL model. We computed these constraints numerically and analytically for different sets of boundary data, e.g. corresponding to non-closing tetrahedra and tetrahedra of different shapes. From both we found that they are (almost) Gaussian peaked on shape matching and closure of the coherent data on both ends, the critical points of the gluing constraints. Crucially, we employed H\"ormander's theorem to calculate the gluing constraints beyond those critical points, confirming their Gaussian shapes and peak locations. Indeed, the asymptotic formula provides an excellent approximation for the SU$(2)$ gluing constraints around spin $j \sim 10$, whereas the EPRL one is well approximated beyond spins $j\sim 50$. The fact that we can analytically compute the gluing constraints (far) away from their critical points is an encouraging sign for future numerical simulations, and it might be possible to use this to significantly speed up calculations. Furthermore, our analysis away from critical points may perhaps in the future be applied to more complicated objects such as the coherent vertex amplitude itself.

At this stage we would like to comment on the different types of gluing constraints derived in this article, i.e. the $\text{SU}(2)$ type and the Lorentzian EPRL type, and on which role they might play in the Lorentzian EPRL model. From our perspective, both types can in principle be applied to split vertices in the model. Indeed, the $\text{SU}(2)$ gluing constraints are in agreement with the EPRL construction, since the particular states entering the vertex amplitude all stem from $\text{SU}(2)$ representations embedded into $\text{SL}(2,\mathbb{C})$ ones. Vertex amplitudes in the Lorentzian EPRL model can be split at the level of $\text{SU}(2)$ states, e.g. $\text{SU}(2)$ intertwiners in the spin network basis \cite{Dona:2022dxs}, such that one could straightforwardly apply the $\text{SU}(2)$ gluing constraints, which are independent of the Immirzi parameter $\gamma$. On the other hand, one can argue for the $\text{SL}(2,\mathbb{C})$ gluing constraints as we defined them. Given that $\text{SL}(2,\mathbb{C})$ is the underlying symmetry group of the theory, the $\text{SL}(2,\mathbb{C})$ constraints would directly project onto the relevant gauge-invariant subspace. They moreover have the added benefit of structurally resembling the actual vertex amplitude, such that they can be used as a more tractable toy model. Still, opting for $\text{SU}(2)$ gluing constraints might be numerically more efficient, and it might correctly capture the relevant physics.


There exists another reason that drove our study of the Lorentzian EPRL gluing constraints. In effective spin-foam models \cite{Asante:2020qpa,Asante:2020iwm}, the postulated gluing constraints are motivated from commutation relations of 3d dihedral angles in tetrahedra, and they are related to the weak imposition of simplicity constraints in spin-foams. This imposition should be connected to the Immirzi parameter $\gamma$. One of our goals was therefore to define gluing constraints for the EPRL model with an explicit $\gamma$-dependence, and to compare them to this previous proposal. To this end we have studied the constraints for boundary data corresponding to geometric tetrahedra that differ in their two angles that parametrize them. As mentioned above, we find the EPRL gluing constraints to be almost Gaussian shaped (with slightly too long tails), peaked on shape matching and showing little $\gamma$-dependence. What this implies regarding the weak imposition of constrains in the model, and whether the constraints restrict strongly the allowed geometries, are two interesting questions to investigate in the future. 

With the definition of the gluing constraints and their accurate asymptotic formula on and away from critical points, an important first step towards building the hybrid algorithm is done. However, many open conceptual and technical questions must be addressed. One of the most pressing ones is the regime in which such an algorithm is valid; in this work we have presented some first evidence, but more thorough investigations are needed.
Work on the coherent vertex amplitude for higher valent spin-foam vertices \cite{Allen:2022unb} suggests that the transition between the quantum and semi-classical vertex regimes might be more intricate, e.g. if some spins are large yet others remain small. Moreover, it is vital to generate all critical vertex configurations for a given set of ten representations, which can correspond to degenerate, vector or Regge geometries. Previous work in this direction exists as part of the \verb|sl2cfoam-next| package \cite{Dona:2019dkf,Gozzini:2021kbt}, as does an algorithm that translates from areas to lengths configurations in the framework of effective spin-foams for different signatures \cite{Asante:2020iwm,Asante:2021phx}. We moreover hope that our algorithm will help shed light on the question of what role configurations other than Regge geometries play in the spin-foam path integral, e.g. whether gluing degenerate or vector geometries to Regge geometries is (strongly) suppressed. This might help us identify the most relevant geometries of multiple vertices in the path integral. Eventually, we must develop the hybrid algorithm in detail and demonstrate that we can faithfully approximate the full quantum amplitudes by non-matching semi-classical vertices glued together via gluing constraints.

\section*{Acknowledgements}

The authors would like to thank Francesco Gozzini for his support on how to use the package \verb|sl2cfoam-next| and his help in comparing our conventions to those used in that package. S.St. would also like to thank Parampreet Singh for his clear presentation of the ``Chimera'' algorithm at the workshop ``Quantum Gravity on the computer'' at Nordita in 2018. J.D.S. is grateful to Michael Mandl for an illuminating discussion regarding Maurer-Cartan forms. 

J.D.S. and S.St. gratefully acknowledge support by the Deutsche Forschungsgemeinschaft (DFG, German Research Foundation) - Projektnummer/project-number 422809950. SKA is supported by the Alexander von Humboldt foundation.

\appendix


\section{Parametrizing closing and non-closing SU(2) intertwiners} \label{AppendixS}



It is well known that the geometry of a classical tetrahedron can be determined from its six edge lengths or equivalently from its four triangle areas and two (non-opposite) dihedral angles up to $\rm SO(3)$ rotations.
A unit normal vector $\hat k_i$ associated to a face $f_i$ of a tetrahedron is a vector in the unit two-sphere $S^2$.  
Such a unit normal vector can be parametrized by the spherical angles as
\be\label{nomEuler}
\hat k_i(\theta_i, \phi_i) = \pm \left( \sin \theta_i \cos \phi_i , \sin \theta_i \sin \phi_i, \cos \theta_i  \right) \, .
\ee
The form of the normal vectors follows from the particular parametrization of SU(2) in \eqref{eulersu2}.  Let $k_i, \, i = 1,\cdots 4$ be the outward pointing normal vectors (not normalized) belonging to the face of the tetrahedron $\tau$. Without loss of generality, we can use the freedom under $\rm SO(3)$ rotations to take one of the normal vectors to be along the $z$-axis, and to take a second vector to be in the $y$-$z$ plane. That is, we choose the normal vectors (not normalized) to be of the form
\be
\vec k_1 = \left( 0,0, c_1 \right) \, , \q \vec k_2 = (0, b_2, c_2 ) \, , \q \vec k_3 = (a_3, b_3, c_3 ) \, , \q \vec k_4 = (a_4, b_4, c_4 ) \, .
\ee
The normal vectors are chosen such that the norm of each vector gives the corresponding area of the triangle face.  For a closed tetrahedron the normal vectors satisfy the closure condition
\be \label{clos-con}
{\cal C} := \sum_{i=1}^4 \vec k_i = \vec 0\, ,
\ee
and hence $\vec k_4 = -(\vec k_1 + \vec k_2 +\vec k_3)$. 
These normal vectors determine the geometry of a tetrahedron up to $\rm SO(3)$ rotations. 
Note that the squared volume of the tetrahedron in terms of the normal vectors is $V^2_\tau = -\vec k_1 \cdot (\vec k_2 \times \vec k_3 )$. 

As an example, consider four area square terms $ p_{ii} = \vec k_i \cdot \vec k_i = A_i^2, \, \, i = 1\cdots 4$ and two inner product terms $ p_{1j} = \vec k_1 \cdot \vec k_j$ for $j = 2,3$ (non-opposite edges). We can easily solve for the variables $a_i,b_i,c_i$ in terms of the $p_{ii}$ and $p_{1j}$. 
This results in vectors of the form
\ba\label{nomdih}
\vec k_1 &=& (0,0,\sqrt{p_{11}})\,, \nn \\
\vec k_2 &=& \frac{1}{\sqrt{p_{11}}}\left(0, \sqrt{p_{11} p_{22} - p_{12}^2}, p_{12}\right)\,, \\
\vec k_3 &=& \left( \frac{V_\tau^2}{\sqrt{p_{11} p_{22} - p_{12}^2}},\frac{(2p_{12}p_{13}+p_{11}(p_{11} + p_{22} + p_{33} - p_{44} + 2p_{12}+ 2p_{13})}{2\sqrt{p_{11}(p_{11} p_{22} - p_{12}^2)}} , \frac{p_{13}}{\sqrt{p_{11}}} \right) \,, \nn \q \\
\vec k_4 &=& - (\vec k_1 +\vec k_2 + \vec k_3) \,, \q \nn
\ea
where the squared volume is 
\ba
V_\tau^2 = &&  \bigg(- p_{12}p_{13} \left( p_{11} + p_{22} + p_{33} - p_{44} + 2p_{12}+ 2p_{13} \right)- \frac{p_{11}}{4}\left( p_{11} + p_{22} + p_{33} - p_{44} + 2p_{12}+ 2p_{13} \right)^2 \nn \bigg. \\
&& \q \q + \bigg.\, p_{33}(p_{11} p_{22} - p_{12}^2 ) - p_{13}^2 p_{22}   \bigg)^\frac12 \, .
\ea
Each normal vector is of the form $\vec k_i = \sqrt{p_{ii}}\, \hat k_i$ for all $i$, and hence the inner products can be given in terms of Euler angles associated to the unit normal vectors (see \eqref{nomEuler}). For the vectors in \eqref{nomdih}, we get the inner products 
\be
p_{12} = \sqrt{ p_{11} } \sqrt{ p_{22} } \cos \theta_{2} \,\, , \q \q  p_{13} = \sqrt{ p_{11}} \sqrt{ p_{33} } \cos \theta_{3} \,. \q
\ee
The Euler angles $\theta_{2}, \theta_{3}$ are exactly the dihedral angles of the associated edges $e_{12}, e_{13}$. In general, the four spins $j_i = \sqrt{p_{ii}}$ together with two (non-opposite) 3d dihedral angles $\Phi_i, i=1,2$ parametrize a closing intertwiner with normal vectors given in \eqref{nomdih}. The variables $\Phi_1, \Phi_2$ are related to the Euler angles by $\Phi_1 = \theta_2$ and $\Phi_2 = \theta_3$.

For a non-closing intertwiner, the closure constraint is violated, that is
\be
{\cal C} = \bm \varepsilon = (\varepsilon_1, \varepsilon_2, \varepsilon_3)  , 
\ee
with $\bm \varepsilon \neq \bm 0$. 
The four normal vectors are therefore unconstrained. 
Such non-closing intertwiners can also be constructed starting from a closed intertwiner and performing $\rm SO(3)$ rotations of some or all of the face unit normal vectors $\hat k_i$ while keeping their norms fixed. A rotation of a unit normal vector $\hat k_i(\theta_i,\phi_i)$ by Euler angles $(\delta \theta_i, \delta \phi_i)$ will result in a general vector $\hat k(\theta_i + \delta \theta_i, \phi_i + \delta \phi_i)$\footnote{Certain combinations of $(\delta \theta_i, \delta \phi_i)$ for $i=1,\cdots ,4$ can result in a closing intertwiner. }. Thus rotating one or more normal vectors of a closing intertwiner will result in a non-closing intertwiner as long as the closure constraint is violated. 

\bibliography{references.bib}

\end{document}

%% file: Style.tikzstyles
\tikzstyle{Node}=[fill=black, draw=black, shape=circle, scale=0.3pt]

\tikzstyle{line}=[-, fill=none, line width=1pt]
\tikzstyle{blockline}=[-, fill=black, line width=5pt]
\tikzstyle{boost}=[-, fill={rgb,255: red,128; green,128; blue,128}, draw={rgb,255: red,128; green,128; blue,128}, tikzit fill={rgb,255: red,128; green,128; blue,128}, tikzit draw={rgb,255: red,128; green,128; blue,128}, line width=5pt]
\tikzstyle{specialsu2}=[-, line width=5pt, fill=black, draw={rgb,255: red,0; green,0; blue,189}, tikzit fill=white, tikzit draw=black]
\tikzstyle{boxdash}=[-, line width=5pt, fill=black, dash pattern=on 2pt off 1pt]

%% file: gluenocoh.tikz
\begin{tikzpicture}
	\begin{pgfonlayer}{nodelayer}
		\node [style=none] (8) at (-1, 0.75) {};
		\node [style=none] (9) at (-1, 0.25) {};
		\node [style=none] (10) at (-1, -0.25) {};
		\node [style=none] (11) at (-1, -0.75) {};
		\node [style=none] (12) at (1, 0.75) {};
		\node [style=none] (13) at (1, 0.25) {};
		\node [style=none] (14) at (1, -0.25) {};
		\node [style=none] (15) at (1, -0.75) {};
		\node [style=none] (32) at (0, 1) {};
		\node [style=none] (33) at (0, -1) {};
	\end{pgfonlayer}
	\begin{pgfonlayer}{edgelayer}
		\draw [style=line] (8.center) to (12.center);
		\draw [style=line] (9.center) to (13.center);
		\draw [style=line] (10.center) to (14.center);
		\draw [style=line] (11.center) to (15.center);
		\draw [style=blockline, in=90, out=-90] (32.center) to (33.center);
	\end{pgfonlayer}
\end{tikzpicture}

%% file: resolution.tikz
\begin{tikzpicture}
	\begin{pgfonlayer}{nodelayer}
		\node [style=none] (1) at (-1, 0) {};
		\node [style=none] (5) at (1, 0) {};
		\node [style=Node] (6) at (-0.25, 0) {};
		\node [style=Node] (7) at (0.25, 0) {};
	\end{pgfonlayer}
	\begin{pgfonlayer}{edgelayer}
		\draw [style=line] (1.center) to (6);
		\draw [style=line] (7) to (5.center);
	\end{pgfonlayer}
\end{tikzpicture}

%% file: cstate.tikz
\begin{tikzpicture}
	\begin{pgfonlayer}{nodelayer}
		\node [style=none] (0) at (-0.25, 0) {};
		\node [style=Node] (2) at (0.5, 0) {};
	\end{pgfonlayer}
	\begin{pgfonlayer}{edgelayer}
		\draw [style=line] (0.center) to (2);
	\end{pgfonlayer}
\end{tikzpicture}

%% file: vertex_left.tikz
\begin{tikzpicture}
	\begin{pgfonlayer}{nodelayer}
		\node [style=Node] (0) at (3.5, 0.75) {};
		\node [style=Node] (1) at (3.5, 0.25) {};
		\node [style=Node] (2) at (3.5, -0.25) {};
		\node [style=Node] (3) at (3.5, -0.75) {};
		\node [style=Node] (4) at (2, -2.5) {};
		\node [style=Node] (5) at (1.25, -2.75) {};
		\node [style=Node] (6) at (0.5, -3) {};
		\node [style=Node] (7) at (-2.5, -2.25) {};
		\node [style=Node] (8) at (-2.75, -1.75) {};
		\node [style=Node] (9) at (-3, -1.25) {};
		\node [style=Node] (10) at (-3, 1.25) {};
		\node [style=Node] (11) at (-2.75, 1.75) {};
		\node [style=Node] (12) at (-2.5, 2.25) {};
		\node [style=Node] (13) at (-2.25, 2.75) {};
		\node [style=Node] (14) at (-0.25, 3.25) {};
		\node [style=Node] (15) at (0.5, 3) {};
		\node [style=Node] (16) at (1.25, 2.75) {};
		\node [style=Node] (17) at (2, 2.5) {};
		\node [style=Node] (18) at (-2.25, -2.75) {};
		\node [style=Node] (19) at (-0.25, -3.25) {};
		\node [style=none] (20) at (2.75, 1) {};
		\node [style=none] (21) at (2.75, -1) {};
		\node [style=none] (22) at (-0.75, 2.75) {};
		\node [style=none] (23) at (2, 1.75) {};
		\node [style=none] (24) at (-1.5, 2.5) {};
		\node [style=none] (25) at (-2.5, 0.75) {};
		\node [style=none] (26) at (-2.5, -0.75) {};
		\node [style=none] (27) at (-1.5, -2.5) {};
		\node [style=none] (28) at (-0.75, -2.75) {};
		\node [style=none] (29) at (2, -1.75) {};
	\end{pgfonlayer}
	\begin{pgfonlayer}{edgelayer}
		\draw [style=line, in=-30, out=-105, looseness=2.00] (14) to (13);
		\draw [style=line, bend left=60, looseness=2.00] (10) to (9);
		\draw [style=line, in=180, out=105, looseness=2.00] (4) to (3);
		\draw [style=line, in=180, out=-105, looseness=2.00] (17) to (0);
		\draw [style=line, in=105, out=-105] (16) to (5);
		\draw [style=line, in=-30, out=105] (6) to (11);
		\draw [style=line, in=30, out=-105] (15) to (8);
		\draw [style=line, bend right=15] (12) to (1);
		\draw [style=line, bend right=15] (2) to (7);
		\draw [style=line, in=30, out=105, looseness=2.00] (19) to (18);
		\draw [style=blockline] (20.center) to (21.center);
		\draw [style=blockline] (24.center) to (25.center);
		\draw [style=blockline] (26.center) to (27.center);
		\draw [style=blockline] (29.center) to (28.center);
		\draw [style=blockline] (21.center) to (20.center);
		\draw [style=blockline] (23.center) to (22.center);
	\end{pgfonlayer}
\end{tikzpicture}

%% file: gluing.tikz
\begin{tikzpicture}
	\begin{pgfonlayer}{nodelayer}
		\node [style=Node] (8) at (-1, 0.75) {};
		\node [style=Node] (9) at (-1, 0.25) {};
		\node [style=Node] (10) at (-1, -0.25) {};
		\node [style=Node] (11) at (-1, -0.75) {};
		\node [style=Node] (12) at (1, 0.75) {};
		\node [style=Node] (13) at (1, 0.25) {};
		\node [style=Node] (14) at (1, -0.25) {};
		\node [style=Node] (15) at (1, -0.75) {};
		\node [style=none] (32) at (0, 1) {};
		\node [style=none] (33) at (0, -1) {};
	\end{pgfonlayer}
 	\begin{pgfonlayer}{edgelayer}
 		\draw [style=line] (8) to (12);
 		\draw [style=line] (9) to (13);
 		\draw [style=line] (10) to (14);
 		\draw [style=line] (11) to (15);
 		\draw [style=blockline, in=90, out=-90] (32.center) to (33.center);
 	\end{pgfonlayer}
\end{tikzpicture}

%% file: vertex_right.tikz
\begin{tikzpicture}
	\begin{pgfonlayer}{nodelayer}
		\node [style=Node] (0) at (-3, 0.75) {};
		\node [style=Node] (1) at (-3, 0.25) {};
		\node [style=Node] (2) at (-3, -0.25) {};
		\node [style=Node] (3) at (-3, -0.75) {};
		\node [style=Node] (4) at (-1.5, -2.5) {};
		\node [style=Node] (5) at (-0.75, -2.75) {};
		\node [style=Node] (6) at (0, -3) {};
		\node [style=Node] (7) at (3, -2.25) {};
		\node [style=Node] (8) at (3.25, -1.75) {};
		\node [style=Node] (9) at (3.5, -1.25) {};
		\node [style=Node] (10) at (3.5, 1.25) {};
		\node [style=Node] (11) at (3.25, 1.75) {};
		\node [style=Node] (12) at (3, 2.25) {};
		\node [style=Node] (13) at (2.75, 2.75) {};
		\node [style=Node] (14) at (0.75, 3.25) {};
		\node [style=Node] (15) at (0, 3) {};
		\node [style=Node] (16) at (-0.75, 2.75) {};
		\node [style=Node] (17) at (-1.5, 2.5) {};
		\node [style=Node] (18) at (2.75, -2.75) {};
		\node [style=Node] (19) at (0.75, -3.25) {};
		\node [style=none] (20) at (-2.25, 1) {};
		\node [style=none] (21) at (-2.25, -1) {};
		\node [style=none] (22) at (1.25, 2.75) {};
		\node [style=none] (23) at (-1.5, 1.75) {};
		\node [style=none] (24) at (2, 2.5) {};
		\node [style=none] (25) at (3, 0.75) {};
		\node [style=none] (26) at (3, -0.75) {};
		\node [style=none] (27) at (2, -2.5) {};
		\node [style=none] (28) at (1.25, -2.75) {};
		\node [style=none] (29) at (-1.5, -1.75) {};
	\end{pgfonlayer}
	\begin{pgfonlayer}{edgelayer}
		\draw [style=line, in=-150, out=-75, looseness=2.00] (14) to (13);
		\draw [style=line, bend right=60, looseness=2.00] (10) to (9);
		\draw [style=line, in=0, out=75, looseness=2.00] (4) to (3);
		\draw [style=line, in=0, out=-75, looseness=2.00] (17) to (0);
		\draw [style=line, in=75, out=-75] (16) to (5);
		\draw [style=line, in=210, out=75] (6) to (11);
		\draw [style=line, in=150, out=-75] (15) to (8);
		\draw [style=line, bend left=15] (12) to (1);
		\draw [style=line, bend left=15] (2) to (7);
		\draw [style=line, in=150, out=75, looseness=2.00] (19) to (18);
		\draw [style=blockline] (20.center) to (21.center);
		\draw [style=blockline] (24.center) to (25.center);
		\draw [style=blockline] (26.center) to (27.center);
		\draw [style=blockline] (29.center) to (28.center);
		\draw [style=blockline] (21.center) to (20.center);
		\draw [style=blockline] (23.center) to (22.center);
	\end{pgfonlayer}
\end{tikzpicture}

%% file: gluing0.tikz
\begin{tikzpicture}
	\begin{pgfonlayer}{nodelayer}
		\node [style=Node] (8) at (-1, 0.75) {};
		\node [style=Node] (9) at (-1, 0.25) {};
		\node [style=Node] (10) at (-1, -0.25) {};
		\node [style=Node] (11) at (-1, -0.75) {};
		\node [style=Node] (12) at (1, 0.75) {};
		\node [style=Node] (13) at (1, 0.25) {};
		\node [style=Node] (14) at (1, -0.25) {};
		\node [style=Node] (15) at (1, -0.75) {};
		\node [style=none] (32) at (0, 1) {};
		\node [style=none] (33) at (0, -1) {};
	\end{pgfonlayer}
 	\begin{pgfonlayer}{edgelayer}
 		\draw [style=line] (8) to (12);
 		\draw [style=line] (9) to (13);
 		\draw [style=line] (10) to (14);
 		\draw [style=line] (11) to (15);
 	\end{pgfonlayer}
\end{tikzpicture}

%% file: gluing2.tikz
\begin{tikzpicture}
	\begin{pgfonlayer}{nodelayer}
		\node [style=Node] (8) at (-1, 0.75) {};
		\node [style=Node] (9) at (-1, 0.25) {};
		\node [style=Node] (10) at (-1, -0.25) {};
		\node [style=Node] (11) at (-1, -0.75) {};
		\node [style=none] (32) at (0, 0) {};
		\node [style=Node] (33) at (1.5, 0.75) {};
		\node [style=Node] (34) at (1.5, 0.25) {};
		\node [style=Node] (35) at (1.5, -0.25) {};
		\node [style=Node] (36) at (1.5, -0.75) {};
		\node [style=none] (37) at (0.5, 0) {};
		\node [style=none] (40) at (0.25, 0.5) {\small $\iota$};
		\node [style=none] (41) at (-1.75, 0) {\small $ h_i$};
		\node [style=none] (42) at (2.25, 0) {\small $ k_i$};
	\end{pgfonlayer}
	\begin{pgfonlayer}{edgelayer}
		\draw [style=line, bend left=45] (8) to (32.center);
		\draw [style=line, bend left=15] (9) to (32.center);
		\draw [style=line, bend right=15] (10) to (32.center);
		\draw [style=line, bend right=45] (11) to (32.center);
		\draw [style=line, bend right=45] (33) to (37.center);
		\draw [style=line, bend right=15] (34) to (37.center);
		\draw [style=line, bend left=15] (35) to (37.center);
		\draw [style=line, bend left=45] (36) to (37.center);
	\end{pgfonlayer}
\end{tikzpicture}

%% file: cohboost1.tikz
\begin{tikzpicture}
	\begin{pgfonlayer}{nodelayer}
		\node [style=Node] (8) at (-1.75, 0.75) {};
		\node [style=Node] (9) at (-1.75, 0.25) {};
		\node [style=Node] (10) at (-1.75, -0.25) {};
		\node [style=Node] (11) at (-1.75, -0.75) {};
		\node [style=Node] (12) at (1.75, 0.75) {};
		\node [style=Node] (13) at (1.75, 0.25) {};
		\node [style=Node] (14) at (1.75, -0.25) {};
		\node [style=Node] (15) at (1.75, -0.75) {};
		\node [style=none] (32) at (-1, 1) {};
		\node [style=none] (33) at (-1, -1) {};
		\node [style=none] (34) at (1, 1) {};
		\node [style=none] (35) at (1, -1) {};
		\node [style=none] (36) at (0, 1) {};
		\node [style=none] (37) at (0, -1) {};
	\end{pgfonlayer}
	\begin{pgfonlayer}{edgelayer}
		\draw [style=line] (8) to (12);
		\draw [style=line] (9) to (13);
		\draw [style=line] (10) to (14);
		\draw [style=line] (11) to (15);
		\draw [style=boost, in=90, out=-90] (32.center) to (33.center);
		\draw [style=boost, in=90, out=-90] (34.center) to (35.center);
		\draw [style=boxdash, in=90, out=-90] (36.center) to (37.center);
	\end{pgfonlayer}
\end{tikzpicture}

%% file: cohboost2.tikz
\begin{tikzpicture}
	\begin{pgfonlayer}{nodelayer}
		\node [style=Node] (0) at (-2.5, 0.75) {};
		\node [style=Node] (1) at (-2.5, 0.25) {};
		\node [style=Node] (2) at (-2.5, -0.25) {};
		\node [style=Node] (3) at (-2.5, -0.75) {};
		\node [style=none] (4) at (-1.5, 0) {};
		\node [style=Node] (5) at (2.5, 0.75) {};
		\node [style=Node] (6) at (2.5, 0.25) {};
		\node [style=Node] (7) at (2.5, -0.25) {};
		\node [style=Node] (8) at (2.5, -0.75) {};
		\node [style=none] (9) at (1.5, 0) {};
		\node [style=none] (12) at (0, 0.75) {};
		\node [style=none] (13) at (0, 0.25) {};
		\node [style=none] (14) at (0, -0.25) {};
		\node [style=none] (15) at (0, -0.75) {};
		\node [style=none] (16) at (1, 0) {};
		\node [style=none] (21) at (-1, 0) {};
		\node [style=none] (22) at (0, 1) {};
		\node [style=none] (23) at (0, -1) {};
		\node [style=none] (24) at (-1.25, 0.75) {\small $ \iota$};
		\node [style=none] (25) at (1.25, 0.75) {\small $ \iota'$};
	\end{pgfonlayer}
	\begin{pgfonlayer}{edgelayer}
		\draw [style=line, bend left=45] (0) to (4.center);
		\draw [style=line, bend left=15] (1) to (4.center);
		\draw [style=line, bend right=15] (2) to (4.center);
		\draw [style=line, bend right=45] (3) to (4.center);
		\draw [style=line, bend right=45] (5) to (9.center);
		\draw [style=line, bend right=15] (6) to (9.center);
		\draw [style=line, bend left=15] (7) to (9.center);
		\draw [style=line, bend left=45] (8) to (9.center);
		\draw [style=line, in=105, out=0] (12.center) to (16.center);
		\draw [style=line, in=150, out=0] (13.center) to (16.center);
		\draw [style=line, in=-150, out=0] (14.center) to (16.center);
		\draw [style=line, in=-105, out=0] (15.center) to (16.center);
		\draw [style=line, in=180, out=75] (21.center) to (12.center);
		\draw [style=line, in=180, out=30] (21.center) to (13.center);
		\draw [style=line, in=180, out=-30, looseness=0.75] (21.center) to (14.center);
		\draw [style=line, in=-180, out=-75] (21.center) to (15.center);
		\draw [style=boxdash] (22.center) to (23.center);
	\end{pgfonlayer}
\end{tikzpicture}